# Coupling of strain and magnetism in manganite-based complex oxide heterostructures

Surendra Singh[1,2,*]

[1]Solid State Physics Division, Bhabha Atomic Research Centre, Mumbai 40085, India

&

[2]Homi Bhabha National Institute, Anushaktinager, Mumbai 400094, India

*Email: surendra@barc.gov.in

**Abstract:**

Complex oxide perovskite thin films and heterostructures offer a wide range of properties originating from the intrinsic coupling between lattice strain and magnetic/electronic ordering. Altering properties of mixed doped manganites by lattice strain is very interesting not only for understanding solid-state phenomena but also for potential device applications. The strain has also been used to control the magnetoresistance and magnetic anisotropy effects in colossal magneto-resistive manganite thin films. This article reviews experimental, phenomenological, and theoretical analyses of the coupling of strain with electronic and magnetic properties of mixed valence manganite heterostructures. The influence of epitaxial strain on the magnetic properties of manganite films is measured using macroscopic magnetization measurements and shown mixed reports suggesting, both, an increase and decrease in ferromagnetic phases on the application of the strain. Using polarized neutron reflectivity (PNR), a simultaneous measurement of transport and magnetic properties of manganite thin films showed direct evidence of modification in the magnetic properties on the application of bending strain. The coupling coefficient of strain and magnetism of manganite heterostructures was estimated using PNR, which not only helped to understand the correlation of elastic strain with magnetism but also explained the condition of magnetic phase order change (second order to first order) in the phase-separated systems within a phenomenological Ginzburg–Landau theory. An overview is also provided of the current perspectives and existing studies on the influence of strain on structure, electronic, magnetic, magnetic anisotropy, phase coexistence and magnetocaloric properties of mixed valence manganite heterostructures. Based on the understanding of a diverse range of perovskite functionalities, detailed perspectives on how the coupling of strain and magnetism open up pathways toward the emergence of novel device design features including the different ways of applying uniform strain, are discussed.

**Keywords**: Manganites; strain and magnetism; polarized neutron reflectivity; complex oxide interfaces

# CONTENTS

# 1. Introduction

Strain is a widespread phenomenon related to the synthesis, fabrication, and application of all types of materials. For many centuries, strain has been applied via hydrostatic pressure and its effects on the material properties extensively investigated [1]. Strain fields in materials have always been one of the main features that influence their mechanical, electronic, and magnetic properties. Bulk oxides are brittle and crack under moderate strains ~ 0.1%. Chemical pressure in the form of isovalent cation substitution in materials is an approach to applying internal strain in complex oxide systems [2]. However, it may introduce disorder and local distortions in the materials. Complex oxide systems are so-called because their magnetic and electronic properties are closely coupled with the atomic structure and strain fields, and strain can be used to manipulate the quantum states of matter [3, 4].

Strain is one of the key parameters for understanding many physical phenomena at the nanoscale. Using epitaxy in thin oxide films, it is possible to achieve a strain far beyond where they would crack in bulk. The term 'epitaxy' is derived from the Greek word 'epi' means 'top' and 'taxis' means 'ordered'. It refers to the deposition of crystalline layers on the top of the crystalline substrate. The elasticity theory [5] of epitaxial growth suggests the development of tremendous stress of the order of 1 GPa (1000 N $mm^{-2}$) for a film with a lattice mismatch of 1% and Young modulus of order $10^{11}$ N $m^{-2}$. The film stress of that order produces an elasticity energy of the order of 0.1 eV per atom of hetero-epitaxial films. However, the epitaxial strain in thin films provides a potentially large biaxial strain. Under such strains, the properties of oxide films can be significantly altered, resulting in emerging phenomena at interfaces. A strain-induced shift in the magnetic [6-9], ferroelectric [10-19], and superconducting [20-22] transition temperatures of thin complex oxide films has been reported.

Complex oxides are a broad class of materials with charge, spin, orbital, and lattice degrees of freedom. Manganites are one of the best examples of complex oxides, which exhibit many interesting macroscopic properties, such as phase coexistence [2, 23-26], metal to insulator transition [27-29], colossal magnetoresistance [28, 30] etc., due to the strong coupling of these degrees of freedom [2, 23-25, 27-29, 31-35]. Recently, hybrid complex oxide (manganite) interfaces have attracted considerable attention because they provide a platform for exploring phenomena with new physics and functionalities, which cannot be realized with individual constituents [36-52]. The broken symmetry, charge density, repulsion energy, and bandwidth at the interface of correlated oxide heterostructures are important parameters that critically control the properties of correlated oxides, to tailor the electronic and magnetic structure of the heterostructures.

Reconstruction of bonds and orbitals at the interfaces of complex oxides is another important phenomenon that plays a key role in determining the exchange interaction and subsequently the magnetism in heterostructures. Epitaxial strain is a vital parameter in determining orbital occupancy at interfaces and can modify the occupation of orbitals at the interface [53]. In transition metal oxide systems (like manganites), the in-plane tensile and compressive strains promote different orbital occupations [53]. Thus, modification in the orbital occupancy due to epitaxial strain influences spin-spin interaction and contributes to the enhancement of magnetism in complex oxide heterostructures. Epitaxial strain-induced effects in heterostructures have been a powerful tool for the creation of new ground states because they affect properties such as the electronic band gap [54], the behaviour of correlated systems [55], thermal conductivity [56], multiferroicity [57, 58], catalytic properties [59-62], charge transport [63-65], and magnetic properties [66-71].

Controlling the electronic and magnetic properties of complex oxide materials employing external stimuli, such as strain, is critical for functional devices [72-75]. Magnetoelastic coupling is an important phenomenon used to control the magnetic properties of materials through a stress/strain field [76, 77]. Strain-induced reductions in magnetization and order temperature for oxide films were observed earlier [65, 66], suggesting that magnetoelastic coupling can also induce phase transitions in oxide compounds [78-81]. The influence of strain on the magnetic and magnetotransport properties of complex oxide heterostructures, especially the coupling of epitaxial strain and magnetism in manganite films, is dramatic and well-documented in the literature. However, experiments measuring the response of magnetism to epitaxial strain in manganite films have provided mixed results regarding the coupling of strain and magnetism [82, 83]. There has been a successful effort to apply stress to create laterally uniform and well-controlled strain fields in manganite oxide systems and to examine how their magnetic and electronic properties are affected [84-88].

The coexistence of distinct metallic (ferromagnetic, FM) and insulating (antiferromagnetic, AFM and charge order) electronic phases within a chemically homogeneous sample of perovskite manganite and its dependence on the strain field is another important study that convincingly shows the sensitivity of strain in manganite films [23, 26, 29, 89-94]. Using the electron holography imaging technique, Marin *et al.*,[92] reported the epitaxial strain-driven segregation of the non-ferromagnetic layer (AFM character) at the top surface of a ferromagnetic layer of $La_{1-x}Ca_xMnO_3$ (LCMO) ($x$ = 0.33) films, which was chemically and structurally homogeneous at room temperature. Theoretical analysis has shown that the strain field can tune the metal-insulator phase fraction in the $La_{1-x-y}Pr_yCa_xMnO_3$ film [29]. Another theoretical study suggested that the observed micrometer scale phase separation (PS) owes its existence to extrinsic causes such as strain [94]. Experimentally, it is observed that unstrained $La_{1-x-y}Pr_yCa_xMnO_3$ films grown on $NdGaO_3$ (NGO) behave quite similarly to bulk material, but the strained films grown on $SrTiO_3$ (STO) show melting of the insulating phase to the metallic phase at low temperatures [93]. However, large-scale PS and a metastable glass-like state are observed in all films despite differences in substrate-induced strain.

The development of state-of-the-art thin film growth techniques for growing chemically homogeneous complex oxide heterostructures and the availability of advanced nondestructive characterization tools, such as x-ray reflectivity (XRR) and polarized neutron reflectivity (PNR), have made it possible to investigate the structure and magnetic properties of buried interfaces with a depth resolution of sub-nanometer length scale, not probed earlier [95-104]. PNR, in particular, a unique technique to reveal the magnetic structure of interfaces [105-110], which can adopt a complex sample environment [99, 100], was successfully used to study the coupling of strain (applied by bending the film) and magnetism by simultaneously measuring the PNR signal and transport data for manganite films as a function of strain, temperature, and magnetic field [85-87]. The coupling coefficient of strain and magnetism (change in magnetization on the application of bending strain) of manganite films estimated using PNR has helped to understand the correlation of elastic strain with magnetism [85] and also explained the condition of magnetic phase order change (second order to first order) in the phase-separated systems within a phenomenological Ginzburg–Landau theory [111].

This article reviews recent experimental findings in the realization of strain in a controllable manner in perovskite complex oxides, focusing particularly on mixed valence manganite heterostructures, and the influence of strain on the electronic and magnetic properties of these systems. The strain-centric design of perovskite oxide material systems for the discovery of novel properties not attainable in the absence of strain has been employed to understand the coupling of strain and magnetism. In addition to the coupling of strain with the electronic, magnetotransport, and magnetic properties of manganite oxide heterostructures, the effect of strain on magnetic anisotropy [112, 113], phase coexistence [29, 114] and magnetocaloric effect [115-118] in manganite thin films and heterostructures are also discussed in this article.

This review article begins with a brief overview of the correlation of the structure-electronic-magnetic properties of complex oxide (manganite) systems, describing the strain-driven interaction and its influence on these properties. The influence of strain (hydrostatic) on the structure and electronic and magnetic properties of bulk manganites is also briefly compared. Subsequently, the importance of complex oxide heterostructures and characterization techniques, especially those suited for interfaces, are described. In addition, different modes of applied strain used in the literature for studying the influence of strain on the magnetic properties of complex oxide heterostructures are reviewed. Next, the coupling of strain with transport and magnetic properties of manganite heterostructures are discussed, which have been categorized mostly for three different strain methods employed, e.g., epitaxial (biaxial strain), piezoelectric (controlled strain by phase transition or electric field on the ferroelectric substrate), and bending (uniform uniaxial strain). This includes defining the coupling of bending strain and magnetism and its correlation with the phenomenological approach for defining different magnetic phase transitions in these systems. The last part of this review will be devoted to highlighting the influence of strain on phase coexistence, anisotropy, and magnetocaloric properties of manganite heterostructures, which are important for understanding the basic mechanism of these phenomena as well as for achieving improved properties for possible applications. Briefly, conclusions and future outlooks are then presented.

## 2. Structure and electronic-magnetic interactions of complex oxides

Complex transition metal oxides show several crystalline structures with an incredible variety of physical phenomena. In particular, perovskite [119] complex oxides offer high dielectric permittivity, piezoelectricity, pyroelectricity, ferroelectricity, high-temperature superconductivity, and colossal magnetoresistance etc., due to strong interaction between different degree of freedoms [6, 9, 10, 16, 19, 22]. In complex oxides, the magnetic and electronic properties are closely coupled with the atomic structure and strain, which produces a drastic nonlinear response to the environment. Recent advances in growth techniques for oxide heterostructures not only showed improved structural quality but also enabled the fabrication of artificial multifunctional materials (heterostructures and multilayers). Simultaneously, the interfaces of these artificial heterostructures have exposed a wealth of phenomena that can be influenced by varying the strain of the heterostructures. The exciting sensitivity of the electronic and magnetic properties of complex oxide heterostructures to structural distortions and crystal chemistry offers many routes for controlling and engineering new functionalities in these materials. Thus, it is only natural to expect that interfaces between different complex oxides will give rise to interesting and unexpected emerging phenomena. This section is devoted to a review of the structural properties of perovskite complex oxides, including strain-induced distortion in these structures, which influences their electronic and magnetic properties.

## 2.1 Strain-structure properties correlations of perovskite oxides

Here we discuss the correlation between strain and structural properties of complex oxides with a perovskite [30, 119] ($ABO_3$, with different *A* and *B* cations, Fig. 1) crystal structure, where the tunability of the physical properties can be enforced through lattice distortions (strain), and microstructural/morphological changes (e.g., grain size, porosity) [53, 120-122]. An appropriate selection of *A*- and *B*-site cations can also dramatically affect structural, electronic, magnetic, optical, polar, and other properties. In addition, the wide range of physical phenomena manifested in these complex oxide materials is highly controlled by the electronic structure and coordination chemistry of the cationic species.

In the perovskite structure, *A*-site cations, which are typically alkali, alkaline, or lanthanide metals, are surrounded by twelve oxygen anions and *B*-site cation of smaller size, typically a transition metal is surrounded by six oxygen anions forming $BO_6$ octahedra [121, 122]. In general, the spherical shape of the valence orbitals of the *A*-site cations (e.g. alkali and alkaline), the majority of the electronic and magnetic properties of the perovskites are related to the physics associated with the *B*-site cation and oxygen anions of the $BO_6$ octahedra. However, the size of the *A* and *B*-site cations induces structural distortion and tilting of the $BO_6$ octahedra, and the effective deformation is determined by Goldschmidt's rule [123], which is described as the Goldschmidt tolerance factor, $t = (r_A + r_O)/\sqrt{2}\,(r_B + r_O)$, where $r_A$, $r_B$, and $r_O$ are the ionic radii of the *A*, *B* (cations), and O (anion) atoms. The tolerance factors range between 0.96 < $t$ < 1.0 for a cubic arrangement of the ions and no tilting of the $BO_6$ octahedra occurs [124]. For example the STO perovskite structure, with $r_{Sr}$ = 1.44 Å, $r_{Ti}$ = 0.605 Å, and $r_O$ = 1.40 Å, has a tolerance factor of $t$ = 1.00 [125]. The smaller tolerance factor ($t$ < 1), advises a greater deviation of the *B*-O-*B* angle (Fig. 1 (b)) from 180°. The structure-properties correlation lies in the key role of the *B*-O-*B* bond. The modulation of the octahedral shape and size is associated with a change in the bond length and thus leads to highly versatile changes in the properties of perovskite complex oxides [119, 120].

Another important parameter of perovskites is the Jahn-Teller (JT) distortion, which is at the heart of the various electronic and magnetic properties exhibited by perovskites. In 1937,

Jahn and Teller used the symmetry approach and showed that geometrical distortion of nonlinear molecules exhibiting orbital degeneracy led to a lower-energy structure with nondegenerate orbitals [126]. It is intimately related to electronic degrees of freedom, as it manifests to remove electronic degeneracy, resulting in orbital ordering, which in turn can affect magnetic ordering. In transition metal (*B*) cations, the splitting of the $d$-orbitals (two $e_g$ and three $t_{2g}$) can often lead to degenerate electron configurations that are subject to the JT effect. In practice, ions with degenerate occupancy of the $e_g$ orbital usually exhibit strong JT distortions. The JT phenomena for solids are extensively given by Goodenough [127]. It plays an important role in exhibiting and understanding the colossal magnetoresistance in doped manganites [128], superconductivity [129, 130], etc. Therefore, choosing an appropriate cation and various structural distortions occurring in perovskites provides enormous opportunities to explore multifunctional properties based on the structure-property correlation.

Depending on the type of transition metal cations, the perovskite complex oxides can be categorized into different groups, e.g., manganites, nickelates, cobaltites, ruthenates, etc. The perovskite manganites are represented by the general formula $RE_xA_{1-x}MnO_3$ (*RE* and *A* are rare earth metal and alkaline earth metal cations, respectively), and they exhibit magnetic, colossal magnetoresistance (CMR), half-metallic, charge-ordering, metal-insulator transition (MIT), and phase coexistence behaviours [28, 53, 120, 121, 125, 128]. Perovskite nickelates ($RE_xA_{1-x}NiO_3$) exhibit MIT and AFM-paramagnetic phase transitions, and thin films of nickelates are interesting systems for many technological applications [131-133]. Perovskite RE cobaltites ($RECoO_3$) and ruthenates ($RERuO_3$) are also of great interest because of their unusual magnetic and transport properties [134, 135].

Manganites are one of the archetypes in the field of functional complex oxides because they combine numerous exotic properties discussed above and their perovskite crystal structure allows the growth of epitaxial thin films and the fabrication of high-quality heterostructure devices. A large number of manganite-based systems, such as parent (or undoped) manganite, $REMnO_3$ (*RE* = La, Pr, Tb, Nd, Ho, Dy, Y) [136-139], layered (single and double layer) manganites [140, 141], and doped manganites ($RE_xA_{1-x}MnO_3$), etc., have been studied extensively in the literature [28, 30, 31, 33, 136, 142]. One important aspect of the perovskite structure of these manganites ($RE_xA_{1-x}MnO_3$) is their capacity to accommodate both divalent and trivalent cations in the *A* position. The substitution of divalent (e.g. Ca, Ba, Sr, Pb etc.) and trivalent (e.g. Cs, Sb, Te etc.) cations in *manganites* donates holes and electrons, respectively; thus, the resultant manganites are called hole-doped and electron-doped manganites [143-145].

Doping with both divalent and trivalent cations gives mixed-valence manganites [136], such as $(La_{1-y}Pr_y)_{1-x}Ca_xMnO_3$. Considerable experimental and theoretical studies on the electronic and magnetic properties of parent (or undoped) manganite, $RE$MnO$_3$ ($RE$ = La, Pr, Tb, Nd, Ho, Dy, Y), have been reported, showing an orthorhombic (*o*) perovskite structure with different magnetic phases and mostly AFM structures [136-139]. Investigation of multiferroic properties of thin films of *o*-$RE$MnO$_3$ manganites, particularly $TbMnO_3$, $YMnO_3$, etc. [139, 146-149], have been explored earlier and have shown both ferroelectric and AFM properties. The influence of strain on the multiferroic properties of *o*-$RE$MnO$_3$ manganite films has also been studied both experimentally and theoretically to understand the mechanism underlying the development of magnetoelectric coupling and multiferroic order [146-149]. In this article, we will primarily discuss the effect of strain on the electronic and magnetic properties of mixed-valence manganite ($RE_xA_{1-x}MnO_3$) films, which show low-temperature FM order compared with the multiferroic $RE$MnO$_3$ manganite films that show mostly AFM order.

Mixed valence manganites present a rich phase diagram because of the close competition of different interactions, and one of the features of these compounds is the electron bandwidth (*W*). *W* is characterized by the overlap between the Mn-3*d* and O-2*p* orbitals and can be described as [150]:

$$W \propto \frac{\cos[0.5(\pi - <\beta>)]}{d_{Mn-O}^{3.5}} \qquad (1)$$

where $\beta$ and $d_{Mn\text{-}O}$ are the Mn-O-Mn bond angle and Mn-O bond length, respectively. Half-doped manganites (with 0.5 electrons per Mn), such as $Pr_{1-x}Ca_xMnO_3$ (PCMO) ($x$ = 0.5) [151], $La_{1-x}Ca_xMnO_3$ (LCMO) ($x$ = 0.5) [152], $Pr_{1-x}Sr_xMnO_3$ (PSMO) ($x$ = 0.5) [153], and $Nd_{1-x}Sr_xMnO_3$ (NSMO) ($x$ = 0.5) [154], usually show the charge/orbital ordering. These manganite systems show a drastic change in resistivity and phase coexistence on the application of small external perturbations like magnetic/electric field, temperature, strain, and light. Figure 2 represents the schematic of the phase diagram of a manganite as a function of hole doping and bandwidth (*W*) [155]. In contrast to manganite with a narrow bandwidth (*W* ~ fraction of eV [136]) (e.g., LCMO ($x$ = 0.5), PCMO ($x$ = 0.5), etc.), manganite with a relatively wider bandwidth (*W* ~ eV [136]) (e.g., PSMO ($x$ = 0.5), NSMO ($x \geq$ 0.5), etc.) generally shows the following sequence of spin/charge ordering on hole doping: insulating *A*-type AFM → (metallic) FM → (metallic) *A*-type AFM → insulating *C*-type, and finally insulating *G*-type AFM state.

The influence of strain on manganite thin films is important to distinguish its properties from those of bulk samples. Strains modify both in-plane and out-of-plane lattice parameters and thus affect electronic properties through orbital ordering. The strain-induced distortion also produces asymmetry in the local structure of the manganite films. Using polarized X-ray absorption spectroscopy (XAS), the strain-induced local modification around the manganese ions in $La_{1-x}Sr_xMnO_3$ (LSMO) ($x = 0.3$) films grown on different substrates was studied [156]. XAS measurements showed significant modifications in the coordination shell around the manganese atoms and suggested that biaxial strain is locally accommodated in the coordination shell, by distortion of the $MnO_6$ octahedron, without change in the tilt angle. In addition, the modifications of the X-ray absorption near-edge spectra were correlated to modifications in the average Mn-O bond distance and distortion of the $MnO_6$ octahedra. Strain-induced crystal structure change in ultrathin PSMO ($x = 0.3$) films grown on (001) $LaAlO_3$ (LAO) substrates were studied using transmission electron microscopy (TEM) [157]. It was found that the films are highly uniform and defect-free and that they are coherently strained, resulting in a tetragonal expansion perpendicular to the film plane. A change in the crystal structure from the ordered orthorhombic of bulk materials to a simple tetragonal perovskite in the thin film was observed, and the variation of the tetragonality with distance from the interface was also determined from high-resolution TEM images [157]. Millis *et al.* [158], made theoretical predictions about the dependence of the ferromagnetic Curie temperature ($T_c$) of the manganese perovskites with strain and suggested that $T_c$ is extremely sensitive to strain. This study also provided evidence of electron-phonon coupling to the CMR phenomena and suggested that strain is an important variable for the design of devices based on manganite thin films.

## 2.2 Electronic interaction of perovskite oxides and strain

Electron filling of 3*d*-orbitals of transition metal in $ABO_3$ perovskite oxides and tailoring electron occupancy with relative orientation in crystal lattices play a crucial role in determining the electronic properties of these oxides [53, 73]. Figure 3(a) shows a schematic of the energy splitting of the *d* orbital of a single cation in the oxygen octahedral, which has two important energy terms [136], (a) spin-orbit coupling and (b) crystal field ($\Delta_{cf}$ ~ 1 eV). The spin-orbit coupling term originates from the relativistic effect in quantum physics and describes the interaction of the electron's spin with its motion (orbital)[159]. The orbital energy level for 3*d* transition metal is mainly determined by the second term, the crystal field, because the spin-orbit coupling term is very weak and is usually neglected. The interaction between the positively charged metal cation and the negatively charged ligand is responsible for the crystal field. The crystal field interaction partially lifts the 5-fold degeneracy of the *d*-orbitals and splits the 3*d*-orbitals into a degenerated $t_{2g}$ triplet and a degenerated $e_g$ doublet with (*xy* / *yz* / *xz*) and ($x^2$-$y^2$ / 3$z^2$-$r^2$) symmetries, respectively [53, 159]. Furthermore, lowering the crystal field symmetry (by distortion) from cubic $ABO_3$ perovskites breaks the degeneracy of $x^2$-$y^2$ and 3$z^2$-$r^2$ orbitals (~ 0.1 eV). A common driving force for such symmetry breaking is the half-filling of $e_g$ orbitals and strong electron-phonon coupling, which leads to a JT distortion of the $BO_6$ octahedron. The distortion of the $BO_6$ polyhedron produces asymmetric *B*-O bonds that break the ($x^2$-$y^2$ / 3$z^2$-$r^2$) degeneracy.

The hole doping level (*x*) and the strength of the JT distortion in bulk LSMO determine the electron occupancy in the ($x^2$-$y^2$ / 3$z^2$-$r^2$) orbitals [53]. The partially filled ($x^2$-$y^2$ / 3$z^2$-$r^2$) orbitals in bulk LSMO ($x$ = 0.33) are responsible for the metallic character and FM coupling. Feng *et al.*, [160] studied the effect of strain on the transport and magnetic properties of LSMO thin films and observed the splitting of $e_g$ ($x^2$-$y^2$ / 3$z^2$-$r^2$) orbitals, similar to JT distortion, which shows a drastic change in these properties. Figure 3(b) shows the schematic of the splitting of $e_g$ orbitals and the $MnO_6$ octahedral distortions under strain. Using synchrotron-based resonant X-ray techniques; it has been shown that tensile strain in LSMO thin film favours $x^2$-$y^2$ occupancy, whereas compressive strain favours 3$z^2$-$r^2$ occupancy [161-164]. These strain-driven electronic structures have been attributed to the elongation and compression of metal-oxygen bond distances to the equatorial and apical oxygen, respectively [165].

Systematic experimental studies on the effect of thickness (strain) dependence on the magnetotransport properties of manganite thin films suggested that MIT is suppressed in very thin (highly strained) films [166-169]. Using element-specific XAS at the Sr *K*-edge, Qian *et al.*, [168] observed a change in local structure and electronic properties in the form of a reduction of the Mn-O-Mn bond angles in thin films of thickness below 100 nm. The spin-polarized measurements revealed a splitting of the Mn 3*d* $e_g$ state in the strained region of the films. X-ray absorption near edge structure (XANES) measurements from PSMO ($x$ = 0.33) films having different thicknesses on (001) LAO substrate showed in-plane compressive and out-of-plane tensile strains, which changes both the lattice constant and $MnO_6$ octahedral rotation and thus influences the electronic properties of the films [169]. The strain tunes the bond-angle/bond length, which contributes to reconstructing and inducing point/planar defects and therefore affects the electronic, orbital states at the interfaces. The variation of the metal-oxygen bond length through *B*-O-*B* angle distortion also leads to changes in electronic structures in the form of the width of the *d*-states near the Fermi level [170]. Strain can induce metallicity in an insulating $LaTiO_3$ (LTO) film by tuning the *B*-O-*B* angle [171]. A compressively strained ($\varepsilon$ ~ -1.6 %) LTO film epitaxially grown on (001) STO substrate shows metallic behaviour in contrast to bulk LTO ($\varepsilon$ ~ 0%) and LTO films grown on $GdScO_3$ (GSO) ($\varepsilon$ ~ 0%) [171]. Thus, the strain-driven modification of the electronic bandwidth of perovskite oxides leads to a variety of electronic and magnetic properties, which may open new pathways for applications.

## 2.3 Magnetic interactions in perovskite oxides and strain

The magnetic moment of a single transition metal cation, a basic unit of magnetic order and, magnetic interactions between different cations, which determine the long-range ordering, are important for understanding the overall magnetism of perovskite oxides. Considering the spin degree of freedom, in addition to the splitting of *d*-orbitals due to the crystal field, as described in Subsection 2.2, the degeneracy associated with spins (spin-up and spin-down state) can be lifted by Hund's rule, which is attributed to the repulsion of electrons in the same orbital [172]. Thus, the spin occupation states of the 3*d* transition metal cations are determined by the competition between Hund's coupling ($J_{\text{Hund}}$) and crystal field ($\Delta_{\text{cf}}$) effects [Fig. 3(c)]. Depending on the dominance of these factors, the electrons occupy the high/low spin state. In the case of $Mn^{3+}$ and $Fe^{3+}$, Hund's coupling is dominant; hence the electrons occupy higher spin states. The crystal field favours the low spin state in the case of $Ni^{3+}$. However, these low and high spin states are tunable through strain due to the modification of the crystal field [9]. The occupancy of atomic electronic *d*-orbitals for perovskites, $t_{2g}$ and $e_g$, with electrons that are consistent with Hund's and crystal-field splitting rules, has been described in several review articles and textbooks [28, 53, 173, 174].

The magnetic coupling between the two nearest cations (transition metal) in perovskites is associated with the oxygen anion, and is known as superexchange, which determines the long-range magnetic ordering in these systems. The concept of superexchange was initiated by Anderson [175], Goodenough [176], and Kanamori [177] by employing an indirect exchange mechanism between moments mediated by nonmagnetic atoms and considering their orbital, hybridization, and symmetry contributions. The fundamental principle of superexchange theory is electron hopping between the cation *d* electron and anion *p* electron. Depending on the number of *d* electrons of the cation and the spin configuration (high or low), the sign of the coupling between the *d* and *p* electrons of anions can be negative (antiparallel) or positive (parallel). Thus, the effective coupling between two nearest cations can be ferromagnetic or antiferromagnetic, and a detailed combination of $d^n$ states can be found in reference [177].

Zener [178] proposed a magnetic coupling in mixed valence manganites (both $Mn^{3+}$ and $Mn^{4+}$ cations exist), called double exchange (DE) interaction, which involves the simultaneous transfer of one electron from the $e_g$ orbital of $Mn^{3+}$ to a 2$p$ orbital of oxygen, and from that of 2$p$ orbital of oxygen to an $e_g$ orbital of the neighbouring $Mn^{4+}$, as shown in Fig. 3 (d). The electron is thus delocalized over the entire structure, and thus the hopping of electrons via DE leads to the conducting channel in the manganites. Following Hund's rule, the hopping electron has the same spin orientation because the spin flips are not allowed during the hopping process, resulting in a long-range ferromagnetic order. The DE theory describes the electron hopping rate between the two neighbouring Mn sites as [179] $t_{eff} = t_0 \cos(\theta/2)$, where $\theta$ is the angle between two neighbouring spins. This advises that hopping is the largest for $\theta = 0^o$ (parallel spin) and zero for $\theta = 180^o$ (antiferromagnetic alignment of spins). However, to understand the physics of manganites better, in-depth knowledge of the polaron effect due to strong electron-phonon coupling is required, as considering only DE interaction cannot explain some experimental results quantitatively [180]. Figure 3(e) shows a schematic of another possible magnetic superexchange interaction between two magnetic cations, which results in antiferromagnetism. In the case of manganites, the electron is shared between an occupied O 2$p$ orbital and a vacant Mn $e_g$ orbital, which can be regarded as a virtual transfer of the electron. Thus, in superexchange coupling, the electrons do not hop between different atomic sites, and it is only suitable to explain magnetic coupling without charge transfer.

Strains in the perovskite complex oxide structure can modify octahedral tilting and rotations, which facilitate/stabilize structural deviations from ideal cubic symmetry [119]. Thus, the modification of electronic and magnetic interactions in complex oxides under strain is primarily caused by two different but competing effects: (a) the shortening of bonds, which increases the hybridization and transfer integral, and (b) the rotation of bonds, which decreases the transfer integral. Manganites are good candidates among complex oxides for tuning magnetic interactions by controlling the electron occupancy of the Mn 3$d$ orbitals via strain [181]. The strain-induced elongation, compression, or rotation of the $MnO_6$ units leads to crystal field splitting of the $e_g$ ($x^2$-$y^2$ and $3z^2$-$r^2$) orbitals and complex orbital reconstructions [53, 162]. Strain-dependent studies on LSMO films suggested that tensile strain favours $x^2$-$y^2$ occupancy ($A$-type AFM structure) and compressive strain favours $3z^2$-$r^2$ (C-type AFM structure) [160, 182, 183].

## 3. Strain-electronic-magnetism correlation in bulk complex oxides

In perovskite complex oxides, oxide ions ($O^{2-}$) are bonded with multiple cations, and their chemical flexibility and internal structural distortions provide a variety of electronic and magnetic properties that are highly tunable [120, 174]. In rare-earth-based complex oxides, the substitution of the rare earth element with another element yields a chemical pressure, which can affect the magnetic properties. The coupling between strain and structural or magnetic properties in bulk complex oxides (manganites) has been mostly studied as (a): a correlation of strain and defect formation energies and (b): hydrostatic pressure-induced effects. Defects cause isotropic or anisotropic changes in volume [184] and form non-stoichiometry in substances. The most important defect-induced expansion in solid materials is chemical expansion or chemical strain, which arises from oxygen vacancies [185-187]. The extra electron remaining after the removal of a neutral oxygen atom reduces the surrounding cations and increases their radii. In perovskites, oxygen vacancies lead to a reduction in the valence state of the *B* cation and produce chemical expansion. The lattice strain ($\varepsilon$) due to this chemical expansion is given as follows [188]:

$$\varepsilon = \frac{\Delta d}{d_0} = \alpha_c \Delta\delta \qquad (2)$$

where $d_0$, $\Delta d$, $\alpha_c$, and $\Delta\delta$ are lattice constant, change in the lattice constant of perovskite, coefficient of chemical expansion, and oxygen deficiency, respectively. The chemical pressure is believed to induce strong JT distortion in PCMO and suppress double exchange, leading to a dramatic difference in the magnetic structure [189]. The detailed structural analysis of manganite complex oxides showed that the chemical or internal pressure only reduces the Mn-O-Mn bond angle ($\beta$) and the Mn-O distance ($d_{Mn\text{-}O}$) does not change significantly [190]. Thus, chemical pressures involve a change in interaction resulting in different magnetic phases.

Uehara *et al.* [2], studied the effect of chemical pressure on $T_c$ in $La_{5/8-x}Pr_xCa_{3/8}MnO_3$ manganites by varying the concentration of trivalent rare-earth ions (Pr) of different sizes in the perovskite structure of manganite [191, 192], without affecting the valency of the Mn ions. The low-temperature phase diagram of manganites showed the dominance of two phases: a ferromagnetic metal (FMM) and an AFM charge-ordered insulator (COI). This study directed that the competition between these two phases gives rise to the observed sensitivity of these materials to external perturbations such as magnetic, pressure, and strain [2]. Collado *et al.*

[193], synthesized $La_{5/8-x}Pr_xCa_{3/8}MnO_3$ with $0.0 \leq x \leq 5/8$ manganites by the ceramic method and observed the evolution of the magnetization and resistivity curves with the smooth variation of the cell parameters upon doping the manganites with Pr atoms. This study suggested that a change in chemical pressure on doping Pr elements in manganite provided a rich magnetic phase with an increase in the FMM phase to a higher temperature. The reports suggested that the electronic phase separation (EPS) and ground states in (La, Pr)$CaMnO_3$ manganite systems can be well-tuned by both chemical doping and epitaxial strain. Thus, the doping level of Pr (chemical pressure) can be used to smoothly tune the bulk materials from the FMM to the antiferromagnetic insulating (AFI) ground state [193, 194]. Recently, Torres *et al.* [195], correlated the concentration of Pr, Co, and Ni elements in $La_{0.67-x}Pr_xCa_{0.33}MnO_3$ at $0.13 \leq x \leq 0.67$ and $LaMn_{1-x}(Co/Ni)_xO_3$ at $0.1 \leq x \leq 0.5$ manganites with the strain developed in these systems. An increase in strain strength with increasing concentration of these elements in the manganites was observed. The study also suggested that an increase in Pr concentration in $La_{0.67-x}Pr_xCa_{0.33}MnO_3$ manganite reduces the compressive biaxial strain in the system, whereas an increase in Co and Ni concentration in $LaMn_{1-x}(Co/Ni)_xO_3$ increases the biaxial tensile strain. The strain values corresponding to different concentrations of the dopant in these manganites were estimated using XRD measurements [195]. Figures 4 (a) and (b) show the variation of $T_c$ with strain (dopant concentration) for bulk $La_{0.67-x}Pr_xCa_{0.33}MnO_3$ at $0.13 \leq x \leq 0.67$ and $LaMn_{1-x}Co_xO_3$ at $0.1 \leq x \leq 0.5$ manganites, respectively, and indicate that an increase in biaxial tensile strain increases the $T_c$ and MIT temperature ($T_{MIT}$) [195].

While the chemical pressure induced by doping in manganite bulk oxides showed modification in transport and magnetic properties, Hwang *et al.*, [196] studied the effect of applied hydrostatic pressure on the electric transport and magnetotransport properties of bulk PCMO ($x$ = 0.3) [Figs. 5 (a) and (b)]. The resistivity as a function of temperature at different applied pressures was measured in the absence of a magnetic field, which showed a decrease in resistivity and a shift in the $T_{MIT}$ to a higher temperature with increasing pressure [Fig. 5 (a)]. The study also showed that the $T_{MIT}$ further shifted to a higher temperature at a magnetic field of 5 T than that at 0 T [Fig. 5 (b)]. In addition, a strong temperature hysteresis was observed at $T_{MIT}$, suggesting a first-order transition. Inset of Fig. 5 (b) represents a decrease in resistivity by eight orders of magnitude in the 5 T field for PCMO bulk manganite. A strong pressure dependence of $T_{MIT}$ in both the 0 and 5 T fields was observed, which was accompanied by an extremely large change in resistivity or magnetoresistance as a function of applied pressure. The effect of quasi-hydrostatic-pressure on electrical resistivity for a series of manganese perovskites $L_{2/3}A_{1/3}MnO_3$ (L = Pr, Sm, Nd, Y, La; A = Ca, Sr) up to 11 kbar was reported earlier [197], suggesting a strong modification in properties. The quasi-hydrostatic pressure was applied using a clamp-type piston-cylinder cell with silicon-organic liquid as the pressure-transmitting medium. Temperature dependence of the normalized resistance *R*(*T*)/*R* (300 K, 1 bar) at several pressures for $Sm_{1-x}Sr_xMnO_3$ (SSMO) ($x$ = 0.33) [Fig. 5 (c)] and LCMO ($x$ = 0.33) [Fig. 5 (d)] bulk samples were measured and found that the application of pressure shifts $T_{MIT}$ towards higher temperature (d$T_c$ /d$P$ is positive), suggesting that the pressure stabilizes the metallic phase in these systems. These results were attributed to the particular dependence of the bandwidth on the Mn-O-Mn bond angle and its compressibility under pressure.

Patrick [198], in his pioneering paper, studied the effect of hydrostatic pressure on the $T_c$ of ferromagnetic systems and proposed the following relation for quantifying the change in $T_c$ with pressure.

$$\frac{1}{T_c}\frac{dT_c}{dP} = \frac{1}{T}\frac{dV}{dH}\bigg/\left(\rho\frac{d\sigma}{dT} - \frac{3\alpha}{K}\frac{dV}{dH}\right) \tag{3}$$

where *V*, *P*, *T*, *H*, *ρ, α, σ*, and *K* are fractional changes in volume, pressure, temperature, magnetic field, density, linear thermal expansion coefficient, magnetization per gram, and compressibility, respectively. Both sizable negative and positive pressure variations ($dT_c/dP$) for $T_c$ were obtained for ferromagnetic materials with different exchange interactions. A list of the bulk magnetic systems [192, 197, 199-211] and the variation of $T_c$ with applied pressure are given in Table 1.

The case of perovskite manganites, where the DE interaction is the dominant interaction for long-range magnetism, showed a positive shift [192, 203, 204] in $T_c$ on the application of hydrostatic pressure with a $dT_c/dP$ = +36.7 K/GPa for the bulk LCMO ($x$ = 0.21) manganite [203]. Figure 6 (a) shows the variation of $T_c$ with applied hydrostatic pressure using a standard piston-clamp pressure cell for a bulk LCMO ($x$ = 0.40), which suggests a positive shift in $T_c$ of 16 K/GPa [203]. $T_c$ vs *P* curves for a few other manganite bulk samples [197] are also shown in Fig. 6 (b). A reduction in resistivity and shift in $T_{MIT}$ on the application of pressure were also observed in these manganites [192, 196, 203-205]. These changes are attributed to changes in the Mn-O-Mn bond angle and bond length of manganites arising from the application of pressure. In DE perovskite ferromagnets, the applied pressure increases the Mn-O-Mn bond angle and compresses the Mn-O bond length ($d_{Mn\text{-}O}$), which subsequently increases the bandwidth (*W*) [150]. Thus, the application of pressure increases the $T_c$ ($T_{MIT}$) and reduces the resistivity of the system [150, 203].

While the influence of external hydrostatic pressure on the electrical and magnetic properties of manganese perovskites has mostly been studied in bulk samples and single crystals [196-198, 203, 205, 207, 212, 213], manganite thin films are expected to show a more complicated behaviour under external hydrostatic pressure due to residual epitaxy stress originating from the substrate. Moreover, the systematic application of external hydrostatic pressure in thin-film samples is difficult, and studies related to the effect of pressure on electronic and magnetic properties are scarce [214, 215]. For thin films, there are other modes to apply strain, as discussed later. However, studies on bulk manganite systems suggest that upon the application of external pressure $T_{MIT}$ shifts to a higher temperature.

## 4. Complex oxide interfaces and investigation of interface magnetization

Considerable work has been done on complex oxide manganite thin films. Superlattices of artificially grown manganites have been intensely investigated for some time and continue to reveal emerging phenomena at interfaces with potential applications as a route to advanced functional materials [37, 216, 217]. Modern physical deposition techniques, e.g., pulse laser deposition, molecular beam epitaxial (MBE), etc., have enabled the controlled growth of thin films at the atomic scale and shown the possibility of making new types of materials by alternating the thin-film growth of two or more materials on top of each other. Thus, the ability to fabricate alternating layers of materials periodically, or superlattices, provides additional levels of compositional control with variable degrees of freedom, e.g., interface and strain control, which have exhibited novel physical phenomena, especially at interfaces [37, 216]. Figures 7(a) and (b) show schematics of the oxide superlattices consisting of periodic alternating layers of different oxides ($ABO_3$ and $A'B'O_3$) forming different interfaces. Figure 7(c) shows the high-angle annular dark-field (HAADF) Z-contrast image of the superlattice consisting of two different $ABO_3$ perovskites. It can be seen that the superlattice has well-defined interfaces between two perovskites.

The interfacial region between perovskite manganites with dissimilar *B*-site cations forms the *B*-O-*B′* bonds and provides a deviation in exchange interaction from that found in the adjoined perovskite materials. Thus, perovskite heterostructures are one of the most attractive research topics because the properties of interfaces can either be fundamentally different from those in the bulk or absent in the bulk. These phenomena are termed emergent phenomena or emergent phases. For example, high-mobility electron gas is observed at the interface of two insulating oxides, the LAO/STO heterostructure [218] and emergent ferromagnetism at the interfacial region of an antiferromagnet $LaMnO_3$ (LMO) grown on STO [219], antiferromagnet $CaMnO_3$ (CMO) and paramagnetic $CaRuO_3$ (CRO) [220], paramagnetic $LaNiO_3$ (LNO) and CMO [221], multiferroic $BiFeO_3$ (BFO) and ferromagnetic LSMO ($x$ = 0.33) [40-42, 222], LNO and LSMO ($x$ = 0.33) [50] heterostructures. Reduced dimensionality, broken symmetry, spin-orbit coupling, quantum confinement, strain, charge transfer, and atomic or electron reconstruction are phenomena that have been attributed to these emerging properties at interfaces. The growth of heterostructures generally involves numerous surface processes, such as adsorption, surface diffusion, association, dissociation, and the propagation of the strain field. For ideal interfaces, the breaks in translation and inversion symmetry also produce various effects.

The distortion of the crystal lattice as achieved by strain in thin films leads to significant modulation of its inherent properties [21, 22, 36, 37, 53-56, 216, 223]. The approach to changing interatomic distances in thin films by applying strain and controllably tuning material properties is known as strain engineering. Strain field engineering has emerged as a powerful approach for modifying the functional properties of manganite oxide thin films. The coupling of the strain field and structure in manganite oxides produces an interesting nonlinear response to the environment [224]. Manganites show a huge change in resistivity near the transition from the FM to the AFM order suggesting the coexistence of FM/conducting and AFM/insulating phases [191]. The coexistence of these phases in manganite thin films has been understood by the elasticity model that combines the continuum staining and collective displacements of atoms [225]. These interface-dependent emerging phenomena that are highly influenced by external stimuli such as strain etc., require a comprehensive characterization, which becomes even more important if the interface is buried. Thus, to characterize the structure and magnetism of interfaces, we require tools that can probe the interfaces with good depth resolution. PNR is non-destructive and intrinsically sensitive to interfacial magnetism, and it has been successively used to resolve the origins of magnetism that arise at interfaces of complex oxides [50, 103, 220, 226, 227]. In this section, two important techniques, PNR and X-ray resonant magnetic scattering (XRMS), used for understanding the coupling of interface strain and magnetism in complex oxide heterostructures, are briefly described [38, 39, 85, 220, 221].

## 4.1 Polarized neutron reflectometry

Due to the rapid growth of large-scale scattering facilities such as synchrotron and neutron sources, two non-destructive techniques, x-ray reflectivity (XRR) and PNR, have been successfully used to investigate the interface structure and magnetism, with sub-nanometer depth resolution, of a variety of heterostructure systems including metallic and complex oxides [95-110, 228-233]. PNR can simultaneously provide both nuclear and magnetization depth profiles across the interfaces because the reflection of neutrons occurs when it sees a change in the nuclear or magnetic structure at the interfaces [95-100, 102-104, 231-233]. A detailed review of the PNR technique can be found in references [95-100, 102-104, 231-233]. Unlike conventional macroscopic magnetization techniques, e.g., vibrating sample magnetometer (VSM) and superconducting quantum interference device (SQUID) magnetometer, which are only sensitive to the macroscopic magnetization of the whole sample (including substrate), PNR is capable of probing weak magnetization from heterostructures and interfaces with little influence from the substrate [99, 100]. Contamination of magnetic signals from the substrate is a serious concern in macroscopic magnetometry [99]. In addition, the interaction between neutrons and most of the material is weak, which helps to implement a sophisticated sample environment for PNR experiments [99]. PNR experiments on heterostructures were successfully carried out considering several external stimuli, e.g., high magnetic field and low temperatures [38], mechanical jig to apply bending strain [85-88], electric field [234], and optical irradiation of samples [235].

Experimentally, specular (angle of incidence = angle of reflection) PNR is measured as a function of wave vector transfer $Q$ (= $\frac{4\pi}{\lambda} sin\theta$, where $\theta$ is the angle of incidence and $\lambda$ is the wavelength of the neutron) for different neutron polarizations [Fig. 8(a)] [103]. The $Q$ can be varied either by changing the angle of incidence (adopted mostly in the continuous neutron source, reactor source, which uses monochromatic neutron beam) or by using different wavelengths (adopted mostly in pulse source, spallation neutron source). Figure 8 (a) shows a schematic of the specular PNR experiment, where the polarization ($P$) direction of the neutron beam is either parallel or antiparallel to the applied field ($H$) direction and the corresponding reflectivity is termed $R^+$ (spin-up) and $R^-$ (spin down). Neutrons interact with both nuclei and magnetization (because the neutron carries a spin ½ and a magnetic moment, which interacts with unpaired electrons responsible for magnetism.) of the material, and the nuclear and magnetic neutron scattering lengths have comparable magnitudes [100]. Polarized neutrons with spin parallel (spin-up, +) or antiparallel (spin-down, −) to the direction of the applied external magnetic field experience the same nuclear-scattering potential, but opposite magnetic scattering potentials and contribute to spin-dependent reflectivities, $R^{\pm}$. Theoretically, specular reflectivity is related to the square of a Fourier transform of the scattering length density (SLD) depth profile, $\rho$(z) (normal to the film surface or along the z-direction), averaged over the lateral dimension of the film. In the case of PNR $\rho$(z) consists of nuclear and magnetic SLDs [NSLD and MSLD] such that $\rho^{\pm}(z) = \rho_n(z) \pm \rho_m(z) = \rho_n(z) \pm CM(z)$, where $C = 2.9109\times10^{-9}$ $Å^{-2}$ $cm^3$/emu, and $M$(z) is the magnetization (emu/$cm^3$) depth profile [97, 102]. The sign +(-) is determined by the condition when the neutron beam polarization (Fig. 8) is parallel (opposite) to the applied field and corresponds to reflectivities, $R^{\pm}(Q)$. Thus, by measuring $R^{+}(Q)$ and $R^{-}(Q)$, $\rho_n(z)$ and $\rho_m(z)$ [or $M$(z)] can be obtained separately. The difference between spin-dependent reflectivities [$R^{+}(Q)$ and $R^{-}(Q)$] provides the magnetization depth profile [$M(z)$] in the magnetic heterostructure. In the case of XRR, $\rho$(z) is the electron scattering length density (ESLD), and it is a very good complementary technique to PNR for studying the depth-dependent structure of heterostructures [97, 102]. Figure 8 (b) shows the spin-dependent reflectivity profiles from a Ni thin film (thickness ~ 20 nm) on a Si substrate. The inset represents the NSLD and MSLD depth profiles of the thin film. In addition to measuring the magnetization depth profile, PNR with polarization analysis provides information about the projection of the net magnetization vector (depth-dependent magnetic structure) onto the sample plane [102, 236, 237]. Employing spin-dependent neutron reflectivity with polarization analysis provides the direction of in-plane magnetization along the depth of heterostructures and interfaces [97, 102]. Thus, PNR is a unique technique for providing a detailed depth-dependent in-plane magnetic structure of magnetic heterostructures, especially across buried interfaces. PNR has been successfully used to study the influence of applied bending strain on the magnetic properties of $(La_{1-y}Pr_y)_{1-x}Ca_xMnO_3$ (LPCMO) ($x$ = 0.33 and $y$ = 0.6) films grown on (110) NGO substrates [85-88].

## 4.2 Resonant X-ray Scattering

Interaction of X-rays with matter through scattering from both the charge and magnetic moment of electrons. Scattering from charge is the dominant term for X-rays, making it a commonly used probe for condensed matter. Scattering from magnetic moment, although small, is sufficient to extract valuable information on the magnetic structure in condensed matter [238]. Tuning the energy of X-rays to selected resonances of the magnetic element of materials enhances the sensitivity to magnetic moments and hence is applied to study magnetism in both absorption (X-ray magnetic circular dichroism, XMCD) [239] and scattering (XRMS) [240] channels. These resonant X-ray techniques, in combination with X-ray absorption spectroscopy (XAS) [241, 242], have been used to study the electronic and magnetic properties of complex oxides both in the bulk and thin film phases [243-250]. The polarization-dependent absorption of x-rays in magnetic films provides the element-specific electronic and magnetic structures because the absorption edges have characteristic energies for each element. Thus, XAS uses the energy-dependent absorption of X-rays to obtain information about the elemental composition of the sample. The magnetic properties of transition metals are largely related to the *d* orbitals, which are best probed by *L*-edge absorption. XAS involves excitation by photon absorption of a core-level electron (*p*-orbital in transition metal) into unoccupied states (*d*-orbital) near or above the Fermi level. The subsequent filling of the core hole by an electron with a lower binding energy results in either the emission of a fluorescence photon or the radiationless emission of an Auger electron. In the case of thin films, the XAS measurement [Fig. 9 (a)] can be performed by two methods (i) total electron yield (TEY): a measure of drain current (flow Auger electron and escaping photoelectrons), which is sensitive to the surface for soft x-rays involved in the *L*-edge energy range and can probe a depth of a few nanometer; (ii) fluorescence yield (FY): a measure of photons and probes the deeply buried interface or the bulk contribution up to a few hundred nanometers at the *L* edges of 3*d* ions. The difference in the absorption of X-rays upon passing through the material in two different polarizations, i.e., dichroism, provides the magnetic properties of the material. The common practice is to use either circularly or linearly polarized X-rays, and using these two types of polarization has the advantage of investigating the detailed contribution of magnetization from the magnetic atom [242]. X-ray magnetic circular and linear dichroism (XMCD and XMLD) measure the dependence of XAS on the helicity of circularly and linearly polarized X-rays, respectively, by a magnetic material. More details on magnetism studies using polarized X-ray techniques can be found in [241, 242, 244, 246, 247].

The XAS spectra of transition metals consist of two main peaks that arise from the spin-orbit interaction of the 2*p* core-shell, and the total intensity of the peaks is proportional to the number of empty 3*d* valence states. Thus, XAS is a great tool to probe the valence state of a particular element. The angular momentum of the circularly polarized X-ray is transferred to the excited photoelectron upon absorption. The angular momentum of the X-ray can be transferred in part to the spin via spin-orbit coupling. The spin polarization is opposite at the $p_{3/2}$ ($L_3$) and $p_{1/2}$ ($L_2$) levels because they have opposite spin-orbit couplings. Experimentally, the sum ($I^+ + I^-$) of the left and right circular polarized signals is a measure of XAS, which provides information on the electronic environment of the 3*d* electrons of transition metals while magnetic information is contained in the difference ($I^+$ - $I^-$) in absorption is referred as XMCD [242, 251].

The sum ($R^+$ + $R^-$) and difference ($R^+$ - $R^-$) of the LCP and RCP X-ray intensities in the reflection/scattering process from the thin film contribute to the absorption and XRMS signals, respectively [Fig. 9 (a)] [242]. Figures 9(b) and (c) show the Mn $L_3$ and $L_2$ edge total electron yield (TEY) average absorption ($I^+$ + $I^-$) spectra (XAS) and the XMCD signal from an LPCMO ($x$ = 0.33 and $y$ = 0.6) film at 20 K [252]. Figures 9(d) and (e) show the XAS and XRMS spectra, respectively, from the same sample at 20 K in the scattering mode. A small change in the XAS spectra near the Mn $L_3$ edge may have resulted from additional modifications in the surface electronic properties. The depth length scale probed using these synchrotron-based techniques is limited, however, they provide element-specific information.

## 5. Mode of strain in the thin films

Both the intrinsic and extrinsic stresses exist in the thin films. Defects such as dislocations in the thin film primarily contribute to intrinsic stress. The adhesion of thin films to the substrate gives extrinsic stress. Stress can be introduced in a thin film due to the different thermal expansion between the film and substrate due to lattice mismatch. Achieving a high critical temperature of functional materials is a major challenge for the exploitation of many exciting physical phenomena, such as CMR, phase coexistence, MIT, and multiferroicity in strongly correlated systems. The effects of strain on manganite films and heterostructures have been studied extensively as a means of achieving high transition temperatures and understanding phase coexistence. Here we mainly discuss three-way of applying strain in thin films, [253] e.g. (a) epitaxial, (b) piezoelectric, and (c) bending, which have been extensively used in the literature to study the influence of strain on the macroscopic properties of thin films. In addition, epitaxially grown thin films of vertically aligned nanocomposites (VANs) [254] with vertical 1D structures (e.g., nanopillars, nanosheets, nanowires, etc.) of one phase, embedded in a film matrix of another phase have also attracted much attention and are discussed here as another mode of applying strain in thin-film systems. These four modes of applying strain on a thin film of complex oxide have been employed to study the correlation of strain with the electronic and magnetic properties of thin films. A brief discussion and comparison of these four modes of strain in thin films is described below.

## 5.1 Epitaxial strain

The influence of substrate strain in manganite films, termed epitaxial strain, has been extensively studied. The epitaxial strain in thin oxide films is achieved by synthesizing perovskite thin films on single crystal substrates that have similar structural, chemical, and thermal properties but with different in-plane lattice parameters than the film material [92]. The lattice constant difference between the film and substrate induces epitaxial (biaxial) strain in the film. The epitaxial strain due to lattice mismatch is defined as $\varepsilon = \frac{d_{substrate} - d_{film}}{d_{film}}$, where $d_{substrate}$ and $d_{film}$ are the lattice constants of the substrate and the film (Fig. 10), respectively. For the $\varepsilon \approx 0$, one obtains lattice-matched epitaxial heterostructure known as homoepitaxy. A strain of $\varepsilon > 0$ corresponds to a film with tensile strain in the *a*–*b* plane (in-plane), while $\varepsilon < 0$ indicates that the film will be compressively strained in the *a*–*b* plane. For compressive (tensile) strains, the perovskite crystal structure contracts (stretches) in the plane, and thus the lattice parameter increases (decreases) in the out-of-plane direction, resulting in polarization along the out-of-plane direction [34].

A biaxial strain of several percent can be achieved in the thin film form by carefully selecting substrates, thicknesses, and buffer layers [34, 92]. This epitaxial strain is not only a way to tune their functionalities but also an interesting approach to stabilize phases that are unstable under ambient conditions. The epitaxial strain gradually changes with the film thickness in perovskite manganites. Upon increasing the film thickness, the film starts to relax (e.g., via misfit dislocations) and the in-plane lattice parameter starts to deviate from the substrate. The effect of compressive and tensile strain on the $MnO_6$ octahedral is schematically shown in Figure 3(b). Strain can also affect the octahedral rotations in a perovskite material, where compressive strain enhances out-of-plane rotations and tensile strain in-plane rotations [255]. Various single-crystal perovskites and perovskite-related substrates [15, 17, 256-276] are commercially available for growing complex oxide thin films. Bulk LSMO (manganite) and other manganite films have a pseudo cubic lattice constant of $d$ ~ 3.87 Å [277]. Table 2 shows the list of single-crystal substrates [278] with their lattice constants and the corresponding strain values for overgrown manganite films. A large tunability in the transport and magnetic properties of manganite films can be achieved by controlling the epitaxial strain. It is noted that the strain values provided in Table 2 are estimated using the epitaxial strain discussed above. Many of these oxide substrates show structures different from cubic and thus provide additional strain along the different in-plane axis (lattice mismatch along two in-plane perpendicular directions). The in-plane anisotropic epitaxial strain provided by the substrate to the film exhibits anisotropic electronic and magnetic properties [279] and will be discussed later.

## 5.2 Vertical strain engineering in nanocomposites

VAN thin films with ordered two phases, grown epitaxially on substrates have attracted considerable interest in the past two decades because they provide growth of unique nanostructured composite thin films with a large vertical interfacial area and a controllable vertical lattice strain [280]. Figure 11 (a) shows a schematic of a self-assembled VAN film (vertical nanocolumns) grown on a substrate with complex oxide, *A*, embedded in a film matrix of another complex oxide, *B*. Several VAN thin films of nanocomposites of manganite and other oxides have been grown and studied for their structure and magnetotransport properties [254, 280-284]. Figure 11 (b) shows the cross-sectional TEM measurements of a self-assembled LSMO: ZnO VAN film, grown on an STO substrate with the corresponding selected area electron diffraction pattern in the inset. The TEM image illustrates the nanocolumn vertical growth of the LSMO and ZnO domains. The VAN of LSMO ($x = 0.3$): ZnO grown on single-crystal STO was discussed for understanding the origin of the vertical lattice strain in this system [281]. The system was later studied for magnetotransport properties and exhibited tunable and enhanced low-field magnetoresistance [283]. While the epitaxial strain relaxes in thick films, the heteroepitaxial VANs provide interfacial strain in thick films with large vertical interface areas, and strain exists at both lateral and vertical interfaces. The vertical strain stems from the elastic coupling of the vertical interfaces between the two phases. The vertical lattice strain state in VANs is directly related to nanopillar size, density, and spacing. For each pillar in a VAN system, stress can be aligned along the radial and axial directions. However, the strain distribution strongly depends on the nanopillar density and spacing.

In VANs, two possible lattice matchings, i.e., direct lattice matching and domain lattice matching, exist along the vertical interface. In the case of direct lattice matching of strained lattices, the *m* lattices of phase *A* match the *km* lattices of phase *B* (*m* and *k* are positive integer numbers), which can produce large strain in VANs if the right components are selected. In the case of domain matching, *m* lattices of phase *A* match with $m$+1 lattices of phase *B*, and lattice mismatch, in this case, is often quite small [280]. Thus, in VAN thin films, a large strain can be designed by judiciously considering the epitaxial framework, lattice parameters, elastic constants, and thermal expansion coefficients of two phases. A vertical strain of ~ 2.0% was observed for two-phase VAN systems of LSMO ($x$ = 0.3): MgO [254] and LCMO ($x$ = 0.3): MgO [284].

## 5.3 Piezoelectric strain

The epitaxial strain in complex oxide heterostructures has been varied either by using substrates with a different lattice mismatch or by varying the film thickness. However, the systematic study of intrinsic strain effects using these methods is difficult because it is not possible to discriminate the effect of many other variables such as lattice relaxation, stoichiometry, crystalline quality, and microstructure. However, in the case of piezoelectric strain, it is possible to vary the strain state of thin films *in situ* while keeping the negative variables fixed. In the literature, piezoelectric strain on epitaxial films has been applied by two methods. In the first method, the structural phase transitions of ferroelectric (FE) (especially $BaTiO_3$, BTO) substrates are used to generate intrinsic strain in manganite films [83, 285-287]. A large biaxial strain can be produced at the transition temperatures, which modify the electronic and magnetic properties of manganite thin films purely due to strain. In another method, manganite films are deposited on a piezoelectric single-crystal substrate, and strain is induced via the converse piezoelectric effect (strain is generated due to an applied electric field in the piezoelectric substrate). The most commonly used piezoelectric materials for this purpose are $Pb(Zn_{1/3}Nb_{2/3})O_3$(PZN), $Pb(Zn_{1/3}Nb_{2/3})O_3$-0.08$PbTiO_3$ (PZN-PT), 0.7$Pb(Mg_{1/3}Nb_{2/3})O_3$–0.3$PbTiO_3$ (PMN-PT), and $PbZr_{0.52}Ti_{0.48}O_3$ (PZT) [288-295]. However, the piezoelectric substrate PMN-PT offers important advantages such as (a) it is nearly cubic even in the ferroelectric state [296], and (b) an extremely high strain level >0.6% can be achieved using this method [288].

The ferroelectric BTO crystal exhibits three different structural phase transitions [83] at different temperatures, as depicted in Fig. 12 (a). At high temperatures (> 400 K), BTO exhibits a cubic structure, which changes to a tetragonal phase at 393 K while cooling from a higher temperature. Further cooling of the crystal shows a transition to a monoclinic structure at 278 K and finally to a rhombohedral structure at 183 K. The variation of lattice parameters with temperature for different structural phase transitions is depicted in Fig. 12 (a). These structural phase transitions were used to vary the epitaxial strain of the complex oxide film grown on the BTO crystal. The BTO crystal, in particular, can offer additional tunability of strain because the lattice structure of BTO can be influenced by an electric field [297]. Using this factor, the magnetic and transport properties of the manganite film grown on the BTO substrate can be controlled by an electric field [298]. For the strain-induced converse piezoelectric effect, Park *et al*. [288] studied the piezoelectric properties of relaxor-based ferroelectric single crystals [Fig. 12 (b)], such as $Pb(Zn_{1/3}Nb_{2/3})O_3$–$PbTiO_3$ and $Pb(Mg_{1/3}Nb_{2/3})O_3$–$PbTiO_3$, for electromechanical actuators and observed strain levels up to 1.7%, an order of magnitude larger than those available from conventional piezoelectric and electrostrictive ceramics. The higher strain achieved in these crystals was attributed to an electric field-induced phase transformation. A linear variation of out-of-plane strain with applied electric field in the PMN-PT substrate at 300 K was also observed in another study, suggesting a higher value of converse piezoelectric effect induced lattice strain [292]. Using electric-field-induced in-plane strain fields in manganite/FE (substrates) hybrid systems, coupling of the two ferroic phases (FE and FM) has also been demonstrated [287, 299], where an electric field controls the magnetic anisotropy and remnant magnetization of FM films grown on FE substrates.

It is noted that magnetoelectric coupling across manganite oxide/FE interfaces, which normally produces an in-plane strain, is also associated with charge carrier modulation/charge transfer coupling in manganite films induced by electrostatic charge modulation in ferroelectric substrates [300, 301]. In the case of charge-mediated/transfer coupling as a response to the electric field in an LSMO/PMN-PT heterostructure, a change in the magnetization of the LSMO film was observed upon switching the polarization of the FE (PMN-PT) substrate because an induced reverse charge accumulates in the LSMO layer near the interface, which changes the ratio of $Mn^{3+}$ and $Mn^{4+}$ and consequently the magnetization in the LSMO [302]. The effects of both interfacial charge and strain-mediated coupling have been addressed in the literature for LSMO/FE systems [301-305]. However, thickness-dependent magnetoelectric coupling studies on LSMO ($x$ = 0.3)/PZT[301] and LSMO ($x$ = 0.3)/PMN-PT [302, 303] suggested that interfacial-charge driven coupling is often observed in much thinner films (thickness up to ~ 4 nm). For thicker LSMO films (thickness ~ 20-50 nm), the strain-driven magnetization modulation in LSMO/FE (substrate) heterostructures [294] is more dominant. In this article, we will discuss manganite/FE heterostructure systems with thicker manganite films in which strain plays an important role in modulating magnetic behaviour.

## 5.4 Bending strain

Experiments measuring the response of magnetism to an epitaxial strain of manganite thin films have been extensively studied by measuring the variation in $T_c$, saturation magnetization ($M_s$) or remnant magnetization ($M_r$), $T_{MIT}$, etc., which are indirectly related at best to magnetism. However, the influence of epitaxial strain on the electronic and magnetic properties of manganite thin films has provided a distinctly mixed picture. Chen *et al.* [306, 307] and Gillaspie *et al.* [82] reported that compressive (tensile) strain strengthens (weakens) the ferromagnetic phase in manganite films. Other studies report that either tensile strain produces the opposite behaviour [83, 308] or compressive strain has a negligible effect on magnetism [82]. Attempts have been made to clarify the influence of epitaxial strain on the electronic and magnetic properties of manganite thin films using structural phase transformation [83] or the piezoelectric properties [292, 293] of a substrate. However, these studies also yielded contradictory results [83, 293]. It is noted that both epitaxial and piezoelectric strains depend on the choice (property) of the substrate, and it is difficult to distinguish the effect of strain on the electronic and magnetic properties of manganite thin films in the presence of other factors, e.g., misfit dislocations, defects, film morphology, and crystalline quality, which can arise from the choice of substrates.

Bending strain is an important approach for delivering variable uniaxial/biaxial strain on the scale of several microns to centimetres compared with nanoscale stress using other techniques. Bending strain can be applied using three [84, 309] or four [85, 310] point beam bending (mechanical jig) techniques [Fig 13 (a)]. In the case of three-point bending measurement, the maximum stress occurs in the center of the sample to be tested, which can lead to unwanted failure and non-uniformity over the sample, thus providing irregular stress [309]. Compared to three-point bending, four-point bending has many advantages. It ensures a uniform bending stress between the two supports, which is desirable in many cases, and the errors caused by misalignment remain small [310]. Thus, the strain from bending the film using a four-point mechanical jig is coherent and uniform [85, 310] and does not affect other structural characteristics of the film [85-87]. Kelman *et al.* [311], employed a wafer bending method (equivalent to four-point beam bending) to impose strains on $Pb(Zr_{0.35}Ti_{0.65})O_3$ thin films and observed that strains as small as 0.08% can reversibly reduce the remanent polarization of piezoelectric films by 12% -14%. Thus, the application of bending stress offers an opportunity to examine the exclusive role of elastic stress on the transport properties of films and it is also reversible. A schematic representation of the applied bending stress for a thin film under different conditions of bending strains, i.e., no strain ($\varepsilon = 0$), compressive ($\varepsilon$ = -ve), and tensile ($\varepsilon$ = +ve), using a four-point mechanical jig is shown in Fig. 13(b). Either tensile or compressive uniaxial strains can be imposed on the film by flipping it in the mechanical jig. Figure 13 (c) shows a picture of the mechanical jig mounted on the cool head of the cryostat.

The applied strain (stress) in the film is a measure of change in the bending (curvature) of the sample and is defined as $\varepsilon = \frac{t_s}{R}$ [312], where $t_s$ and $R$, are the thickness of the substrate and the radius of curvature of the film on the application of strain, respectively. To measure the radius of curvature of the thin film system, the laser scanning method can be used successfully [313]. The ray diagram of the reflected laser beam from the curved surface is depicted in Fig. 13(e), and the radius of curvature of the curved surface can be measured as $R = \frac{2L\Delta x}{\Delta d - \Delta x}$ , where $L$, $\Delta x$ and $\Delta d$, are distance between the sample and image and, the sizes of the images before and after bending (on applying strain), respectively. Measurements of the size ($\Delta x$ and $\Delta d$) of the laser beam reflected from the sample surface before (zero stress) and after bending (strained) are shown in Fig. 13 (d). Thus, using this method, one can measure the applied strain on the film quantitatively. Mechanical bending stress measurements have been performed for various complex oxide materials [84-88, 311, 314, 315] to study the correlation between strain and the physical properties of the systems.

## 5.5 Other approaches for applying strain

In addition, a few other studies exist in the literature where attempts were made to generate strain in the film. Ion implantation [316-318] and ion irradiation [319, 320] are two techniques that have been used to change the strain and morphology of manganite films, which are directly correlated to their electronic and magnetic properties. Lee *et al.* [318] studied the effect of $Ar^+$ ion implantation on the magnetic properties of the LSMO ($x$ = 0.3) film and found a uniaxial expansion of the *c*-axis upon ion implantation that led to a transition from in-plane to PMA. Rajyaguru *et al*. [319] performed swift heavy ion (SHI) irradiation of the LMO/NSMO($x$ = 0.3)/STO heterostructure using 100 MeV $O^{7+}$ ions with different fluences and found changes in the strain field at interfaces. Bhatt *et al.* [320] investigated the structure and morphological changes of LPCMO ($x$ = 0.375 and $y$ = 0.625) upon SHI irradiation using $Ag^{15+}$ (120 MeV) ions. This result indicated the development of tensile stress along the out-of-plane direction of the LPCMO film upon ion irradiation, which increased with increasing ion fluence. Zhang *et al.* [321] found a maximum strain of 3-4 times along the *c*-axis as compared to the *a*-/*b*-axis in $LiTaO_3$ single crystal substrate after irradiation with Au ions of energy 3 MeV. These studies suggest that ion implantation and irradiation can be used to change the strain of both the substrate and film. However, parasitic effects due to ion implantation/irradiation, such as defects, dislocations, and annealing of defects, also change the macroscopic properties of films.

Ultrafast strain engineering in complex oxide heterostructures using time-dependent terahertz (THz) spectroscopy [322-325], which is not covered in this review, has recently been explored to investigate the change in electronic and magnetic phase behaviour by optical excitation of either a film (manganite) [323, 325] or substrate [322, 324] lattice. I believe that a combination of this technique with PNR will help quantify the change in the magnetic phase (AFM to FM) and to understand the mechanism of induced magnetization in the film due to photodoping [325].

# 6. Coupling of strain with transport and magnetism in manganite thin films

Manganese perovskite oxide heterostructures exhibit a half-metallic character and CMR response, which make them ideal materials for the development of novel oxide-electronic devices and the study of fundamental physical interactions. In particular, the role of strain in manganite thin films has been widely studied [29, 82, 83, 158, 293, 307] and identified as an interesting perturbation not only for a basic understanding of physical interactions but also proposed as a more adequate mechanism to describe the multiscale multiphase coexistence [29] observed in manganites. The coupling of strain and magnetism of manganite films was studied theoretically [53, 158, 326-328], which showed that the possible modifications produced by strain in manganite thin films are twofold: (a) the reduction (increase) of the lattice parameter of manganite film in a particular direction would lead to an increase (reduction) of the hopping amplitude, and (b) a distortion of structural symmetry leads to a splitting of the Mn $e_g$ levels, which may produce orbital ordering via the JT interaction. The Mn $e_g$ shell doubly degenerates in a regular octahedron environment, although it is partly occupied. The $Mn^{3+}$ ions are thus JT active, and a small distortion of the oxygen octahedra can lift the degeneracy and stabilize one of the $e_g$ orbitals to the other. Thus, by imposing a specific geometry (e.g., by applying strain) around the Mn ions, it is possible to selectively occupy either $d_{3z^2-r^2}$ or $d_{x^2-y^2}$ $e_g$ orbitals. If the $d_{3z^2-r^2}$ is occupied, the double exchange will take place along the *c* direction (out-of-plane direction in manganite film) and it will reduce the in-plane magnetism in the manganite films and also affect the $T_C$ [326]. Thus, a compressive strain [see Fig. 3 (b)] would increase the hopping within the *xy* plane and decrease it in the *z*-direction favouring the $d_{x^2-y^2}$ $e_g$ orbitals (rather than the $d_{3z^2-r^2}$ orbitals) to be occupied [53, 326-330]. On the other hand, a tensile strain would produce a lowering of the $d_{3z^2-r^2}$ orbitals with respect to the $d_{x^2-y^2}$ ones. The $T_c$ of manganite films will be maximized [326] if: (i) the $d_{x^2-y^2}$ orbital stabilizes (lowering of orbital) over the $d_{3z^2-r^2}$, and (ii) there is a large overlap of the in-plane effective exchange integral between the $d_{x^2-y^2}$ orbitals by preventing the tilt of the octahedra, and the $T_c$ can be scaled as:

$$T_c \sim J \sim t_{dp}^2 \sim S_{dp}^2 \qquad (4)$$

where $J$, $t_{dp}$, and $S_{dp}$ are the effective exchanges between adjacent Mn ions, the transfer integral between Mn $d$ – $O_2$ $p$ and the overlap between the Mn $d_{x^2-y^2}$ and the $O_2$ $p$ orbitals, respectively. Thus, depending on the strain applied to manganite films, a strong coupling between the strain and magnetism can be obtained in these systems. However, unlike bulk manganites, in which compressive stress mostly favours the ferromagnetic phase, theoretical and experimental studies of manganite films paint a decidedly mixed picture, and the role of strain and magnetism in manganite films is far from understood. Here we will study the effect of strain, applied using different techniques, on the magnetism of manganite thin films.

## 6.1 Epitaxial strain-driven transport and magnetism

The response of the magnetic properties of manganite films to epitaxial strain has been studied extensively [82, 112, 158, 183, 279, 306, 308, 326, 327, 330-332]. Hwang *et al.* [13] argued that because of the strong correlation among different degrees of freedom in manganite heterostructures, strain can profoundly affect magnetism. However, the epitaxial strain in the film, resulting from either growing the film on different substrates or growing films of different thicknesses, decays as a function of distance from the film/substrate interface. Thus, the strain is nonuniform and the measurement of bulk magnetization, e.g., using macroscopic magnetization techniques (i.e. SQUID and VSM), will provide a strain-averaged result. Manganite thin films are epitaxially grown on insulating substrates, and the lattice mismatch between the different layers gives rise to a strain that affects the bulk properties of thin manganite [160, 326, 332-335] and/or the electronic reconstruction at its interface [336]. The in-plane epitaxial strain may range from ~ −2.3% to ~3.2% depending on the type of substrate and growth direction [332]. This strain modifies the relationship between the lattice parameter in the direction perpendicular to growth ($c$) and that in the parallel plane ($a$). Millis *et al.* [158], developed an analytical model to describe the effects of biaxial strain ($\varepsilon_{xx}$ and $\varepsilon_{yy}$) on the magnetic properties (Curie temperature, $T_c$) of CMR manganites. In this model, $T_c$ depends on two parameters: (i) bulk compression $\varepsilon_B = \frac{1}{3}(2\varepsilon_{xx} + \varepsilon_{zz})$ (with $\varepsilon_{xx} = \varepsilon_{zz}$), and (ii) biaxial distortion $\varepsilon^* = \frac{1}{2}(\varepsilon_{zz} - \varepsilon_{xx})$, where $\varepsilon_{xx} = (a_{xx} - a_{bulk})/a_{bulk}$, and $\varepsilon_{zz} = (a_{zz} - a_{bulk})/a_{bulk}$ are the pseudo-cubic in-plane and out-of-plane strains, respectively. Using these parameters, the influence of strain on the $T_c$ are described as [158, 332]:

$$T_c(\varepsilon_B, \varepsilon^*) = T_c(0,0)[1 - \alpha\varepsilon_B - b\varepsilon^{*2}] \qquad (5)$$

with $\alpha = (1/T_c)(dT_c/d\varepsilon_B)$ and $b = (1/T_c)(d^2T_c/d\varepsilon^{*2})$ are two important experimental quantities. $T_c(\varepsilon_B, \varepsilon^*)$ and $T_c(0,0)$ are transition temperatures in the presence of strain and without strain, respectively.

The parameter $\varepsilon_B$ is mostly linked with a compressive (tensile) strain and it increases (decreases) the electron hopping probability. Therefore, depending on the sign of the strain (+ve for tensile and -ve for compressive), the change in $T_c$ will be positive (for compressive) or negative (for tensile). The other source of strain, which is related to the biaxial distortion $\varepsilon^*$, will increase the Jahn–Teller splitting of the $e_g$ electron levels and will cause only a decrease in $T_c$. Typical values for $\alpha$ and $b$ in manganites were predicted to be around 6 and $1.4\times10^3$, respectively [158]. The value of *b*, a second-derivative effect, is exceedingly large and usually, is of the order of the square of the first derivative effect, thus of the order of 36. The large value indicates a dramatic sensitivity of magnetization to strain in manganites and suggests that a 1% biaxial strain would cause about a 10 % shift in $T_c$ [158].

The LSMO ($x$ = 0.3) film is a well-studied and prominent family member of semi-metallic ferromagnetic manganites [31]. The application of LSMO in thin-film form is determined by its relatively high $T_c$ ~360 K and its structural compatibility with many functional perovskite-type materials such as oxide superconductors, ferroelectrics, and dielectrics [31]. Coherently strained LSMO thin films can be grown on various single crystal oxide substrates and allowed to accommodate the strain caused by distortion and/or rotation of the $MnO_6$ octahedron, which dictate the magnetic and magnetotransport properties of LSMO films [164, 337-339]. Kwon *et al.* [337] observed an in-plane easy magnetization direction for tensile-strained films (*c/a* ratio < 1) grown on (001)-oriented STO substrates, where compressively strained films (*c/a* ratio > 1) grown on (001)-oriented LAO substrates exhibited an out-of-plane easy axis, confirming strained-induced magnetic anisotropy in LSMO films. Subsequently, the effect of substrate-induced biaxial strain from -2.3% to +3.2% on thin films of (001) LSMO grown on different single crystalline substrates was studied [332], suggesting strong coupling of epitaxial strain on the electronic and magnetic properties of the LSMO film. Figure 14 (a) shows the θ-2θ XRD scans of the (001)- oriented LSMO thin films on different substrates. The arrows indicate the shift of (002) reflection of the film in response to the in-plane spacing of the substrates upon which the films are grown. Figures 14 (b) and (c) show the temperature variation of the transport and magnetization data, respectively, from LSMO films grown on different substrates. These magnetization and electrical transport measurements [332] reveal that compressive (tensile) strain increases (decreases) the $T_c$ of the LSMO films, which is in excellent agreement with the theoretical predictions (Eq. 5) of Millis *et al.* [158]. However, there exists a mixed report in the literature, about the value of $\alpha$ and $b$ for LSMO films [112, 181, 340] of different thicknesses grown on different substrates. Tsui *et al.* [112] found $\alpha$ and $b$ values around 2 and 70, respectively, for LSMO ($x$ = 0.33) film of thickness ~25 and 50 nm grown on LAO, NGO, $(LaAlO_3)_{0.29}$-$(SrAl_{1/2}Ta_{1/2}O_3)_{0.71}$ (LSAT), and STO substrates. Ranno *et al.* [181] found $b$ values around 2200 for LSMO films grown on STO, whereas Angeloni *et al.* [340] reported $b$ values around 103 for LSMO films (thickness ~ 16 nm) grown on LAO. In addition, a larger strain effect on the magnetic properties of LSMO films with reduced doping ($x$ = 0.1) grown on STO substrates was observed [306]. The thinnest manganite film (larger compressive strain) showed enhancement in both saturation magnetization and $T_c$, suggesting that the strain dependence of $T_c$ mainly results from strain-induced electron-phonon coupling.

The influence of strain on the magnetotransport properties of PSMO ($x$ = 0.33) films grown on different substrates and with varying film thicknesses [341] was studied. It was found that the biaxial tensile strain affects $T_{MIT}$ more than the biaxial compressive strain in PSMO films with a relatively large JT strain contribution. A strong epitaxial-induced strain-dependent ferromagnetic metallic to antiferromagnetic insulator phase transition was observed for PSMO ($x$ = 0.5) films grown on different substrates [342]. The effects of thickness-dependent strain relaxation on PSMO films grown on LAO substrates were also studied [169], and an increase in $T_c$, and saturation magnetization of PSMO films with an increase in thickness (decrease in in-plane compressive strain) was observed. The observed thickness-dependent change in the magnetization and transport properties of PSMO films were explained using polarized XANES measurements, which suggested an electronic anisotropy with charge transfer (between Mn 3$d$ and O 2$p$ orbitals) along the in-plane and out-of-plane directions, resulting in different absorption energies $E_r$ [169].

The LCMO ($x$ = 0.33) thin film is another important doped manganite thin film that has been studied for great magnetoresistance (MR) effects with a maximum ferromagnetic $T_c$ of ~

267 K [191]. Several groups have reported the effect of strain on $T_c$ and the MR of LCMO thin films, which have shown mixed results [158, 166, 343-348]. Biswas *et al.* [343] studied the effect of strain on $T_c$ in LCMO films grown on (001) LAO and (110) NGO substrates and found a reduction in $T_c$ when the film was grown under compressive strain (on LAO). However, in another study [344, 349], no connection was observed between the change in $T_c$ and the JT effect, although the $T_c$ of the films showed strong film thickness dependence. Another study [166] showed that the MR is maximum when the film thickness is approximately 100 nm, which is a thicker film and cannot be explained by the strain effect. The substrate-induced strain effect on the magnetic and electrical transport properties of LCMO thin films showed that an increase in the in-plane tensile strain reduces the magnetization, $T_{MIT}$, and $T_c$ [348]. However, the thickness dependence of the magnetotransport properties of LCMO thin films epitaxially grown on single-crystalline STO, LAO, and NGO substrates showed that on decreasing thickness, the resistivity and the low-temperature MR of films increases, while the $T_{MIT}$ and $T_c$ are lowered [350]. These studies suggested an enhancement of the disorder in the Mn-O bond length and Mn-O-Mn angles due to strain, which promotes a decrease of the double exchange transfer integral and, consequently, accounts for the reduction of the Curie temperature for the thinnest film [350].

More detailed strain-induced properties of LCMO films, including the in-plane tensile stress-dependent structure, transport, and magnetic properties of (001) and (110) LCMO films of different thicknesses grown on (001) and (110) STO substrates, were studied [334]. The study showed the dependence of the ferromagnetic ordering temperature ($T_c$) on the biaxial strain (Fig. 15) and confirmed that the elastic deformation of the manganite lattice can describe the response of the less strained films, typically the (110) ones, whereas other effects should be invoked to account for the properties of the most strained films, typically the (001) ones. The study further indicated that PS, which is likely triggered by orbital ordering, may account for both the suppression of $T_c$ and the higher concentration of nonferromagnetic and nonmetallic regions in the (001) films than in (110) films [334]. Figure 15 shows the thickness dependence of $T_c$ and saturation magnetization ($M_s$) for all (001) and (110) LCMO films grown on (001) and (110) STO substrates. The variation indicates that the (110) LCMO films have higher $T_c$ and $M_s$ than (001) LCMO films of similar thickness, suggesting an anisotropic behaviour of LCMO films under strain. Later, a competition between charge-orbital ordered insulating (COI), and FM metal (FMM) phases and transport properties in $Pr_{1-x}(Ca_{1-y}Sr_y)_xMnO_3$ (PCSMO) ($x$ = 0.3 and $y$ = 0.2) thin films grown on different substrates were studied [351], revealing that the large tensile strain in films grown on STO and a smaller tensile strain in films grown on LSAT substrates enhance the stability of the long-range COI phase, while the compressively strained films grown on LAO favour short-range FMM phases. The study also showed that a decrease in in-plane compressive strain increases the $T_c$ of the PCSMO films [351].

LPCMO is a prototypical EPS manganite system that displays large-scale EPS in the FMM phase and charge-ordered antiferromagnetic insulating (CO/AFMI) phase at low temperatures [2, 26, 39, 82, 90, 142, 279, 331, 352, 353]. Gillaspie *et al*. [82], and Wu *et al*. [331], observed that the in-plane tensile strain stabilizes the CO/AFI phase but disfavors the FMM phase for LPCMO ($x$ = 0.375 and $y$ = 0.48) films, epitaxially grown on STO and LAO substrates, respectively, whereas Jeen and Biswas [353], and Ward *et al*. [279] reported that the in-plane tensile strain destabilizes the CO/AFI phase but favours the FMM phase for LPCMO ($x$ = 0.33 and $y$ = 0.60) film, and LPCMO ($x$ = 0.375 and $y$ = 0.48) film, on NGO and $SrLaGaO_4$ (SLGO) substrates. Later, a tuning of the sharpness of MIT with epitaxial strain was reported in epitaxial LPCMO ($x$ = 0.33 and $y \leq 0.35$) films [354], and it was observed that the lattice-matched LPCMO/ (110) (NGO) films show a sharp MIT near the $T_c$, with an FMM ground state [354], as compared to multiple MITs found in anisotropically strained LPCMO/(001)(LSAT) films with a phase-separated ground state. Thus, the biaxial epitaxial strain in LPCMO systems [82, 331, 354], especially the anisotropic one [279, 355], can subtly induce phase competition in LPCMO films with an FMM ground state, which shifts to a higher temperature upon the application of a large strain.

Anisotropic epitaxial strain (AES) induced phase stability in manganite films, especially LPCMO films, shows a preferential orientation of electronic phase domains driven by anisotropic long-range elastic coupling between manganite films and substrate [279, 355, 356]. Simultaneous temperature-dependent resistivity measurements for the LPCMO ($x$ = 0.375 and $y$ = 0.48) film along the two in-plane perpendicular axes $[10\bar{1}]$ and [101] of the film parallel to the in-plane direction of (101) NGO substrate showed substantial differences in the $T_{MIT}$ and extraordinarily high anisotropic resistivities, which indicate that percolative conduction channels open more readily along one axis [279]. Remarkably, it was found that the $T_{MIT}$ measured along the higher-strained $[10\bar{1}]$ direction was 24 K higher than that measured along the [101] direction. In addition, the difference between the resistivities measured at the respective $T_{MIT}$ values was 0.31 Ω cm, which corresponds to a relative variation of ~200 % [279]. The magnetization measurements along two in-plane perpendicular directions suggested that the strained direction $[10\bar{1}]$ is a magnetization easy axis and indicates that the FMM phase domains are elongated along the direction of the greater strain field. Further, to confirm the anisotropic elongation of the FMM phase, the variation of resistivities and $T_{MIT}$ for both the $[10\bar{1}]$ and [101] directions as a function of the magnetic field were also studied. It was found that at all magnetic fields, the direction with the highest strain field, i.e., the $[10\bar{1}]$ direction, has a consistently lower $T_{MIT}$ value, and on increasing the applied magnetic field, the MIT temperatures converged [279]. Jhang *et al*. [355] also studied the effect of AES on LPCMO ($x$ = 0.33 and $y$ = 0.20) films, grown on cubic (001) LSAT and orthorhombic (001) and (110) NGO substrates having a negligible lattice constant mismatch but various anisotropic strains in the film plane. A bulk-like FM ground state on (110) NGO and a coexisting AFMI state with high phase instability over a wide temperature range on (001) LSAT and (001) NGO substrates were observed. The study also showed that after annealing under the same conditions, epitaxial LPCMO films grown on lattice-closely-matched substrates exhibit strikingly different magnetotransport behaviour.

Recently, a cooperative effect of AES and chemical doping (internal strain) on the transport and magnetic properties of LPCMO (x = 0.33 and $0 \leq y \leq 0.4$) films grown on (001) LSAT, (001) NGO, and (110) NGO substrates was reported [356]. The results showed a reduction in the average Mn–O–Mn bond angles on increasing the Pr doping ($y$) in the LPCMO film from 0.0 to 0.4, (Fig. 16 (a)), which tends to localize the charge carriers, thus facilitating the COI phase [331, 354]. While the substrates showed a progressive increase in the in-plane AES for these manganite films (Fig. 16 (b)). The enhanced AES improves the orthorhombicity of the films by increasing the rotations and deformation of the $MnO_6$ octahedra [356], which also favours the COI phase. Thus, the cooperative effect of these parameters (AES and chemical doping) was confirmed by the enhanced magnetic fields needed to melt the COI phase as a function of AES and the doping level [356]. The strain-induced transport data from three substrates (different epitaxial strains) for the LPCMO ($x$ = 0.33 and $y$ =0.20) film (Fig. 16 (c)) show a variation in MIT temperature as well as the co-existence of different phases, which occur at different temperatures in different substrates [356]. The LPCMO films grown on (110) NGO and (001) LSAT show an FM ground state with phase coexistence with AFM-COI near MIT. The LPCMO film grown on (001) NGO showed insulating behaviour with a sharp growth of resistivity at 266 K, and no MIT was observed at low temperatures [356]. The effect of Pr doping on the transport properties of LPCMO films grown on different substrates was also studied (Figs. 16 (d to f)), indicating that the chemical pressure also facilitates the growth of the COI phase. The study found that AES and chemical doping can cooperatively facilitate the COI phase in LPCMO films, and it is more robust in LPCMO films grown on NGO(001) than the films grown on the other two substrates [356].

Further, the effect of epitaxial strain on the transport and magnetic properties of different manganite films are given in Table 3, where the values of $T_c$ or $T_{MIT}$ of a few manganite films grown on different substrates [82, 112, 181, 279, 329, 332, 334, 338, 348, 349, 351, 354] as a function of in-plane epitaxial strain are listed. The variation of $T_c$ and $T_{MIT}$ as a function of epitaxial strain for different manganite films (by growing manganite films on different substrates or different thicknesses) is plotted in Figs. 17 (a) and (b), respectively, which indicates mixed results on the influence of epitaxial strain on $T_c$ and $T_{MIT}$ of manganite films. Therefore, the coupling of epitaxial strain and magnetism of different manganite films has been varied i.e., in a few cases, in-plane tensile strain favours magnetism, whereas in other cases the compressive strain favours the ferromagnetism (compressive strain increases the $T_c$ or $T_{MIT}$ of manganite films).

## 6.2 Vertically aligned nanocomposite thin films and magnetotransport

In contrast to epitaxial strain, lattice strain in much thicker films can be maintained in two-phase VAN films because of vertical interface coupling. Other factors such as the density of the secondary phase, column dimensions, and morphologies also play a major role in the strain tuning of VAN films. A wide range of multifunctionalities, such as multiferroicity, ferroelectricity, low field magnetoresistance (LFMR), and anisotropic electrical/ionic transport properties, have been demonstrated in VAN systems. Lebedev *et al.* were the first to demonstrate vertical nanopillar microstructures in an LCMO ($x$ = 0.33): MgO nanocomposite thin-film system [357]. The MR effect in several manganite composite thin films has been reported to be more favourable for device applications. A phase transition–induced large CMR in LCMO ($x$ = 0.33): MgO films were observed [284], where the MgO boundary played a significant role in magnetotransport. The MgO phases were concentrated exclusively within the phase boundaries of the LCMO ($x$ = 0.33) phases in these nanocomposite films. An LFMR of 12% at 77 K and 1 T in $[LSMO(x = 0.3)]_{0.5}:(ZnO)_{0.5}$ nanocomposite films on $Al_2O_3$ substrates was reported by Kang *et al.* [358]. Tunable and enhanced LFMR was observed in epitaxial $[LSMO (x = 0.3)]_{0.5}:(ZnO)_{0.5}$ self-assembled VAN thin films grown on (001) STO substrates by PLD [283]. These results were attributed to spin-polarized tunneling across the artificial vertical grain boundaries introduced by ZnO nanocolumns and the enhancement of spin fluctuation depression at the spin-disordered phase boundary regions.

A comparison of the magnetic and MR properties of the LSMO ($x$ = 0.3) single-phase and VAN films of $(LSMO)_{0.5}$:$(CeO_2)_{0.5}$ grown on STO substrates [282] suggested an enhancement in the coercivity and LFMR properties. Later, Fan *et al.* [359] investigated the tunable and enhanced LFMR properties near the RT of VAN $[LSMO (x = 0.3)]_{1-y}$:$(CeO_2)_y$ thin films with varying compositions, $y$ = 0, 0.3, 0.4, 0.45, 0.5, and 0.55. Further, an increase in the LFMR properties of LSMO ($x$ = 0.3): $CeO_2$ nanocomposite films was achieved by implementing a three-dimensional (3D) strain scheme in epitaxial thin films, which was achieved by combining 2-phase VAN with thin interlayers to effectively couple the vertical interface strain with the lateral interface strain [360]. Thus, these results on manganite-based VAN thin-film systems provide strain-driven magnetotransport properties by growing a functional oxide (manganite) in another oxide matrix with a vertical heteroepitaxial form. This may be a promising approach for achieving new functionality that may not be easily realized in the single epitaxial phase.

## 6.3 Piezoelectric strain-driven transport and magnetism

The magnetoelectric effect (ME) [361] observed in multiferroic materials, that simultaneously exhibit magnetism and ferroelectricity, opens a way to electrically control magnetization [57]. However, magnetism and ferroelectricity are unlikely to coexist because the former breaks time-inversion symmetry and the latter breaks space-inversion. An alternative to magnetoelectricity is to combine FE and FM materials in two-phase heterostructures, usually in the form of epitaxial thin films [287, 362]. The strain-mediated ME coupling is one of the main coupling mechanisms between the FE and FM order parameters [115, 294, 362, 363], and the ME coupling constant ($\alpha$) can be defined as [287, 363]:

$$\alpha = \mu_0 \, \Delta M / \Delta E \qquad (6)$$

where $\Delta M$ is the change in the magnetization along the direction of the applied magnetic field, $\Delta E$ is the applied electric field perpendicular to the FM/FE interface [287] and $\mu_0$ is the permeability of free space. For strain-mediated ME heterostructures, an ME coupling constant as high as $2\times10^{-7}$ s/m has been reported [287]. A striking example of device design, where strain plays a fundamental role, is the piezoelectric/manganite heterostructure, which is the prototype of the ferroelectric-gate field-effect device [364].

Strain and electric field control of magnetism in manganite films is achieved by growing epitaxial manganite films on ferroelectric (single crystal) substrates [83, 285-288, 290-294, 363, 365-370]. These reports showed that the BTO ferroelectric substrate is a suitable candidate for providing dynamical strain to modulate the magnetic and electric transport properties of manganite films. Strong magnetoelectric coupling was induced in the LCMO ($x$ = 0.3)/BTO heterostructures because of dynamical strain by applying an electric field to the BTO substrates [363]. Magnetization changes of approximately 30%–40% [Figs. 18 (a) and (b)] were obtained in the LCMO ($x$ = 0.3) manganite film upon applying an electric field of the order of 1 kV/cm to the BTO substrate, corresponding to a magnetoelectric coupling constant ($\alpha$) of the order of ~ (2 to 5)×$10^{-7}$ s/m [363]. A change in the magnetic anisotropy of this film by the electric field-induced strain was also observed, which resulted in a considerable reduction of the coercive fields [Fig. 18 (b)]. In addition, the ME coupling in the LCMO ($x$ = 0.3)/BTO heterostructure was found to be the same for positive and negative electric fields, demonstrating a negligible contribution to the otherwise dominant role of interface reconstruction-based coupling mechanisms [363].

The influence of dynamical strain also led to giant sharp magnetoelectric effects in the LSMO ($x$ = 0.33)/BTO multiferroic heterostructure [287], and this study reported electrically induced giant, sharp, and persistent magnetic changes over a wide range of temperatures, including room temperature. Dale *et al*. [370], reported a prominent influence of dynamical strain on the electrical transport properties of the LSMO ($x$ = 0.5) film and found that the resistivity of LSMO films can be modulated by the electric field (a change of ~12%) as well as by the phase transition-induced strain of BTO (a change of ~ 30%). Burton *et al.* [371] reviewed theoretical studies on the control of magnetization by the application of an electric field at the interface between LSMO ($x$ = 0.5) and BTO. Using first principle calculations, it was shown that a change in the electron density at the interface due to the reversal of the ferroelectric polarization of the BTO leads to a change in the magnetic order at the interface from FM to AFM.

Therefore, the electric field ($E$-field) control of resistivity (transport) and MIT transitions for manganite films has been widely studied by growing manganite films on piezoelectric (PZT) substrates [372-375]. Wu et al. [372], observed a large change in resistance (76%) for the LCMO ($x$ = 0.3) film on ferroelectric PZT at a modest $E$-field of ~ 4×$10^5$ V/cm. A similar change in electro resistance (ER) and a 35 K shift in $T_c$ were observed for the LSMO (x = 0.2) film, using the polarization field of the PZT in the LSMO/PZT heterostructure [373]. The effect of electric and magnetic fields on the ER of the LCMO ($x$ = 0.2) film was studied in an LCMO/PZT heterostructure [374], and a maximum modulation in the ER of 20% after applying an electric field poling of $1.5\times10^5$ V/cm, and 50% under a magnetic field of 1 T near the $T_{MIT}$ of the LCMO film was observed. The $E$-field control of the metal-insulator transition at room temperature for $La_{1-x}Ba_xMnO_3$ (LBMO) ($x$ = 0.10 and 0.15) films in LBMO/PZT heterostructures was also studied and found a ~5% change in both ER and $T_{MIT}$ by ferroelectric polarity [375].

Thiele et al. [293] demonstrated the principle of reversible control of film strain by applying an electric voltage to a piezoelectric (PMN-PT (001)) substrate in an epitaxial LSMO ($x$ = 0.3)/PMN-PT (001) heterostructure. The epitaxial strain in the manganite films grown on the piezo crystal (PMN-PT) was controlled by applying an electric field across the piezo crystal. Figure 18(c) shows the reversible in-plane strain of a PMN-PT crystal (substrate) recorded by laser interferometry [294]. The substrate (PMN-PT) shrank approximately linearly with increasing $E$-field, applied along both in-plane perpendicular directions [294]. Reversible tuning of the lattice strain and, hence, the physical properties (resistance, magnetization) of the films by application of an electric field were also realized for LSMO ($x$ = 0.3) and LCMO ($x$ = 0.3) films epitaxially grown on PMN-PT due to a huge converse piezoelectric effect of the substrate [294]. An enhanced magnetization of the LSMO film [Fig. 18(d)] by approximately 25% at 330 K was observed on the application of E = 10 kV $cm^{-1}$ on LSMO/ PMN-PT heterostructure. The study also reported strain-induced shifts of the $T_c$ of up to 19 K for LSMO and LCMO films and an effective ME coupling coefficient of ~ $6\times10^{-8}$ $sm^{-1}$ at ambient temperature in the manganite film/PMN-PT heterostructure systems, which mainly resulted from the strain-induced magnetization change. The large ME coupling coefficient in this LSMO/PMN-PT heterostructure was attributed to different properties [294]: (a) the strong strain dependence of magnetization in the ferromagnetic manganites, (b) the large piezo-strain of the PMN-PT substrates, and (c) effective elastic coupling at the film-substrate interface. Later, a 10 K shift in the magnetic $T_c$ was reported for the LSMO ($x$ = 0.3) film grown on a PMN-PT substrate by reversing the electric polarization of the FE substrate [376]. Therefore, piezoelectric substrates help to set two distinct remanent strain states in the substrate by poling the FE polarization out-of-plane or parallel to the surface, which influences the electronic and magnetic properties of the manganite film grown on the FE substrate.

The electrical transport properties of LCMO ($x$ = 0.25) thin films grown on ferroelectric PMN-PT single-crystal substrates also indicated a decrease in resistance and an upward shift in $T_{MIT}$ of the LCMO films, which was attributed to the ferroelectric polarization-induced lattice strain effect [292]. The resistance of the LCMO films was modulated by applying dc/ac electric fields across the polarized PMN-PT substrates, and the lattice strains in the PMN-PT substrates were induced by the electric field via the converse piezoelectric effect, which subsequently changed the strain state and resistance of the LCMO films [292]. In addition, the *E*-field control of PS and memory effect for PCMO ($x$ = 0.40) films in PCMO/PMN-PT heterostructures showed that the magnetization of the PCMO films can be tuned dramatically by electric fields via piezo-strain, and the effect was dominated by the change in PS [377]. More interestingly, the electric-field control of magnetization is non-volatile, suggesting a memory effect of strain in the film.

Zhang *et al.* [369] reported *in situ* control of EPS in the LPCMO ($x$ = 0.375 and $y$ = 0.8) film in the LPCMO/PNM-PT heterostructure using ferroelectric-poling-induced strain. It was found that the ferroelectric poling reduces the in-plane tensile strain and enhances the out-of-plane tensile strain of LPCMO film, which decreases the resistance and the charge ordering transition temperature ($T_{CO}$) but raises the low-field-magnetization of films. These results were attributed to the strain-induced change in the volume fraction of coexisting phases, *e.g.*, FM, AFM, etc. Using reversible uniform biaxial strain from piezoelectric (PMN-PT) substrates [293, 295], Dekker *et al.* [378] investigated the direct response of the electrical transport and the magnetization of epitaxial LPCMO ($x$ = 0.3 and $0.0 \leq y \leq 1$) films to a reversible biaxial strain in LPCMO/PMN-PT heterostructures. This study revealed a drastic reduction in the resistance/a "colossal" ER, a higher saturation magnetization, and an increase in $T_c$ of LPCMO films upon controlled release of the piezoelectrically tensile strain. The finding of both magnetization and resistance data in controlled strain states in LPCMO/PMN-PT heterostructures demonstrate a strong suppression of the ferromagnetic double exchange interaction by tensile strain in the films [378]. Low-temperature persistent photoconductivity (PPC) for LPCMO ($x$ = 0.37 and $y$ = 0.38) films in LPCMO/PMN-PT heterostructures, when illuminated by visible light of 532 nm, was also observed [379]. Thus, this study indicated that both the resistance and the PPC were modulated by electric fields across PMN-PT, which were attributed to the change of lattice deformation in LPCMO films on the application of electric fields via the converse piezoelectric effect. Photoinduced transport properties of the $Pr_{0.65}(Ca_{0.75}Sr_{0.25})_{0.35}MnO_3$ manganite film grown on a PMN-PT substrate were also observed, which was modulated by the applied electric field across the heterostructure [380]. These results confirmed that manganite films show a decrease in resistance up to five orders of magnitude by enhancing applied electric fields, which are combined with an electric-field-induced $T_{MIT}$. A reversible bi-stability in the photoinduced change in resistance at T < 80 K was also observed on increasing the voltages [380].

Application of *in-situ* in-plane compressive strain to LPCMO films grown on ferroelectric PMN-PT substrate suggested that the in-plane compressive strain, as a result of the piezoelectric strain, favours the FMM phases [369, 378, 379]. Attempts were made to understand and separate the influence of strain and magnetic field on the relative strength of EPS evolution in the phase-separated LPCMO ($x$ = 0.33 and $y$ = 0.5) films, which were epitaxially grown on (001)-oriented PMN-PT, and (111)-oriented $0.31Pb(In_{1/2}Nb_{1/2})O_3$-$0.35Pb(Mg_{1/3}Nb_{1/2})O_3$-$0.34PbTiO_3$ (PIN-PMN-PT) ferroelectric single-crystal substrates [278]. A dramatic decrease in resistance, up to 99.98%, and complete melting of the charge-ordered phase was observed in the manganite film by the poling-induced strain along the [001] or [111] direction. The observed modification in transport properties was attributed to the poling-induced strain effects, which can be effectively tuned by a magnetic field and mediated by EPS. In particular, this study showed that the evolution of the strength of EPS against temperature and magnetic field can be determined by measuring the strain-tunability of resistance under magnetic fields [278].

While electric field-induced strain (piezoelectric response) in manganite/FE hybrid structures was well studied for controlling the possible ME effect and its influence on the transport and magnetic properties of manganite thin films. There have been considerable efforts to study the influence of epitaxial strain on the electronic and magnetic properties of manganite/FE hybrid systems, where strain was generated due to pure structural phase transition of the FE crystal substrate, in particular the BTO. This approach provides a controlled way to study strain-driven properties of the films at the transition temperatures of the FE crystal by keeping parameters fixed, e.g., growth conditions, film thickness, and morphology. The temperature-dependent volatile ER was observed in LSMO ($x$ = 0.33)/BTO heterostructure, which first increases with an increase in the temperature and then decreases, and this phenomenon is related to the FE phase transition of the BTO substrate [286]. Lee *et al*. [83] studied the effects of induced biaxial strain on the electrical transport and magnetic properties of epitaxial thin films of $SrRuO_3$ and LSMO ($x$ = 0.33) using structural transitions of ferroelectric BTO substrates. Large and abrupt changes in resistivity and magnetization of both $SrRuO_3$ (5% change in resistivity) and LSMO films (12% change in resistivity and ~ 70% in magnetization) were found at the phase transition temperatures of BTO. This study demonstrated that for manipulating lattice strain of epitaxial thin film heterostructures, the use of phase transitions of single crystal ferroelectric substrates can be a useful way to modify the properties of perovskite oxides. Recently, Schmitz *et al*. [285], reported the observation of strain- and magneto-electric coupling in LSMO ($x$ = 0.5 and 0.3) on a ferroelectric BTO substrate. Temperature-dependent magnetization for LSMO ($x$ = 0.5 and 0.3) [Figs. 19 (a) and (b)] in the field-cooled cooling (FC) and field-cooled warming (FW) cycles in an applied in-plane magnetic field of 10 mT shows sharp steps at the structural phase transition temperatures (emphasized by grey boxes) of BTO [Fig. 19]. These pronounced magnetization steps at the BTO structural phase transitions are attributed to a strain-induced change of the magnetic anisotropy and the exchange interaction of the LSMO layer induced by the BTO substrate. A temperature-dependent magneto-electric coupling was also found by exploring the magnetic response to an applied AC electric field in all ferroelectric phases of the BTO substrate [285]. Using complementary methods, such as angle-dependent magnetization, XAS, and transport measurements, Panchal *et al.* [365] concluded that strain is the decisive factor controlling magnetic anisotropy switching in epitaxial LSMO ($x$ = 0.3) film on BTO substrate. A magneto-elastic coupling in the LSMO/BTO heterostructure with electric-field tunability was observed, which was manifested in a discontinuous change in magnetization and resistivity data at different temperatures where BTO undergoes successive structural phase transitions [365]. It was inferred that upon the application of an electric field, large magnetization changes occur via the strain-mediated converse magnetoelectric effect in the LSMO/BTO heterostructure [365]. Strain-induced tunable anisotropic magnetoresistance (AMR) in LCMO ($x$ = 0.33) films epitaxially grown on (001) BTO single crystal was investigated by Xie *et al*. [381]. This study demonstrated that not only the strain value due to structural phase transition in BTO but also the distortion type can considerably tune the AMR of LCMO films. Therefore, ME composite systems with FM/FE compounds, e.g., LSMO/BTO, LSMO/PZTO, etc., not only help in understanding the strain-induced macroscopic (transport and magnetic) properties of manganite (FM) films but also stimulate the ever-increasing research activities for their technological applications in multifunctional devices, such as sensors, transducers, and memories, etc., where strain-driven magnetic and transport properties play an important role.

## 6.4 Bending strain and magnetism

Bending strain in the form of the cantilever bending beam technique [382, 383] was used long ago to study the effect of stress (strain) on magnetization and magneto-elastic coupling of magnetic thin films. Pellegrino *et al*. [384] employed ‘*strain-generator devices*’ for studying the effect of strain and electric field on the electrical transport properties of manganite films, where a crystalline suspended bridge of STO substrate was used as a flexible substrate for the deposition of functional epitaxial LSMO thin films. Using this approach, the film was bent both mechanically by an AFM tip and electrically by applying an electric field. Recently, the application of flexible electronics in the area of flexible displays, sensors, artificial skins, etc., is growing steadily because it provides the electronics with mechanical flexibility, portable, lightweight, biocompatible, and high scalability [385-387]. The flexible electronic consists of electronic components such as ultra-thin chips on flexible substrates [Fig. 20 (a)] (typically a plastic substrate), which can be easily bent. Bending strain is one of the main factors for testing the mechanical reliability of these flexible electronics. Therefore, attempts have been made to study the effect of bending strain on the physical and magnetic properties of magnetic films by direct deposition of magnetic materials on flexible substrates and then fixing the flexible magnetic system on a bending object [Fig. 20 (b)] [388-392].

The influence of bending strain on the magnetic anisotropy in the $Co_2FeAl$ (CFA) Heusler thin film grown on Kapton flexible substrate, where 20 nm thick CFA film was grown on Kapton, was investigated for the magnetoelastic coupling [388]. The bending strain in this flexible magnetic system was achieved by glueing the sample on curved aluminium blocks of different known radii [388]. This study reported a drastic change in magnetic anisotropy by bending the thin films, which was attributed to the interfacial strain transmission from the substrate to the film and the magnetoelastic behaviour of the CFA film [388]. A good quality epitaxial LSMO ($x = 0.33$ ) film with excellent ferromagnetic and magnetoresistance properties was realized by Huang *et al*. [391] on growing LSMO films on flexible mica substrates. Interestingly, this study did not show any deterioration in the magnetic properties of LSMO film under mechanically bending conditions, suggesting the good mechanical stretchability and stability of the LSMO thin films on flexible mica substrates. To investigate the robustness and stability of the LSMO films under bending conditions, the mica substrate with the LSMO film was glued onto a curved plastic mode under tensile (convex) and compressive (concave) strain geometry [inset of Fig 20 (c)] and the in-plane *M*(H) curves were measured [Fig. 20 (c)] for both bending conditions (after 1 or 10 times of bending). This experiment revealed no significant change in *M*(H) curves for different repeats and indicated robust magnetic performance with mechanical flexibility [391]. In contrast, Fan *et al*. [392] observed an increase in $T_C$ [inset of Fig. 20 (d)] upon bending of the LSMO ($x = 0.2$) film grown on flexible mica substrate with BTO as a buffer layer in the LSMO/BTO/mica heterostructure. Figure 20 (d) shows the *M*(*T*) curves from LSMO/BTO/mica heterostructure before and after the cycling bending $10^4$ times with an in-plane strain of ~ 0.07%. The results from this study suggested an increase in $T_c$ to 143 K for the film after bending, which is 6.0 K higher than that of the film without cycling bending. The *M*(*H*) curves for the heterostructure before and after cyclic bending [Fig. 20 (e)] also indicated that the ferromagnetism in the film is increased through cycling bending.

In recent ground-breaking experiments [393, 394], membranes of the perovskite manganites deposited on a flexible polymer layer were studied to investigate the correlation of strain and magnetism in these systems. Lu *et al.*, [393] grew a freestanding membrane of LSMO ($x = 0.3$) by etching the water-soluble $Sr_3Al_2O_6$ (SAO) layer grown between manganite film and substrate. The structural evolution and correlation of the transport and magnetic properties of LSMO film before etching and freestanding LSMO film after release from the substrate were characterized. The study demonstrated that the physical structure of the LSMO membrane was preserved and showed improved magnetic (higher $T_c$) and transport (low resistivity) properties [393]. A similar approach was adopted by Hong *et al.* [394] to get the membrane of LCMO ($x = 0.3$) by growing heterostructures of LCMO/$SrCa_2Al_2O_6$ (SCAO)/STO and the SCAO was "sacrificed" after the growth by dissolving the heterostructure in the water. The extreme tensile (increase in the lattice constants) strain was achieved in LCMO membranes, exceeding 8% uniaxially, a strain never reported before, by transferring the LCMO membrane onto a polymer, used as a flexible substrate. The study suggested that a uniaxial and biaxial strain suppresses the FMM at distinctly different strain values, inducing an insulator that can be quenched by a magnetic field [394]. These experimental results were further tested by a theoretical approach using Monte Carlo simulations of the two-orbital double-exchange model including JT distortions [395] and found a good agreement with experimental findings and theoretical analysis.

The epitaxial and piezoelectric strains (both by phase transition and application of electric field) applied on manganite films are generally of the order of $\geq 0.25\%$ [82, 83, 112, 293, 294, 332], which provided a mixed report regarding the influence of strain on the magnetic properties of manganite films. Magnetic films grown on a flexible substrate and bent on the bending mode will also produce high stress on the film. Considering that Young's modulus of manganites is on the order of 200 GPa [85, 334] and the yield strength of perovskite ceramic (e.g., SrTiO3) is on the order of 120 MPa [396], the maximum elastic strain estimated for the manganite film is ~ 0.06%. Therefore, the applied stress for manganite films by epitaxial and piezoelectric mode (a strain of 0.1% to 0.2%) may not be elastic. The bending strain applied by a mechanical jig (4-point bending) and 3-point beam bending techniques for manganite films grown on a rigid substrate is of the order of ~ 0.01-0.03 % [84-88, 397], which confirms that the applied bending strains are purely elastic. Moreover, the bending of films on a rigid substrate provides a uniaxial strain, which is reversible. Therefore, to understand the influence of (true) elastic strain on the transport and magnetic properties of manganite films, here we discuss the studies on a few manganite films grown on rigid single crystal oxide substrates, which are bent for applying the strain. Since one requires a specific bending device (mechanical jig etc.) to uniformly bend the thin films there are only limited studies reported in the literature including extensive studies on LPCMO films with different compositions [26, 35, 39, 84-87, 251, 252] and LSMO film [88]. In this section, we will discuss these two systems (LPCMO and LSMO) in detail. The properties of these manganite films upon bending stress show similar properties as revealed by other manganite films subjected to other modes of strain (epitaxial, piezoelectric etc.) which are discussed earlier. Bending strain provides other advantages as described earlier. In the following subsections, the free energy phenomenological approach for understanding the coupling of bending (uniaxial) strain on the magnetism of manganite films is described and a review of the influence of bending strain on the transport and magnetic properties of manganite (especially LPCMO and LSMO) films is provided.

### 6.4.1 Free energy for coupling of bending strain and magnetism

The interpretation of the electric and magnetic properties of manganite perovskites has been given traditionally based on the DE mechanism [28, 142, 190]. Millis *et al*. [158, 180] demonstrated that the DE picture is not enough to explain the CMR properties and proposed a strong electron-phonon interaction as an additional term in the DE model, which arises from the JT splitting of the outer Mn *d* level. Experimentally manganite materials are found to be phase-separated into nanometer to sub-micrometre length scale mixtures of insulating regions and metallic ferromagnetic domains [2, 23, 398]. The percolation of this inhomogeneous structure of coexisting metallic and more insulating areas gives the CMR effect and explains possible first-order magnetic phase transitions in these manganite systems. Bean and Rodbell [399] developed a theory of first-order phase transition due to magnetic disorder by incorporating the molecular field model and exchange interaction, which is strongly dependent upon lattice spacing, and calculated the minimum of the Gibbs free energy. Banerjee [400] observed the essential similarity between the Bean–Rodbell approach [399] and the classical formulation of Landau–Lifshitz theory [111, 401, 402] and concluded that a negative slope of isotherm plots of $H/M$ vs. $M^2$ ($M$ is the experimentally observed magnetization, $H$ the applied magnetic field) would indicate a first-order phase transition. Nevertheless, despite the clear definition of first- and second-order phase transitions, the experimental determination of such character is usually complex.

Ahn *et al*. [29, 225, 403] developed a theoretical analysis for predicting phase texture and phase coexistence of manganite films, based on an elasticity model, which combined the continuum strain and collective displacement of atoms. In this model, they represented the total energy of the system for metal-insulator phase coexistence as the sum of three terms [29]:

$$E_{tot} = E_s + E_a + E_c = \sum (A\varepsilon_i^2 + B\Delta_i^2 + C\Delta_i^2\varepsilon_i) \qquad (7)$$

The first term is the elastic strain energy, $E_s$, and is proportional to the square of strain ($\varepsilon^2$) through stiffness *A*. The second term, $E_a$, is the energy of collective displacements of atoms within a unit cell. This term is related to the square of the displacements ($\Delta^2$) through a constant *B*. The third term, $E_c$, couples the strain and displacements through a constant *C*. By minimizing $E_{tot}$ with respect to $\varepsilon$ and $\Delta$, they produced different maps representing a landscape of mixed conducting and insulating phases. Moreover, this model reproduces many of the important features of the experimental findings. However, the interplay between strain and magnetization, which pose more complexity in the system was not considered by Ahn *et al*. [29].

Pressure-induced quantum phase transitions in many FM and AFM systems have been studied using a quantum critical point (QCP) approach [404]. Gehring [405] showed that Landau's theory [111, 401, 406] can be successfully applied to account for the first-order transition seen in compressible magnets near the QCP. Ginzburg–Landau theory [111, 401, 406] allows the study of phase transitions in a phenomenological way and it consists of expressing the free energy as a power expansion of the order parameters and their gradients. Milward *et al*. [407] showed that manganites, exhibiting the coexistence of complex phases of the two-order parameters of magnetism and charge modulation, can be naturally explained within a phenomenological Ginzburg–Landau theory. Three contributions of the free-energy density: magnetization, charge modulation, and coupling between them, were considered. In contrast to models where PS originates from disorder [408] or as a strain induced kinetic phenomenon [29], it was argued that magnetic and charge modulation coexist in new thermodynamic phases. However, the coupling of strain and magnetism was not considered explicitly.

The coupling of strain with magnetism under Ginzburg–Landau theory was considered to understand the change of magnetization ($M$) of LPCMO films on the application of bending strain ($\varepsilon$) [85-87]. The free energy density, which contributes to the coupling is given as follows [85]:

$$F_c = \gamma \epsilon M^2 - \frac{A}{2}\varepsilon^2 \qquad (8)$$

The first term is the contribution of applied bending (uniaxial) strain and magnetization (magnetization under uniaxial strain) of the system and the second term is the energy associated with the applied strain to the system. The $\gamma$ is coupling constant, which gives the idea of coupling between strain and magnetization and $A$ is Young's modulus which can be estimated for the system (both bulk and film) using nanoindentation [409] and it was found to be 200 GPa for LPCMO films [85]. At equilibrium, we obtain a relation for the coupling constant ($\gamma$) by minimizing $F_c$ with respect to strain ($i.e.,\ \frac{dF_c}{d\varepsilon} = 0,\ M = M_\varepsilon$), yielding $\varepsilon = \frac{\gamma M_\varepsilon^2}{A}$ and $F_c = -\frac{\gamma^2 M^4}{2A}$. The total magnetization on the application of bending strain (uniaxial) is given by: $M^2 = M_0^2 + M_\varepsilon^2 = M_0^2 - \frac{A\varepsilon}{\gamma}$, where $M_0$ is the saturation magnetization of the system at $\varepsilon$ = 0. Thus, from the slope of $M^2$ vs $\varepsilon$ curve ($\sim -A/\gamma$), the coupling constant is determined experimentally. Considering the strain-magnetism coupling discussed above, the magnetic phase transition (first or second order) of the manganite thin film systems can be predicted by a simple and more transparent phenomenological approach using the Ginzburg–Landau theory. The free energy density for systems with strong coupling between strain and magnetism can be given as:

$$F = \left(\frac{T - T_c}{2}\right) M^2 + \alpha M^4 + \beta M^6 + F_c = \left(\frac{T - T_c}{2}\right) M^2 + \frac{(\alpha - \frac{\gamma^2}{2A})}{4} M^4 + \frac{\beta}{6} M^6 \qquad (9)$$

where $\alpha$ and $\beta$ are positive coefficients. $T_c$ is the transition temperature (ferromagnetic to paramagnetic order) for the system without applied strain, $\varepsilon$ = 0. Thus, depending on the nature of the coupling following cases arise:

(a) If $\alpha > \frac{\gamma^2}{2A}$ i.e., the coefficient of $M^4$ is positive. The magnetic phase transition shows a second-order phase transition with the same transition temperature ($T_c$), $T_c^\varepsilon = T_c$.

(b) If $\alpha < \frac{\gamma^2}{2A}$ i.e., the coefficient of $M^4$ is negative. This is the condition for the first-order phase transition, which can occur in phase-separated manganite thin films due to applied strain. The extrema of $F$ can be found using $\frac{dF}{dM} = 0$; which gives:

$$M^2 = \frac{-\left(\alpha - -\frac{\gamma^2}{2A}\right) \pm \sqrt{\left(\alpha - -\frac{\gamma^2}{2A}\right)^2 - 4(T - T_C)\beta}}{2\beta} \qquad (10)$$

The $M^2$ has two solutions with (±) sign in equation (10) and a negative value of the $M^2$ is responsible for getting the minimum value of energy density. The first-order phase transition happens when $F(M) = F(0) = 0$, and $T = T_c^\varepsilon$ (transition temperature in applied strain), which gives:

$$\left(\frac{T - T_c}{2}\right) M^2 + \frac{\left(\alpha - \frac{\gamma^2}{2A}\right)}{4} M^4 + \frac{\beta}{6} M^6 = 0$$

$$\Rightarrow \quad \left(\frac{T - T_c}{2}\right) + \frac{(\alpha - \frac{\gamma^2}{2A})}{4} M^2 + \frac{\beta}{6} M^4 = 0 \qquad (11)$$

Now using $\frac{dF}{dM} = 0$ from equation (9) we obtain:

$$M^4 = -\frac{\left[(T - T_c) + \left(\alpha - \frac{\gamma^2}{2A}\right) M^2\right]}{\beta} \qquad (12)$$

Using equations (10), (11), and (12) we find the shift in transition temperature as:

$$T_c^\varepsilon = T_c + \frac{3}{16}\left[\frac{\left(\alpha - \frac{\gamma^2}{2A}\right)^2}{\beta}\right] \qquad (13)$$

Thus, the strong coupling of strain and magnetism in a manganite thin films can be studied by measuring the magnetization of the system as a function of strain, and depending upon coupling strength the phase transition can be explained, either by first or second-order transitions using the simple phenomenological approach of Ginzburg–Landau theory. Though measuring a shift in transition temperature and predicting the magnetic phase transition, require the knowledge and estimation of the coefficients $\alpha$ and $\beta$, which may involve measuring magnetization as a function of field and applied strain. However, measurements of the coefficients $\alpha$ and $\beta$, are not straightforward. However, by measuring the magnetization of the manganite films as a function of applied bending strain one can quantitively estimate the coupling constant ($\gamma$). The $\gamma$ of ~0.00029(1) $N/A^2$ was obtained for LPCMO ($x$ = 0.33 and = 0.60) film grown on NGO substrate [85]. This theoretical phenomenological approach suggests that the strain field and its coupling with magnetism can drive the phase transition from second-order to first-order, which is the case for manganite complex oxides.

### 6.4.2 Influence of bending strain on transport properties

The influence of strain on transport properties of manganite thin film systems has been studied extensively by measuring the resistance/resistivity of the film along both in-plane perpendicular directions and on the application of bending strain along these directions by different methods (three-point or four-point bending). Although bending experiments have been performed on several materials including semiconductor nanowires [410] and ferroelectrics [315, 411], it is difficult to perform a bending measurement on manganite thin films grown on oxide substrates due to the brittle nature of oxide substrates. In general, thinner substrate (thickness ≤ 0.25 mm) helps to achieve higher bending in the film.

Bulk LPCMO, manganite systems with different doping concentrations [2, 25, 29, 90, 194] exhibit PS in the range of nanometres to microns for $x$ = 0.33 [23] and 0.375 [2]. Moshnyaga *et al.* [412] predicted EPS in the LPCMO ($x$ = 0.33 and $y$ = 0.40) thin films that develop at the nanometre scale, where adjacent FM nanodomains are antiferromagnetically coupled. The EPS at the nanometre length scale at the surface of LPCMO ($x$ = 0.33 and $y$ = 0.60) film was observed using a conductive atomic force microscope (cAFM) [26]. Transport properties in manganites highly depend on the PS behaviour of materials. The transport properties of LPCMO ($x$ = 0.33 and $y$ = 0.50) films on the application of bending elastic strain using the three-point bending technique were reported earlier, which indicated a change in $T_{MIT}$ of film on the application of strain [84]. This study has also shown that the bending strain has a profound effect on the transport properties of phase-separated manganites (LPCMO) due to the difference in the structures of the FMM and COI phases, which elucidates the underlying mechanism of PS in manganites. The transport properties of LPCMO films with fixed $y = 0.60$ and a small change in $x$ from 0.33 to 0.375 show drastically different behaviours in terms of resistance, $T_{MIT}$, and width of hysteresis in the resistance *vs* temperature [$R(T)$] curves [87], which also depend on the cooling/warming rates of the sample during $R(T)$ measurements.

To separate the effect of bending strain and chemical strain on the transport properties of LPCMO thin films, we will first discuss the effect of change in composition ($x$) (chemical pressure/strain) and applied magnetic field on the transport properties of LPCMO films with $x = 0.33$ to 0.375, without applying a bending strain. Figure 21 (a) shows the temperature-dependent resistance normalized to the resistance at 200 K [i.e., $R$ (T)/$R$ (200 K)] from LPCMO ($x = 0.33$ and $x = 0.375$) films for a fixed $y = 0.60$ grown on NGO (110) substrate, while films were cooled and warmed at a rate of 0.3 K/min and without bending strain. The normalized $R(T)$ curves for LPCMO ($x = 0.33$) show an insulator to metal ($T_{IM}$) (while cooling) and metal to insulator ($T_{MI}$) (while warming) transition temperatures (location of the peak of d$R$/d$T$) of ~70 K and 89 K, respectively, with a difference of $\Delta T$ (hysteresis width) ~19 K. Whereas the MIT for LPCMO ($x = 0.375$) film occurred at higher temperatures ($T_{IM} = 117$ K and $T_{MI} = 130$ K) with a smaller thermal hysteresis ~10 K. The large difference in MIT for LPCMO films of similar thickness but a small change in the $x$ (composition) may be arising due to a change in the chemical pressure [2], as a result of different compositions. Specular and off-specular (diffuse) XRMS studies were also carried out on these samples, which shows that the different transport properties for these compositions arise due to different sizes of in-plane charge-magnetic correlation lengths (different domain sizes) [252]. The behaviour of transport measurements on the application of a magnetic field (~ 6 kOe) is illustrated in Figs. 21 (b) and (c) for LPCMO films with $x = 0.33$ and 0.375, respectively. It is evident from these profiles (Figs. 21(b) and (c)) that the application of the magnetic field shifts both $T_{IM}$ and $T_{MI}$ to a higher temperature. Application of the field also shows a reduction in peak resistance, which contributes to the CMR properties of these manganite films.

The anisotropic strain induced by two perpendicular in-plane directions of the substrate (with different lattice constants) is another important contribution of the strain field which can affect the transport and magnetic properties of the manganite film and thus require proper attention before understanding the effect of bending strain. The LPCMO ($x = 0.33$ and $y = 0.60$) film of thickness ~ 30 nm was grown on (110) NGO substrate. The two in-plane perpendicular directions ($x$ and $y$-axis) of the NGO substrate are $(1\bar{1}0)$ and (001), which are depicted in the inset of Fig. 21 (d). In the absence of bending strain ($\varepsilon = 0$), the LPCMO film of thickness ~ 30 nm showed similar temperature hysteresis (same width and same $T_{MIT}$) upon measuring the transport properties [Fig. 21 (d)] along two in-plane perpendicular directions [*x* (001) and *y* $(1\bar{1}0)$ directions in the inset of Fig. 21 (d)]. The study suggested that despite definite epitaxial strains of 0.49% || $(1\bar{1}0)$ and 0.26% || (001) for the LPCMO film along two in-plane perpendicular directions of NGO substrate [353], due to lattice mismatch, a significant difference in epitaxial strain for the two in-plane directions did not affect $T_{IM}$ or $T_{MI}$. This observation was consistent with earlier measurements on similar systems [353] but it is in contrast to another study [279], where a significant difference in these properties was observed. Ward *et al.* [279] found a large difference in $T_{IM}$ and peak resistivity along two in-plane perpendicular directions of the LPCMO (x = 0.375 and y = 0.48) thin film grown on (101) NGO substrate, which was attributed to anisotropic epitaxial strain. Such strain-induced anisotropy in the resistivity has also been predicted theoretically [328]. The large difference in the transport properties of LPCMO films discussed above [279, 328, 353] is believed to arise due to the length scale associated with the measurements. While the chemical pressure (change in composition *x*) in LPCMO films clearly shows different transport and magnetic properties, the in-plane anisotropic strain does not show a significant difference when one measures these properties from thin films of dimensions larger than the PS length scale. Therefore, comparing the bending strain-dependent transport and magnetic properties from a single compositional LPCMO thin film (i.e. LPCMO with $x = 0.33$ and $y = 0.60$) will help us to understand the influence of bending strain on these properties.

On applying pure bending stress (uniaxial) using the four-point bending mechanical jig set-up, as shown in Fig. 13, a significant modulation in transport properties of LPCMO film was observed [87]. Depending on whether the film is on top (compressive strain, $\varepsilon$ = -ve) or bottom (tensile $\varepsilon$ = +ve) in the jig, a uniform stress over a relatively larger lateral dimension of the film was achieved [85, 87]. A comparison of the resistance as a function of applied stress and temperature for different directions of applied stress and transport measurements from LPCMO film ($x = 0.33$ and $y = 0.60$) is shown in Figs. 21 (e) and (f). The normalized $R$ (T)/$R$ (200 K) curves, plotted on a semi-log scale, were measured along the *y*-axis [$(1\bar{1}0$ direction of NGO] as a function of applied strains (without strain, $\varepsilon = 0\%$, compressive strain, $\varepsilon = -0.006\%$ and tensile strain, $\varepsilon = +0.006\%$), which were applied along two perpendicular in-plane directions, i.e., along the x-axis [(001) NGO direction] [Fig. 21 (e)] and along the y-axis [$(1\bar{1}0)$ NGO] [Fig. 21 (f)]. It is evident from Figs. 21 (e) and (f) that an increase (decrease) in $T_{IM}$ (a vertical dashed line in Figs. 21 (e) and (f)) and $T_{MI}$ with an application of compressive (tensile) strain. However, a larger shift in $T_{IM}$ and $T_{MI}$ was observed when the transport was measured in a direction perpendicular to the axis of bending stress [Fig. 21 (f)] as compared to the direction parallel [Fig. 21 (e)] to the axis of bending stress. Thus, the direction dependence of electrical response to external stress indicated a strong influence of stress on the electronic properties of the LPCMO films. The bending stress produced an anisotropic response to transport properties which could be understood in terms of PS length scale and how stress changes the shape and size of metallic regions (coexistence of phase) in the LPCMO films [87].

Recently, Jin *et al.* [397] studied the effect of uniaxial strain, applied both using strain related to the crystallographic axis of the substrate and bending strain, on the ground state electronic (transport) properties of strongly correlated perovskite oxide LCMO ($x$ = 0.34) films deposited on the different crystallographic orientations of the NGO substrates. The LCMO films of thickness ~ 24 nm were deposited on (100)-, (010)-, (001)-, and (110)-oriented NGO substrates, and it was found that LCMO films on (010) and (110) NGO display a bulklike FMM state. In this study, an emergent COI phase was found in LCMO films on (100) and (001) NGO, which showed clear signatures of PS and a complex phase evolution with temperature and magnetic field. Whereas the bending strain-dependent transport properties of LCMO films on (100) and (001) NGO were carried out for understanding the strain-induced COI phase in this system [397]. The strain was applied by the four-point bending technique described earlier. The resistivity ($\rho$) vs temperature curves for LCMO films in (100) and (001) NGO substrate, while cooling and warming the film under different conditions of applied strain, were measured with the direction of current along the in-plane $b$-axis in both the cases and the data are shown in Figs. 22 (a) and (c) for two thin film samples. The applied bending strain ($\varepsilon$) for tensile and compressive mode was +0.03% and -0.03% respectively. The $\rho$ ($T$) curves in Figs. 22 (a) and (c) clearly indicate that the tensile bending stress, +0.03%, along the $b$-axis, the paramagnetic insulator (PI)-FMM transition vanishes and the uniaxial strain drives the LCMO/NGO (100) or (001) films toward a more robust COI state, while the compressive bending stress, −0.03%, gives rise to a suppressed phase hysteresis. In contrast, when applying bending stress along the in-plane $c$ direction in LCMO/NGO (100) and $a$ direction in LCMO/NGO (001) film, the induced COI phase enhances or suppresses only slightly, as shown in Figs. 22 (b) and (d), respectively. The first principle calculations studies on these systems further revealed that the cooperatively increased JT distortion and charge ordering may be essential for inducing and promoting the COI phase in LCMO films.

Huang *et al.* [391] studied the electrical transport (resistivity vs. temperature) properties of LSMO ($x$ = 0.33) films grown on a flexible mica substrate and observed higher MR values near room temperature on bending the film. The influence of external uniaxial mechanical strains on the transport properties of thin epitaxial LSMO (*x* = 0.17) films, where the uniaxial mechanical stain was applied using a standard isosceles triangle-shaped cantilever, was also studied [413]. The thickness-dependent resistance changes of the LSMO films on the application of uniaxial strain showed opposite behaviour in the change of resistance on the application of compressive and tensile stress (Fig. 23). Thinner films (thickness ~ 80-200 Å), which had residual strain due to lattice mismatch, show higher (lower) resistance on the application of uniaxial mechanical compressive (tensile) strain. In the case of thicker films (> 1400 Å), the resistance of the film decreases on the application of both compressive and tensile strains (Fig. 23).

Thus, the effect of uniaxial (bending) strain on the transport properties of manganite films suggested a strong dependence on the sign and direction of applied strain. The response of the MIT and resistance in the film plane with the current direction showed anisotropic behaviour, which was explained by stress-induced rotation and growth of stripe ferromagnetic (metallic) domains in the manganite films [87]. The manganite films exhibited low (high) resistance to compressive (tensile) strain. The compressive stress stabilized the metallic phase to higher temperatures than tensile stress and the direction dependence of electrical response to external stress also indicated a strong influence of stress on the electronic properties of the manganite films [87, 397].

### 6.4.3 Coupling of bending strain and magnetization depth profile

Here we discuss the experimental determination of the coupling coefficient of bending (uniaxial) strain and magnetism for manganite thin films, especially LPCMO and LSMO systems, using the PNR technique. Before discussing the coupling of strain and magnetism obtained by simultaneous measurements of PNR and transport data from LPCMO film, we will describe the key findings of PNR measurements on LPCMO film without applying a bending strain ($\varepsilon = 0$). LPCMO is a prototype phase-separated material [2]. LPCMO ($y = 0.60$ and $x = 0.33$ and 0.375) thin films (thickness ~ 30 nm) grown on NGO (110) substrate show the in-plane magnetic anisotropy [Figs. 24 (a) and (b)]. The $M(T)$ curves for these thin films measured by macroscopic magnetization techniques (SQUID) mostly did not show a clear indication of the different phase (FMM, COI, etc.) transitions as observed in their bulk counterpart [2], which might be because of the influence of interface properties in these thin films and heterostructures. The macroscopic $M(T)$ data for the unstrained LPCMO films are shown in Figs. 24 (c), and (d), which shows a small hysteresis for LPCMO ($x = 0.375$) film and no hysteresis was observed for LPCMO ($x$ = 0.33) film. However, the transport data from both the LPCMO films show a prominent hysteresis with a larger width for LPCMO ($x$ = 0.33) film [Fig. 24 (c)]. Similar magnetic and transport properties were observed for other manganite thin films [34, 306, 307, 378] grown on different substrates. PNR measurements are performed by applying a magnetic field along the easy axis of the LPCMO film.

PNR measurements for LPCMO ($x$ = 0.33) film (without a bending strain) grown on NGO substrate have shown a quite evident temperature-dependent hysteresis for magnetization [39]. As mentioned earlier, PNR can simultaneously measure the structure and magnetic depth profiles of thin film heterostructures [97, 103]. The depth-dependent structure, i.e., nuclear scattering length density (NSLD) depth profile, of LPCMO film was obtained from PNR measurements [Fig. 25 (a)] and it shows a variation of chemical composition along the thickness of the film. The three-layer structure with different NSLD values (region I: at the air-film interface; region II: film bulk; region III: at the substrate-film interface; as depicted in the inset of Fig. 25 (a)) best fitted the PNR data [39]. To investigate the temperature-dependent magnetization depth profile of LPCMO film, PNR measurements at different temperatures (open green stars on both cooling and warming cycles in $R(T)$ curves in Fig. 25 (b)) and transport (Fig. 25 (b)) measurements were carried out simultaneously [39]. The magnetization at different temperatures for three regions along the depth of the film, extracted from PNR data, are shown in Figs. 25 (c-e), suggesting different magnetization in different regions. PNR results showed that the thermal evolution of the saturation magnetizations for the surface and buried interface (regions I and III in Fig. 25 (a)) are different from the film bulk (region II). Further, a thermal hysteresis of the magnetization, which is indicative of a first-order transition, was observed and the hysteresis width (ΔT) was different for the surface and buried interface as compared to the film bulk [39].

As discussed in section 6.3.1, the coupling of strain and magnetism in manganite films can be estimated by measuring a slop of $M^2$ $vs$ $\varepsilon$ curve. To generate the $M^2$ $vs$ $\varepsilon$ correlation, an experiment was designed for simultaneous measurement of transport and PNR data (for magnetization depth profile) on a thin LPCMO ($x$ = 0.33) film (thickness ~ 25 nm) grown on a thin NGO substrate (thickness ~ 0.25 mm) for different cases of applied bending strain [85]. Figure 26 shows the transport [$R$ ($T$)] data (simultaneously measured with the PNR experiment) for the LPCMO film under an applied strain ($\varepsilon$) of 0% (no strain), -0.011% (compressive), and +0.011% (tensile) along the in-plane $(1\bar{1}0)$ direction of NGO (magnetic easy axis) and in a magnetic field of 6 kOe (applied along the easy axis). The inset of Fig. 26 shows the schematic of the bending of the film and PNR measurements. The film is bent along an axis perpendicular to the neutron (*n*) beam, so the rocking width of the sample is not affected in the plane of specular reflection. The transport data were collected simultaneously along the *y*-axis (easy axis of the film). PNR measurements as a function of applied stress were carried out at constant temperatures as well as for constant ratios of temperatures to transition temperatures ($T/T_{IM}$ = $T/T_{MI}$ = 0.93) while cooling and warming across the MITs [85], where $T_{IM}$ (insulator-to-metal transition) and $T_{MI}$ (metal-to-insulator transition) temperatures are defined as temperatures at which the resistance changed by 95% of its maximum value during the cooling and warming cycles, respectively. The constant temperatures while the cooling and warming cycles were 78 and 89 K. PNR was performed for three different cases of strains [closed star in Fig. 26], at these temperatures. PNR data suggested a three layers structure [thin layer at the air-film interface, thin layer at the substrate-film interface and the middle region of the film] for LPCMO film with different NSLD suggesting variation of composition at interfaces [85].

Figures 27 (a) and (b) show the spin-dependent [spin up ($R^+$) and spin down ($R^-$)] PNR data for LPCMO film at 78 K (while cooling) and 89 K (while warming), respectively, under three values of the applied strains and the corresponding magnetization depth profile obtained from PNR data is shown in Figs. 27 (c) and (d). The PNR analysis revealed that compressive strain increases the average magnetization while tensile strain suppresses the average magnetization. Furthermore, the trends are the same regardless of the chemical composition of the region [e.g., as shown by regions I–III in the inset of Figs. 27 (c) and (d)], however, the change in magnetization for interfacial regions (I and III) was smaller. PNR data suggested an increase (decrease) of ~20% in the average magnetization for a bending compressive (tensile) stress ($\varepsilon$ ~0.011%) at $T$ = 78 K [Fig. 27 (c)]. A similar change in magnetization was observed at T = 89 K (while warming the sample). The $M^2$ vs $\varepsilon$ curves at 78 and 89 K for LPCMO film bulk (region II) are plotted in Fig. 27 (e), which suggests that for the film bulk (region II), the coupling coefficient $\gamma$ is ~ 0.00029(1) N/A$^2$ at T = 78 and 89 K. The study also suggested that coupling coefficients for surface and film-substrate interface regions (regions I and III) are much larger ~0.002(1) for the same temperatures. A smaller value of the coupling coefficient suggests that a small strain produces a large change in magnetization. The coupling coefficients of regions I and III were approximately a factor of seven larger than that of region II, which was attributed to the different chemistry (including Mn valence) and its influence on magnetization.

To investigate the influence of applied elastic bending stress on the magnetic ordering temperature ($T_C$) of a single crystalline LPCMO ($x$ = 0.33) film (thickness ~ 21 nm), PNR measurements were carried out to determine the saturation magnetization ($M_s$) as a function of temperature and applied stress near the $T_c$ of the unstrained film [86]. Figure 28 (a) shows the temperature-dependent variation of $M_s$ and resistance for an applied strain ($\varepsilon$) of 0% (unstrained) and -0.01% (compressive strain), which were measured simultaneously with PNR measurements. The enlarged version of $M_s$ ($T$) curves for different $\varepsilon$ near the $T_c$ (~ 135 K) of the LPCMO film is shown in Fig. 28 (b). Small compressive bending stress (i.e. $\varepsilon$ ~ -0.01%) induces a positive shift of $T_{MI}$ (metal to insulator transition temperature) by ~ 4 K (~ 4% increase in $T_{MI}$) and extraordinarily high relative difference [($R$(0) – $R$(ε))/$R$(ε), where $R$(0) and $R$(ε) are resistances without and with bending stress] in resistance ~ 200,000 % near $T_{MI}$, suggesting an increase in metallic phase on the application of compressive uniaxial stress [86]. Since $M_s$(T) of LPCMO film exhibited thermal hysteresis (Fig. 25 (d)), consistent with a first-order transition, $M_s$ data of the film for the different conditions of applied strain were fitted to a line in the temperature region between 120 K and 130 K for estimation of the $T_c$ [86]. The $T_c$ values for the different conditions of strains obtained from the fits show a shift of $T_c$ by 6 K (~ 4.5% increases in $T_c$) to a higher temperature with -0.01% (compressive) strain, suggesting that the compressive strain promotes magnetism in LPCMO films to higher temperatures compared to the absence of applied stress. The study showed that magnetic ordering occurred at significantly higher temperatures (~ 18 K) than MIT [86]. These findings also suggested that magnetic ordering is not caused by the MIT; rather magnetic ordering first occurs at a higher temperature.

Wang *et al*. [88] also studied the influence of applied elastic bending strain on the magnetization depth profile of an LSMO ($x$ = 0.20) epitaxial film (thickness ~ 50 nm) grown on (001) STO. PNR measurements were performed as a function of applied elastic bending strain ($\varepsilon$) of ~ ±0.03%. The depth profile of structure and magnetism of the film (Figs. 28 (c) and (d)) at 20 K as a function of $\varepsilon$ was obtained from PNR measurements. Magnetization depth profiles (Fig. 28 (d)) as a function of applied strain suggested a small change in the magnetization of the LSMO film. Thus, a negligible coupling of elastic bending strain with magnetization was observed for LSMO film as compared to LPCMO film, which was attributed to weak and negligible phase coexistence in LSMO film [88]. Thus, the study of coupling of bending elastic strain with magnetism in the LSMO and LPCMO manganite films suggested that bending stress has a very weak to negligible effect on a system that shows weak or no PS (LSMO) compared to one that shows strong PS (LPCMO). Therefore, these findings emphasize that the strain-energy landscape is a crucial factor in determining the magnetic properties of phase-separated complex oxide materials. Using a bending elastic strain as a tuning knob, Jin *et al.*, [397] study the ground state of LCMO ($x$ = 0.33) film grown on (100) NGO single crystal substrate. Measuring resistivity ($\rho$)-$T$ and $\rho$-$H$ curves as a function of bending strain, they obtained a $T$-$H$ phase diagram of LCMO/NGO (110) film and found an extended PS in the system induced by bending tensile strain.

The mechanical bending jig can provide a tuneable and reversible bending uniaxial stress, which can easily be deployed in PNR instruments for measuring the magnetization of heterostructures. The bending approach further eliminates the defects and other external factors as possible origins for phase coexistence in manganite thin films. Unlike other studies for controlling the phase-property correlation via changing the film thickness, substrate orientation, interfacial coupling, or chemical doping, this tuneable bending stress provides an effective way to dynamically modify phase competition and understand the complex phase diagram in a single manganite film. In addition, this experimental technique further enables studies of functional materials in which only the conjugate field of elastic stress is perturbed and results from such studies should be most amenable to comparison to theoretical predictions from a variety of interesting systems (e.g., multiferroics).

## 6.5 Strain and magnetic anisotropy

The orientation of magnetization ($M$) within a sample has directional dependence, which is called magnetic anisotropy (MA) and is described by magnetic anisotropy energy (MAE). The magnetization tends to align in the direction, called the easy axis, which minimizes the energy and hinders aligning in the direction, called the hard axis, which maximizes the energy. In general, the contribution of MA in manganite thin films is shape anisotropy (depends mainly on grain growth), magneto-crystalline anisotropy (crystal structure and spin-orbit interaction), magnetostriction (change in shape by applied field), and magneto-elastic anisotropy (lattice mismatch between the thin film and substrate), etc. The MA for manganites can easily be estimated in the absence of external stresses for properly oriented single-crystal samples. However, the strain affects not only $T_c$ and magnetoresistance [166, 337] properties but also the MA in manganite thin films. MA in the manganite films has been measured using several techniques, including Kerr microscopy [414], torque magnetometry [415], magnetotransport [416, 417], ferromagnetic resonance (FMR) [418], and usual magnetometers (SQUID/VSM) [35, 39, 279, 353]. In this section, we will review studies related to the strain-dependent MA in manganite films, which have made use of these techniques to study MA in manganite systems.

Using Kerr microscopy, Lecoeur *et al.* [414] studied the magnetic domain structures and magnetic anisotropy of epitaxial LSMO ($x$ = 0.3) thin films grown on (001) STO substrates and suggested a biaxial MA with (110) direction as an easy axis. Later, Suzuki *et al.* [415] studied the influence of epitaxial strain on MA of LSMO ($x$ = 0.3) films, grown on different STO substrates, using the torque magnetometry technique, where the torque was measured as a function of angle on the epitaxial films. The MA data from epitaxial films on STO indicated that a lattice mismatch between the film and substrate induces a strain anisotropy ($K$) of the order of $\sim 8.4\times10^4$ ergs/cm$^3$ [415]. Variation in MA as a function of the thickness of LSMO ($x$ = 0.3) films grown on different STO ($\varepsilon$ = 1.4%) and $LaGaO_3$ (LGO) ($\varepsilon$ = 1%) substrates suggested the higher value of stress anisotropy constant [113]. Thus, in the case of ferromagnetic LSMO, different studies [112, 113, 414] suggested that the tensile or compressive strain induced by the film-substrate lattice mismatch can induce in-plane or out-of-plane easy magnetization directions, respectively. A well-defined uniaxial anisotropy with an easy axis along the (001) direction was observed for LSMO ($x$ = 0.3) films grown on (110) STO substrates [419], which was attributed to the existence of elongated in-plane [001]-oriented structures. The study also suggested that the strength of MA increases with the LSMO thickness.

The control of the uniaxial MA in LSMO ($x$ = 0.33) film grown on (110) and (010) NGO substrates was achieved by strain engineering [420], where the magnetization of the LSMO films of different thicknesses was measured along the different in-plane angle. The LSMO films grown on (110) and (010) NGO substrates showed an out-of-plane tensile (in-plane compressive) strain of 0.67% and 1.1%, respectively. Figure 29 (a) shows the $M$($H$) curves for LSMO film (thickness = 18 nm) on (110) NGO substrate along two in-plane perpendicular directions (001) and ($1\bar{1}0$) of NGO, which showed different epitaxial strains for the films [420]. These measurements suggested a typical easy axis behaviour with sharp magnetization switching at small fields and large remanence (value of $M$ for $H$ = 0), while for the field applied in the (001) in-plane direction, the film showed very little remanence and much larger saturation field typical for a magnetic hard axis. Figure 29 (b) shows the extracted magnetic remanence ($M_r$) vs. in-plane field angle ($\theta$), measured from one edge of the sample as defined in the atomic force microscopy image shown in Fig. 29 (c). The oscillating behaviour with a periodicity of 180° for $M_r$ (Fig. 29 (b)) implies a uniaxial magnetic anisotropy. The highest and lowest $M_r$ value occurs around $\theta$ ~ 58° and ~148° corresponding to the in-plane ($1\bar{1}0$) (easy axis) and (001) (hard axis) directions, respectively. Interface-induced modifications in the MA were also observed for LSMO (*x* = 0.3) film by making a heterostructure with an ultra-thin layer of $LaCoO_3$ (LCO) [418], which was independent of the substrate. Many bilayers of LCO/LSMO heterostructures grown on STO substrate with change in the sequence of the layer were studied using the FMR technique [418] and a modification in MA of the LSMO layers, triggered by the structural effects of the cobaltite layer responsible for magnetic interactions at the interface was observed. FMR experiment also revealed an enhancement of the out-of-plane anisotropy because of magnetic interactions occurring at the interface with the cobaltite.

Epitaxial strain-induced MA in LCMO (*x* = 0.33) manganite films grown on (110) STO and LAO substrates were studied by measuring the magnetotransport [417] and magnetization along different crystallographic directions [421]. LCMO thin films (thickness ~ 20 nm) grown on (110) STO substrate, which provided a tensile strain to the film, showed an easy axis along the in-plane $(1\bar{1}0)$ direction [417]. While compressively strained LCMO films grown (110) LAO substrates showed an easy axis along the other in-plane (001) direction. The different magnetic easy axes in the LCMO films are because of the different misfit strains from different substrates. The transport results also suggested a reduction in resistivity and magnetoresistance along the corresponding easy axes for the two samples, showing a close correlation between the magnetic and transport properties in the strained LCMO films. Later a correlation of crystal structure and magnetism of LCMO films grown on LAO and STO substrates suggested that strains imposed by these substrates (compressive and tensile by LAO and STO, respectively) lead to a deviation of the crystal structure from orthogonal symmetry [421] and the structural deformation of the LCMO film was due to strain dependent different magnetic easy axis. The results also suggested that a magnetic reversal of the easy axis from the [001] to $[1\bar{1}0]$ axis for the LCMO/LAO films can be achieved using different shear strains. In contrast, no change in the easy axis was observed for LCMO/STO films and it maintained the [001] direction as an easy axis.

The strain-driven low field magnetoresistance (LFMR) anisotropy was also studied in ultrathin PSMO (x = 0.33) films (thickness = 5-15 nm) epitaxially grown on (001) LAO (compressive strain ~ -2%), (110) NGO (smaller epitaxial strain ~0.3%), and (001) STO (tensile strain ~ +1%) substrates [416]. The easy magnetization axis along a perpendicular (parallel) to the PSMO film plane was found for epitaxial compressive (tensile) strains. Further on the application of a magnetic field perpendicular to the film plane, a very large negative (positive) LFMR was observed for the compressive (tensile)-strain films, while a very small negative LFMR was found for the nearly strain-free (grown on NGO substrate) films. These results suggested that strain-induced anisotropic magnetization plays a major role in determining the LFMR properties of PSMO films.

A giant in-plane anisotropy in PCSMO (*x* = 0.35 and *y* = 0.3) thin films grown on (110) NGO substrate was reported, which was driven by strain-engineered double exchange interaction and EPS [422]. PCSMO shows a phase coexistence and competition of COI and FMM phases. The two in-plane perpendicular directions of the film on NGO substrate are (100) and (010) directions of PCSMO, which produced an anisotropic tensile strain with ~0.5% along (100) and ~0.3% along (010) direction of the film. The effect of anisotropic strain on the physical properties of PCSMO film was studied by measuring the magnetic and magnetotransport data along these two in-plane perpendicular directions [422]. The *M* (*H*) curves measured along the two in-plane perpendicular directions (100) and (010) of the PCSMO film at 80 K are shown in Fig. 30 (a). It is evident from Fig. 30 (a) that the (100) direction is the easy axis with a clear in-plane uniaxial MA (because of a larger remnant magnetization and the coercive field in this direction). The uniaxial anisotropy was further confirmed from the *M* (*T*) data (Fig. 30 (b)), which was measured under both the zero-field-cooled (ZFC) and the field-cooled (FC) conditions with an applied field of 100 Oe. The FC *M* (*T*) curve along the (100) direction showed an increase in *M* at $T_c$ due to the onset of the FMM phase. A large splitting between the FC and ZFC curves below $T_c$ also indicates the PS in the PCSMO film. In contrast, The FC and ZFC *M* (*T*) curves along the (010) direction overlap with each other and no abrupt magnetic transition was observed down to 5 K (Fig. 30 (b)). The *T*-dependent resistivity measurements (Fig. 30 (b)) of the PCSMO film exhibited a very sharp MIT at $T_{MI}$ (~ 125 K), as well as a clear anisotropy, i.e., $T_{MI}$ (resistivity), was higher (lower) in the (100) direction compared with those in the (010) direction. The difference of $T_{MI}$ under zero magnetic field between the two in-plane directions was about 3 K and the transport data indicated that the hopping of the $e_g$ electrons was much easier along (100). These anisotropic magnetic and transport behaviours of the film were related to the anisotropic coupling and competition between the double-exchange interaction and the JT distortion as a result of induced anisotropic strain along the two in-plane directions of the film.

Ward *et al*. [279] studied anisotropic in-plane epitaxial strain-induced anisotropy in both transports (resistivity measurements) and magnetic properties of LPCMO (*x* = 0.375 and *y* = 0.48) film (thickness ~ 500 Å) grown on (101) NGO substrate. The LPCMO films grown on (101) NGO substrate experience an anisotropic epitaxial strain of 0.3% and 0.1% along two in-plane perpendicular directions of ($\bar{1}$01) and (010), respectively, of NGO substrate. The anisotropic transport and magnetic properties were attributed to strain-dependent PS in the film which provided percolative conduction (FMM) channels along the highly strained direction and the study also suggested that the origin of phase coexistence was much more strongly influenced by strain than by local chemical inhomogeneity. However, the other studies [39, 87, 353] on LPCMO ($x$ = 0.33 and 0.375; and $y$ = 0.60) film grown on (110) NGO substrate suggested a strong in-plane magnetic anisotropy (Figs. 24 (a) and (b)) due to anisotropic biaxial strain in the plane of the film but no anisotropy in the resistivity measurements. Resistance measured along two in-plane perpendicular directions shows similar MIT transition temperature. The magnetic easy axis was found along the in-plane ($1\bar{1}0$) direction of NGO, which present higher epitaxial strain (+0.49%) to LPCMO film as compared to (001) direction (+0.26%) of NGO substrate (Figs. 24 (a) and (b)). Jeen and Biswas [353] compared the transport and magnetic properties of LPCMO ($x$ = 0.33 and $y$ = 0.60) films grown on (110) NGO and (001) SLGO substrates. While (110) NGO substrate presents an anisotropic lattice mismatch strain for LPCMO film ($d_{film}$ = 3.844 Å) along two in-plane directions (001) with $d_{001}$ = 3.854 Å and ($1\bar{1}0$) with $d_{1\bar{1}0}$ = 3.863 Å. Whereas the substrate-induced strain on LPCMO//SLGO thin films was negligible due to well-matched in-plane lattice parameters of SLGO ($d$ = 3.842 Å) [353]. Temperature-dependent $M$ ($H$) curves for LPCMO films on NGO substrate along two in-plane directions suggested magnetic anisotropy at all temperatures and the highly strained direction ($1\bar{1}0$) was the direction of the magnetic easy axis. While no magnetic anisotropy for the LPCMO film grown on SLGO substrate (which shows negligibly small strain along two in-plane perpendicular directions, (100) and (010) of SLGO) was observed at any temperature of measurements. Thus, the main reason for magnetic anisotropy in the LPCMO thin films was the anisotropic stress exerted by the substrate. The contradictory results [279, 353] regarding the absence of anisotropy in transport measurements along two in-plane perpendicular directions of NGO substrate were believed to be due to different measurement length scales probed in two experiments. A recent study on MA of LPCMO ($x$ = 0.375 and $y$ = 0.50) films grown on (110) NGO also suggested an in-plane magnetic anisotropy (Fig. 30 (c)) due to different in-plane strain, which was further confirmed by measuring $M$($H$) curves as a function of angle of rotation of applied field with respect to the magnetic easy axis (Figs. 30 (d) and (e)) [35]. The in-plane angular dependence of the coercive field ($H_c$) and remanence ratio ($M_r/M_s$) [Fig 30 (e)] of the LPCMO film also suggested the understanding of magnetization reversal in this system. Therefore, the long-range anisotropic strain field is effective in not only influencing the competition between separated phases but also coupled to the spin and orbital degrees of freedom in manganite thin films, resulting in anisotropic magnetic and transport properties.

## 6.6 Strain and Phase Coexistence

As discussed earlier mixed-valence manganites display an extremely rich variety and complex behaviour of noticeable physical properties as a consequence of the interplay among several physical degrees of freedom (e.g., spin, charge, lattice, and orbital), which are simultaneously active in these materials [24, 27, 28, 33, 34, 53, 120, 142, 173, 224]. Even more than fifty years after the discovery of this class of materials, some fundamental aspects like the intrinsic inhomogeneous behaviour of manganites with phase coexistence on a nanometer to micron length scale, which are not been explored completely, attract increasing attention. Phase coexistence or separation (existence of FMM, COI, etc. phases) is a very important phenomenon in these materials to understand the macroscopic properties like MIT, CMR, etc. and it is a consequence of the competition among nearly degenerate phases in the presence of sizeable quenched disorder arising, for instance, from cation substitution, oxygen deficiency or inhomogeneous strain [29, 120, 423]. The reports suggested that there are two leading scenarios responsible for the larger length scale PS (coexistence): (i) the presence of quenched disorder in the vicinity of a first-order transition [120] and (ii) long-range strain field interactions [29, 423]. In the first scenario, the presence of disorder sites pinned the phase boundaries which makes the PS static in nature [120], and the effect of strain at constant temperature should essentially be the sum of the strain effects on the pure FMM and COI phases [29]. However, in the case of the long-range strain interaction scenario, different phases (FMM and COI phases) introduce surface energy between them, and propagation of the phase boundaries is stopped by the long-range strain interactions in the material [29]. Therefore, the quenched disorder is localized in space, whereas the phase boundaries are free to move due to long-range strain interactions on the application of stresses, which gives a clear distinction between these two models. The effect of strain on the transport and magnetic properties of manganites helps to understand the actual mechanism of micrometer scale PS. While the phase coexistence has been experimentally investigated by space-resolved techniques such as cAFM [26, 424], electron microscopy [2, 425], scanning tunnel microscopy/spectroscopy (STM/STS) [23, 89, 426, 427], magnetic force microscopy (MFM) [90, 424, 428-430], photoemission spectroscopy [25, 431], Lorentz electron microscopy [432], which have shown a nanometer to micron length scale network of conducting/insulating (magnetic/nonmagnetic) domains. Ahn *et al.,* [29] proposed a theoretical understanding of the existence of distinct metallic and insulating electronic phases in the manganite thin films, where they found that the electronic inhomogeneities were due to the intrinsic complexity of a system with strong coupling between the electronic and elastic degrees of freedom. The coupling leads to local energetically favourable configurations and provides a mechanism for the self-organized inhomogeneities over both nanometer and micrometer scales.

The direct evidence for the coexistence of metallic (magnetic) and insulating (nonmagnetic) regions which appears compelling for bulk manganite systems [2, 25, 426, 430] has been elusive for thin films because thin films lie at the edge of the FMM phase-field [433], unlike bulk [30], where FMM volume fraction approaches to 100% as compared to thin films (~50 %). Only limited microscopic details of the apparent phase coexistence in thin manganite films have been reported, which may be due to drastically different phase diagrams [433] shown by films as compared to bulk counterparts. The phase coexistence of several micrometer length scales was measured in LPCMO manganites using microscopic techniques. Phase coexistence and its dynamics were inferred from the transport measurements of nanometer and micrometer-size bridges in LPCMO ($x$ = 0.375 and 0.33; and $y$ = 0.5) manganites [434, 435]. However, STM/STS measurements did not show any evidence for electronic PS of LPCMO ($x$ = 0.375 and $y$ = 0.44) films at several temperatures except atomic-scale inhomogeneities [427]. In the same study, STS measurements from the LPCMO film at low temperatures (T < 150 K) showed an energy gap of 0.5 eV. The presence (absence) of an energy gap (PS) in the film also contradicted the metallic behaviour seen in resistivity measurements at low temperatures. Although the STM measurements on LCMO ($x$ = 0.3) thin films purportedly showed direct evidence of EPS with a lateral length scale of a few nanometers [23].

The cAFM is an important mode of atomic force microscopy, which can be used for measuring the conductivity profile on the thin film sample surface. In cAFM, the current image is generated by scanning the film surface with a conductive tip, while a voltage is applied between the tip and the sample and it simultaneously generates a topographic image. Israel *et al*. [424] found the coexistence of metallic and insulating phases using cAFM measurements at a temperature of 77 K well below $T_{IM}$ (~ 145 K) of LCMO ($x$ = 0.4) film grown on (001) NGO substrate. The in-plane length scale for PS for this film was found to range from 30 nm up to several hundred nanometers. An unexpected in-plane elongation along the (100) direction of NGO was observed, which could be due to stress arising from the asymmetry of the epitaxial mismatch between the film and substrate. cAFM image from the film at room temperature showed a drastically reduced phase coexistence. Non-uniform distributions of metallic and insulating domains were found at different temperatures across the MIT temperature (Fig. 31) during cooling and warming of the single crystalline LPCMO (*x* = 0.33 and *y* = 0.6) film (thickness ~ 300 Å), grown on a (110) NGO substrate [26]. The topography and conductivity (Fig. 31) images of the film were simultaneously recorded at selected temperatures while cooling and warming the film across MIT ($T_{MI}$ ~ 81 K). The topographic image of the film showed the surface of the film to be stepped-presumably due to a vicinal surface formed after the NGO substrate was cut and polished before film deposition and it is consistent with the step terrace (flow) growth of the film. The temperature-dependent topographic image did not show significant change, whereas the temperature-dependent current image showed drastically different behaviours [26]. On cooling the film from 140 to 60 K, the current images are mostly composed of insulating (cyan region in Fig. 31) domains with occasional small (< 50 nm) circularly shaped metallic (red region) domains. At 50 K, the larger size (100-150 nm) stripe domains were formed, which were orientated at an angle of 135° from [1$\bar{1}$0] NGO and parallel to the orientation of the terraces in the topographic images. This behaviour was consistent with heterogeneous nucleation (as a result of defects, e.g., terrace steps) of metallic domains. The hysteric behaviour of the conductivity measured with cAFM was similar to the resistivity of the film bulk (middle panel of Fig. 31) suggesting a manifestation of competition between metallic and insulating domains with nanometer length scales in the LPCMO film [26].

Earlier, using low-temperature MFM, direct evidence of percolative MIT transition in LPCMO (*x* = 0.33 and *y* = 0.50) thin-film grown in (110) NGO substrate was observed [90]. The MFM image was recorded while cooling the film and it showed a uniform image above $T_{IM}$ (insulator to metal transition temperature ~ 120 K) and growth of the FM phase below $T_{IM}$, which became stronger on decreasing the temperature. While warming the sample the FM phase regions were observed up to a much higher temperature, $T_{MI}$ (metal to insulator transition temperature ~140 K). The study also suggested that the evolution of magnetic phase coexistence across MIT follows the hysteresis behaviour in transport measurements for the film. Later, using the MFM technique, Rawat *et al.* [429] investigated the nucleation and growth of the FMM/AFI phase during the cooling, warming, and isothermal field cycling across the MIT in LPCMO (*x* = 0.375, *y* = 0.72) thin film (thickness ~ 1500 Å) grown on NGO substrate. Like a previous study [90], the MFM images as a function of temperature, showed AFI to FMM transformation during in-field cooling and FMM to AFI transformation during warming (Fig. 32) across MIT, which cannot be described solely in terms of the broad first-order transition due to quench disorder. The variation of the FMM phase fraction (estimated from MFM images) during 1 T cooling and subsequent warming is shown in the right panel of Fig. 32, suggesting that the FMM phase fraction increases continuously from 125 K to the lowest temperature of MFM measurement, i.e., 60 K, during cooling. While warming FMM phase fraction remains nearly constant up to 125 K and above which decreases rapidly, and the system becomes almost AFI by 150 K. The close evolution of FMM and AFI phases across MIT suggested that chemical inhomogeneity may not entirely be responsible for PS and the evolution may arise due to distribution in strain across the film thickness. The phase coexistence of epitaxial LPCMO ($x$ = 0.375, $y$ = 0.44-0.48) films (thickness ~ 2000 Å) grown on (010) NGO substrate were also examined *in situ* using Lorentz transmission electron microscopy (TEM) and other microscopy methods [432] and observed the competing two-phase coexistence of AF-COI and FMM phases. These phases were obtained with mesoscale PS below the MIT (~ 164 K). The AES field-dependent transport properties of LPCMO ($x$ = 0.375 and $y$ = 0.48) films grown (101) NGO substrate suggested a large anisotropic transport behaviour, which reduces on applying a high magnetic field of ~8.5 T and thus indicated the formation of EPS [279]. These findings suggested that the origin of phase coexistence is much more strongly influenced by strain than by local chemical inhomogeneity.

Epitaxial strain-induced magnetic phase segregation in the LCMO ($x$ = 0.33) films grown on different substrates ((001)-oriented STO, LSAT, and LAO), presenting different epitaxial strains for manganite film, was studied using electron holography and observed phase coexistence in the systems were correlated with structural and magnetic properties [92]. The experimental findings were further compared with the DFT-based calculation for studying the phase fraction of different phases in the system. The experimental and theoretical analysis suggested that the strain engineering of the structurally and chemically homogeneous LCMO film at room temperature produces two spatially segregated layers with different magnetic ordering. The study also suggested that the strain can induce two functional phases, which are inherently coupled through an interface that produces a combined function that was absent in unstrained material [92]. Combining Raman spectroscopy, magnetic, and resistivity measurements, the effect of epitaxial strain on phase coexistence in LCMO ($x$ = 0.5) films of various thicknesses grown on different substrates ((111) and (100) STO, (001) SLAO, and (001) SLGO) substrates were studied and a strong correlation between strain and phase coexistence in this system was reported [114]. Raman spectroscopy as a tool was efficiently used to investigate the charge and orbital ordered phases in LCMO manganites and it was shown that small perturbations due to strain can cause changes in structural properties and manipulate the PS in the manganite thin films.

Therefore, in recent years, remarkable progress has been achieved in understanding the phase coexistence/separation phenomenon in perovskite manganite films, which was made possible by building upon experimental measurements and theoretical approaches. The studies suggested the evolution of phase coexistence as a result of strain propagation in manganite films. The shape and scale of PS are different for different manganite systems, with domain sizes ranging from a few nanometres to several micrometres. The PS describes many macroscopic properties of manganites (e.g., MIT, CMR, etc.).

## 6.7 Strain and magnetocaloric effect

Refrigeration, where thermodynamics plays a central role, is an important area in modern life because of its widespread use in household and industrial applications. This has led to a need for cheap and environmentally friendly processes that might replace the costly approaches based on mechanical work. One promising approach is the development of new magnetic refrigeration (MR) technology, based on the magnetocaloric effect (MCE) which involves the thermodynamic changes associated with the varying magnetization of materials in an external magnetic field. This requires magnetic materials with suitable intrinsic properties that maximize the thermal response associated with the change in magnetization [436]. The MR has already been successfully employed for cooling at very low temperatures (< 1 K), but its application near room temperature is not yet commercially available [437]. The physical origin of the MCE is the magnetic field-dependent entropy of material and it is associated with the change of the magnetic entropy $\Delta S$, which originates in the coupling of atomic magnetic moments. The isothermal entropy change, $\Delta S$ of magnetic material at a temperature $T$, due to applied magnetic field $H$ is obtained from the Maxwell relation: $\mu_0^{-1}(\partial S/\partial H)_T = (\partial M/\partial T)_H$ using:

$$\Delta S(H) = \mu_0 \int_0^H \left(\frac{\partial M}{\partial T}\right)_H dH \qquad (14)$$

Where $\mu_0$ and $M$, are the permeability of free space and magnetization of the material, respectively. Thus, a large gradient in $M(T)$ is desirable but does not usually guarantee giant MCE. Experimentally, $\Delta S$ is determined from a set of magnetization $M(H)$ isotherms, which are ideally obtained in thermodynamic equilibrium [117]. There is another important parameter for such MCE materials termed relative cooling power (RCP), which measures how much heat can be transferred from the cold to the hot end in a refrigeration cycle and it is defined as [118, 438]:

$$\mathrm{RCP} = \int_{T_1}^{T_2} |\Delta S|\ \delta T_{FWHM} \qquad (15)$$

Where $\delta T_{FWHM}$ is the full-width half maximum of the ΔS(T) curve. Thus, an efficient prototype MCE material should exhibit not only the increased and broad ΔS (T) but also a large value of the $\delta T_{FWHM}$ for a significant temperature difference between the hot and the cold ends of the operating refrigeration cycle. Broadly magnetocaloric material is classified based on the order of the magnetic phase transition in the material as first or second-order magnetic phase transition (FOMPT or SOMPT) [439]. While FOMPT magnetocaloric materials exhibit large $\Delta S$ due to the coupling of structural and magnetic transition, SOMPT magnetocaloric materials show smaller MCE responses as compared to FOMPT materials. However, unlike SOMPT materials, FOMPT materials show a thermal/magnetic hysteresis and a limited temperature range for MCE.

Manganite oxides are among the potential materials for magnetic refrigeration due to their MCE values in a wide range of temperatures and their low cost for fabrication [117]. In bulk manganite oxides, both CMR and MCE are often observed around the magnetic-ordering phase transition temperature ($T_c$) [440]. Morelli *et al.*, [441] studied the magnetocaloric properties of doped (with Ca, Ba, or Sr) LMO manganite films grown on (100) LAO substrates and determined the entropy change associated with their corresponding $T_c$ by measuring the magnetization of these films as a function of magnetic field and temperature. The large magnetization of these films resulted in a total entropy change, a factor of five less than that of gadolinium, the prototypical high-temperature magnetocaloric material [436]. A strong correlation between strain and magnetization in manganite films can affect the resistivity, magnetic moment, and $T_c$. Therefore, it is interesting to study the strain-driven MCE properties of manganite thin film. The MCE properties for different manganite films grown on different substrates with different strains and RCP values for different manganite films are given in Table 4 [115, 116, 438, 441-448]. The epitaxial LCMO ($x$ = 0.33) film grown on (001) STO showed a small intrinsic MCE near the $T_c$ (~ 265 K) [442]. Campillo *et al.*, [445] studied the MCE of LCMO ($x$ = 0.33) and $La_{1-x}Ca_xMn_{0.94}Cr_{0.06}O_3$ (LCMCrO, $x$ = 0.33) (thickness ~ 2600 Å) manganite thin films grown on (100) LAO substrates. A change in entropy $(|\Delta S|) \sim -$ 1 $JK^{-1}kg^{-1}T^{-1}$ and -0.1 $JK^{-1}kg^{-1}T^{-1}$ was observed for LCMO film at T = 210 K and LCMCrO at 238 K, respectively.

A strain-engineered manganite heterostructure driven by a low-temperature magneto-structural phase transition also yielded an enhanced MCE. Mayo *et al.*, [115] found a strain-induced giant and reversible extrinsic MCE in LCMO ($x$ = 0.30) films grown on BTO substrate, by exploiting a first-order structural phase transition (near the rhombohedral-orthorhombic transition at $T_{R\text{-}O}$ ~190 K) in BTO, which was well away from intrinsic MCE in the LCMO film due to its ferromagnetic transition temperature ($T_c$ ~ 225 K). A sharp entropic jump in $M(T)$ for LCMO film on BTO was observed near $T_{R\text{-}O}$, primarily due to strain-mediated changes in ferromagnetism and magnetic field-driven jump due to interfacial strain-mediated feedback, even though BTO is non-magnetic. To investigate MCE in BTO/LCMO film over a wide range of temperatures, Mayo *et al.*, [115] used Eq. (14) to calculate the change in entropy by measuring $M(T)$ at selected values of $H$ >0. A spike in entropy was observed near $T_{R\text{-}O}$, which develops at a rate of $|\Delta S| \sim -9$ $JK^{-1}kg^{-1}T^{-1}$ and at nearby temperatures $|\Delta S(H)|$ was an order of magnitude smaller than the peak value. A much smaller intrinsic MCE of ~ -0.7 $JK^{-1}kg^{-1}T^{-1}$ was obtained for the LCMO/BTO film at its $T_c$. Giri *et al*., [443] studied the strain-induced extrinsic MCE in LSMO ($x$ = 0.33) thin films, grown on (001) BTO single crystal substrates. Using the crystallographic structural transformation of the BTO, a large change in the entropy of about -1.03 to -1.95 $JK^{-1}kg^{-1}T^{-1}$ was obtained at the monoclinic–tetragonal transition of BTO crystal at 283 K. The drastic change of the 3D film-lattice strain, which affected the local magnetic anisotropy of the LSMO at the interface, changes the entropy of the film. However much smaller MCE (~ -0.98 $JK^{-1}kg^{-1}T^{-1}$ ) in LSMO ($x$ = 0.33) thin films were observed due to rhombohedral-orthorhombic transition in BTO as compared to the same structural transition in BTO/LCMO film [115]. The strain-dependent magnetocaloric properties of epitaxial LSMO ($x$ = 0.33) thin films deposited on three different substrates ((001) oriented LAO, STO, and LSAT) were investigated under low magnetic fields and found that the substrate (STO) induced tensile strain not only decrease the $T_c$ to the room temperature but also improves the MCE and cooling capacity of the material (LSMO) [444].

Strain-modulated large MCE were observed in epitaxial SSMO ($x$ = 0.45) films grown on (001)-oriented LAO, LSAT, and STO substrates, which imposes an in-plane biaxial compressive stress of -1.0%, the in-plane tensile stress of +1.7% and +0.93%, respectively [116]. $M(T)$ data from SSMO films grown on LSAT and STO substrates showed thermal hysteresis between FCC and FCW magnetization cycles [116]. The temperature-dependent magnetic entropy change, $|\Delta S|$, for STO/SSMO, LSAT/SSMO, and LAO/SSMO films at different fields are shown in Figs. 33 (a), (b), and (c) respectively. The $M(H)$ curves for three samples at different temperatures are shown in the inset of Figs. 33 (a-c). Figure 33 (d) shows the temperature-dependent specific heat curve at zero magnetic fields for LAO/SSMO film, which suggested second-order magnetic phase transition (SOMPT) as compared to the other two films, which showed first-order magnetic phase transition (FOMPT). The epitaxial strain-dependent change in magnetic entropy for SSMO film on LAO, STO, and LSAT was found to be 0.35, 1.67, and 0.95 $JK^{-1}kg^{-1}T^{-1}$, respectively. Thus, a large value of MCE observed in the case of STO/SSMO film might be due to tensile strain-induced FOMPT, whereas a small value but broad $|\Delta S|$ was observed for LAO/SSMO film due to compressive strain-induced SOMPT [116].

The MCE was also found in the superlattice of the ultrathin layer of LSMO ($x$ = 0.3) manganite [thickness ~ 20 unit cell (u.c.)] and SRO (thickness = 2,3 and 6 u.c.) grown on (001) STO substrate [446]. Three superlattices, [LSMO (20 u.c.)/SRO ($n$ u.c.)]$_{\times 15}$ with $n$ =1, 3, and 6, where the bilayer (LSMO/SRO) is repeated 15 times, were examined for MCE and RCP properties and suggested that the change in entropy for superlattices were found to be the same as that of polycrystalline LSMO but the working temperature have increased and so the RCP parameter, which is attributed to the effect of the interfaces (modification of the charge states of the Mn and Ru ions at the interface) at LSMO/SRO and higher nanostructural disorder [446]. The superlattices exhibited a SOMPT with improved magnetic properties. The observed entropy and RCP for these superlattices are given in Table 4. Recently, strain-modulated MCE properties were also studied in LSMO (x = 0.3)/SRO superlattices on (111)-STO crystal by changing the sequence of the stacking order [447]. Two superlattices, [LSMO (11 u.c.)/SRO (3 u.c.)]$_{\times 15}$ and [SRO (11 u.c.)/LSMO (3 u.c.)]$_{\times 15}$ were grown on (111)STO substrates, which provided an epitaxial tensile (+0.64 %) and compressive (-0.64%) strains for these two superlattices. These two superlattices with different stacking orders showed different MCE properties (Table 4) suggesting that the stacking order dependent $\Delta S$ can be explained by the different strains observed for these superlattices [447]. An interface-induced enhanced MCE in an epitaxial LSMO ($x$ = 0.3)/$CoFe_2O_4$ (CFO) heterostructure grown on a single-crystal MgO (100) substrate because of interfacial strain-induced magneto-structural coupling was reported [438]. A magnetic entropy change ($|\Delta S|$) of 0.14 and 0.63 $J\,K^{-1}kg^{-1}T^{-1}$ was found for MgO/CFO and MgO/LSMO/CFO heterostructures, respectively, at a structural transition temperature of 102 and 79 K. The study also suggested that the MgO/LSMO/CFO heterostructure with an RCP of 73.6 $\mathrm{J\,kg^{-1}T^{-1}}$ can be a promising candidate for magnetic refrigeration at low temperatures.

Some of the manganites exhibited a complex MCE because of the coexistence of competing magnetic phases (or EPS) in a wide range of temperatures [31]. Different magnetic phases in the EPS state, as well as the different length scales of the EPS, strongly influence the MCE [449]. Growing manganite systems with a special length scale smaller than the length scale of EPS domains, the systems can transit from the EPS state to a single-phase state, which offers an excellent opportunity to investigate the effect of this transition to MCE. An enhanced MCE in LPCMO ($x$ = 0.375 and $y$ = 0.32) nanodisk arrays fabricated from epitaxially grown LPCMO (thickness ~ 60 nm) thin films on (100) STO substrate was also reported [448], where LPCMO disk arrays (2-μm and 500-nm) were fabricated from the epitaxially grown thin films by using electron beam lithography. Figures 34(a)–(c) illustrate the $M$(H) curves measured in the temperature range of 40–300 K for the LPCMO film, 2-μm LPCMO disk array, and 500-nm LPCMO disk array. It is evident from the $M$(H) curves that by reducing the lateral dimension of LPCMO (film of 1cm$^2$ to 500 nm) higher magnetic field is required to saturate. Figures 34(d)–(f) present temperature-dependent magnetic entropy changes ($-|\Delta S|$) for different magnetic fields deduced from Eq. (14). The $-|\Delta S|$ vs $T$ curves of both the film and 2-μm disk array exhibited two distinct peaks around $T_{CO}$ (CO phase to FMM) and $T_C$. The low temperature peak near $T_C$ was higher than the high temperature peak near $T_{CO}$ for both the film and 2-μm disk array (Figs. 34(d) and (e)). Whereas the 500-nm disk array exhibited one distinct peak around $T_C$. The MCE of the 500-nm-sized LPCMO disk array was found to be more than twice that of its parent film. This study suggested dependent of MCE properties on the phase coexistence which can be controlled by strain in these manganite films.

The phase transition and MCE in NSMO ($x$ = 0.5) epitaxial thin films were tailored by controlling the lattice-mismatch-induced strain by depositing on (011) LSAT and STO substrates [450]. A uniaxial tensile strain of 1.3% along the in-plane [100] direction of STO was exhibited for NSMO film grown on STO, which undergoes paramagnetic to FM transition at ~ 210 K followed by an FM to A-type AFM transition at ~179 K. Whereas the NSMO film experienced anisotropic in-plane tensile strains of 0.36% along [100] and 0.50% along [011] directions of LSAT, which undergoes further transition to CE-type AFM transition at ~145 K. Thus, the NSMO/LSAT heterostructure with such transitions facilitates a strong MCE over a much wider temperature range from ~90 to 170 K providing a comparable magnetic entropy change. The study suggested that controlling appropriate anisotropic strain to realize the successive first-order transitions from FM to A-type AFM and further to CO/CE-type AFM phase in NSMO films can broaden the MCE temperature range and improve the magnetic entropy change and RCP properties [450]. Recently, an enhanced MCE performance of manganite bilayer films was observed [451]. Bilayer films of LSMO ($x$ = 0.2) and LCMO ($x$ = 0.3) with different layer thicknesses and stacking sequences were investigated for magnetic and MCE properties. An enhancement of the RCP by over 40% with an enhanced operating temperature range was observed, which was attributed to the epitaxial strain-induced broadening in the magnetic phase transition in the heterostructure [451]. These studies implicitly suggested that it would be ideal to realize a large MCE over a wide temperature range for a manganite film with improved MCE performance. Growing manganite epitaxial films on single crystalline substrates can induce in-plane strains and phase coexistence, which can tune the phase transition behaviours effectively [115, 452] and tune high-performance MCE. The strain-dependent magnetic entropy and RCP of the manganite films grown on different substrates are shown in Figs. 35 (a) and (b), respectively. Thus, the strain mismatch plays an important role in influencing the magnetic and magnetocaloric properties of manganite heterostructures.

## 7. Application of manganite heterostructures

In addition to magnetic refrigeration applications, as mentioned in the previous section, manganite-based perovskite oxide heterostructures offer features of high sensitivity and tunability by the magnetic field, electric field, light irradiation, and high carrier mobility, suggesting many possible applications including information storage, optoelectronics information processing, and biomedical applications [453-456]. Manganite heterojunction photodetectors have also shown important application in harsh environments with fluctuations of temperature and pressure [457]. Non-volatile memories using an electric-field-induced MIT in the PZT/LSMO, PZT/LCMO, and PZT/LSCO devices were examined earlier. Liu *et al.* [458] developed a new programmable metallization cell based on amorphous LSMO ($x$ = 0.21) thin films for nonvolatile memory applications, where LSMO film was utilized as solid electrolytes for fast Ag+ ion conduction. Owing to the 100 % spin polarization in several FM manganite oxides, efforts have been made to apply these systems for spintronics devices [459-462]. Mixed-valence LSMO is an optimal source of fully spin-polarized carriers, which can be modified by strain and shows a rich physics of magnetic phases and transport mechanisms near room temperature. Therefore LSMO films, showing very well-tailored properties, are considered for many selected applications [462]. Large tunnelling magnetoresistance (TMR) was observed for the current-perpendicular-to-plane (CPP) magnetic tunnelling junction (MTJ) device structures using doped perovskite manganite LSMO ($x$ = 0.33)/STO/LSMO ($x$ = 0.33) and LCMO ($x$ = 0.33)/STO/LCMO ($x$ = 0.33) heterostructures at a moderate applied field of less than 200 Oe [460]. Further improvement in TMR properties for device application was achieved for LSMO/STO/LSMO heterostructures [461].

The CMR properties of perovskite manganite thin films have also been exploited as magnetic sensors at room temperature. The bolometer, an instrument for measuring the radiated energy in the infrared and optical wavelength regions, is a resistive element constructed from a material with a large temperature coefficient of resistance [TCR ~ 1/R (dR/dT)]. Several mixed-valent manganites thin films such as LSMO ($x$ = 0.5), LCMO ($x$ = 0.3), LBMO ($x$ = 0.5), LPCMO ($x$ = 0.33, $y$ = 0.6), grown on single crystal (001) silicon substrate were examined for uncooled bolometric applications [463] and it was found that the TCR and electrical noise ($S_V$) depend on the chemical composition. An optimum performance of TCR (~ 7% / K) and $S_V$ (~ $8.9 \times 10^{-13}$ $V^2$/Hz) was found for LCMO. FM manganite films are also considered as a component for making a multiferroic heterostructure with other non-FM multiferroic oxides to enhance the magnetoelectric effect for possible application [40, 42, 57, 222]. The LFMR effect studied in manganites-based VAN complex oxide thin films also shows great interest for applications in bolometric detectors, magnetic read-heads, and magnetic field sensors or position sensors. Staruch *et al*. [464] studied LSMO ($x$ =0.33): ZnO VAN thin films for possible application of magnetic field sensors and obtained a maximum field sensitivity of 632%/T, which is one of the highest values achieved in these VAN thin film systems.

## 8. Concluding remarks and future direction

The manganite thin films offer an excellent example of complex oxides exhibiting properties like CMR, MIT, magnetic anisotropy, phase coexistence, etc., that arise from the strong coupling between spin, lattice, charge, and orbital degrees of freedom. These systems also provide an ideal platform for the systematic investigation and control of the role of intrinsic interfacial interactions involving these degrees of freedom by application of external stimuli. Strain is one of such external perturbations which is thus found to be a unique way to engineer the functionalities of manganite heterostructures by modifying the energy scales due to interfacial interactions. Elastic strain engineering of perovskite manganite heterostructures provides a new strategy to probe intrinsic properties with novel functionalities and has been the most pursued to date. The strain can enhance the MIT, Curie temperature, and saturation magnetization of manganite interfaces for better FM properties. Further extension of strain engineering of perovskite manganite thin films helps to understand their structure-property relationships and the continued development of the physically accurate design of heterostructures tune material properties for diverse functionalities. Moreover, the advancement of controlled growth and characterization techniques with an ability to achieve atomically precise perovskite manganite oxide interfaces allows for the unprecedented control of the spin, orbital and structural degrees of freedom. This review article has discussed the progress in the studies regarding the influence and coupling of strain with electronic, magnetic, and magnetotransport properties of manganite heterostructures. The coupling of a different mode of strain, i.e., epitaxial (biaxial strain), piezoelectric (controlled strain by phase transition or electric field on the ferroelectric substrate), vertically aligned nanocomposites (for thicker films) and bending (uniform uniaxial strain), with transport and magnetic properties of manganites heterostructures are discussed in the present article. This includes defining the coupling of bending strain and magnetism as well as its correlation with the phenomenological approach for defining different magnetic phase transitions in these systems.

Here, the aim has been to demonstrate the influence of pure bending elastic strain (uniaxial) on the transport and magnetic properties of a single manganite film by simultaneously measuring the transport and PNR data as a function of temperature and magnetic field. The coupling coefficient of strain and magnetism of manganite films were formulated under free energy formalism and was experimentally estimated using the PNR technique, which not only helped to understand the correlation of elastic strain with magnetism but also explained the condition of magnetic phase order change (second order to first order) in the phase-separated manganite systems within a phenomenological Ginzburg–Landau theory. In addition, the overview of the current perspectives and existing studies on the influence of strain fields on structure, electronic, magnetic, magnetic anisotropy, phase coexistence, and magnetocaloric properties of manganite heterostructures have also been discussed. The review also highlighted the PNR technique, which provides a high-resolution (sub-nanometer) depth profile of structure and magnetism over a large sample area of a few $cm^2$ and is compatible with a complex sample environment. Thus PNR, in combination with other direct imaging techniques (SEM, TEM, etc.), which provides local structure, provides a complete set of tools for studying the strain-driven properties in complex oxide heterostructures for a wide range of technological applications.

Advancements in strain engineering in manganite heterostructures offer assurance in the fine-tuning of electronic and magnetic properties, where minor changes in the strain can induce dramatic changes in transport and magnetic properties. Obtaining emerging properties by designing manganite oxide heterostructures and the subsequent new physics resulting from the strain effect have always been an interesting topic of research as they can have a possible application for cheap and environment-friendly oxide-based electronic devices. Despite great progress in the strain-dependent tuning of material properties, there is still a long way to go to fully understand the intrinsic mechanisms and theoretical developments behind these strain-dependent phenomena in manganite heterostructures. It is also noted that the correlation of strain-driven structure and magnetic properties on the nanoscale will be required even more in the future for complex oxide functional layer systems (e.g., multiferroic, paleomagnetic, etc.,). The efficiency of the simultaneous application of different analytical methods (XRR, PNR, TEM, etc.) for the investigation of these complex oxide interfaces and their correlation with strain in material science and engineering will be crucial for designing advanced functional materials.

The manganite-based perovskite oxide heterostructures have shown possible potential applications in the fields of nanoelectronics, spintronics, solar energy conversion, biomedical field, magnetic sensors, solid oxide fuel cells, magnetic refrigeration, and catalysts. Recent efforts have been focused on understanding the strain-driven tuning of structure and magnetic properties as well as potential applications of manganite-based complex oxide heterostructures in the above-mentioned fields. Strain-driven properties of oxide heterostructures discussed here demonstrate the incredible potential that oxide interfaces hold for the discovery of new behaviour as well as possible applications. I believe that this review on the coupling of strain with transport and magnetic properties of manganite-based perovskite oxide heterostructures will provide a platform for researchers interested in this field to gain an understanding of the current state of strain-driven phenomena in these systems. In addition, it will motivate them for exploring strain-controlled studies on other complex oxide-based heterointerfaces e.g. multiferroic, superconducting, etc., for improved properties and possible technological applications on large scales.

**Acknowledgements:** I would like to thank Dr S. Basu for introducing me to the polarized neutron reflectivity technique for studying interface magnetism. I also thank Dr M. R. Fitzsimmons for providing me an opportunity to work in the field of complex oxide heterostructures.

**Table 1: Change in transition temperature ($T_c$) or metal to insulator transition temperature ($T_{MIT}$) with applied pressure ($P$) for bulk magnetic systems.**

| **Magnetic systems** | ***$dT_c/dP$* (K/GPa)** | **References** |
|---|---|---|
| Ni | +3.2 | [199] |
| Fe | 0.0 | [199] |
| Co | 0.0 | [199] |
| $Sc_3In$ | +1.7 | [200] |
| $SrRuO_3$ | -5.7; -7.9 | [201]; [202] |
| $Y_3Fe_5O_{12}$ | +12.5 | [199] |
| $La_{0.60}Ca_{0.40}MnO_3$ | +16.0 | [203] |
| $La_{0.67}Ca_{0.33}MnO_3$ | +15.7; +9.95 | [203]; [197] |
| $La_{0.79}Ca_{0.21}MnO_3$ | +36.7 | [203] |
| $La_{0.60}Sr_{0.40}MnO_3$ | +2.0 | [204] |
| $La_{0.75}Sr_{0.25}MnO_3$ | +6.0 | [198] |
| $La_{0.85}Sr_{0.15}MnO_3$ | +20.0 | [204] |
| $Pr_{0.8}Ca_{0.2}MnO_3$ | +2.4 | [205] |
| $La_{0.67}Ca_{0.33}(Co_{0.03}Mn_{0.97})O_3$ | +23.4 | [206] |
| $Pr_{0.6}Ca_{0.4}Mn_{0.96}Co_{0.04}O_3$ | +34.0 | [207] |
| $Pr_{0.6}Ca_{0.4}Mn_{0.96}Cr_{0.04}O_3$ | +31.4 | [207] |
| $Sm_{0.2}Nd_{0.47}Sr_{0.33}MnO_3$ | +20.0 | [197] |
| $Sm_{0.4}Nd_{0.27}Sr_{0.33}MnO_3$ | +22.0 | [197] |
| $Sm_{0.5}Nd_{0.17}Sr_{0.33}MnO_3$ | +22.0 | [197] |
| $Sm_{0.67}Sr_{0.33}MnO_3$ | +21.0 | [197] |
| $La_{0.645}Y_{0.025}Ca_{0.33}MnO_3$ | +13.0 | [197] |
| $La_{0.47}Y_{0.20}Ca_{0.33}MnO_3$ | +20.5 | [197] |
| $Sm_{0.55}Sr_{0.45}MnO_3$ | +20.0 | [208] |
| $La_{1.2}Sr_{1.8}Mn_2O_7$ | +19.0 | [209] |
| $(La_{1-z}Pr_z)_{1.2}Sr_{1.8}Mn_2O_7$ with z =0.2 | +17.0 | [210] |
| $(La_{0.4}Pr_{0.6})_{1.2}Sr_{1.8}Mn_2O_7$ | +22.2 ($T_{MIT}$) | [211] |

**Table 2: The commercially available single-crystal perovskite substrates. I have also estimated the maximum strain observed by manganite (lattice constant ~3.87 Å) films if grown on these substrates.**

| Single crystal substrates | Lattice constant (Å) | Strain (%) for manganite (lattice constant ~3.87 Å) film | References |
|---|---|---|---|
| $LuAlO_3$ (LuAO) | 3.67 | -5.45 | [256] |
| $YAlO_3$ (YAO) | 3.72 | -4.03 | [257] |
| $LaSrAlO_4$ (LSAO) | 3.75 | -3.20 | [258] |
| $NdAlO_3$ (NAO) | 3.76 | -2.92 | [259] |
| $LaAlO_3$ (LAO) | 3.782 | -2.30 | [260] |
| $LaSrGaO_4$ (LSGO) | 3.835 | -0.91 | [261] |
| $(NdAlO_3)_{0.39}$-$(SrAl_{1/2}Ta_{1/2}O_3)_{0.61}$ (NSAT) | 3.85 | -0.52 | [262] |
| $NdGaO_3$ (NGO) | 3.86 | -0.29 | [263] |
| $(LaAlO_3)_{0.29}$-$(SrAl_{1/2}Ta_{1/2}O_3)_{0.71}$ (LSAT) | 3.87 | 0.00 | [262, 264] |
| $LaGaO_3$ (LGO) | 3.89 | 0.51 | [265] |
| $SrTiO_3$ (STO) | 3.903 | 0.85 | [266, 267] |
| $Sr_{1.04}Al_{0.12}Ga_{0.35}Ta_{0.50}O_3$ (SAGT) | 3.925 | 1.40 | [262] |
| $DyScO_3$ (DSO) | 3.94 | 1.78 | [15, 268]] |
| $TbScO_3$ (TSO) | 3.95 | 2.02 | [269] |
| $GdScO_3$ (GSO) | 3.96 | 2.27 | [268, 269] |
| $EuScO_3$ (ESO) | 3.985 | 2.88 | [17, 270]] |
| $SmScO_3$ (SSO) | 3.99 | 3.00 | [17, 270] |
| $KTaO_3$ (KTO) | 3.996 | 3.15 | [271] |
| $NdScO_3$ (NSO) | 4.00 | 3.25 | [272] |
| $BaTiO_3$ (BTO) | 4.01 | 3.49 | [268] |
| $PrScO_3$ (PSO) | 4.02 | 3.73 | [273] |
| $LaLuO_3$ (LLO) | 4.17 | 7.19 | [274] |
| MgO | 4.21 | 8.07 | [275] |

**Table 3: Variation of Curie temperature ($T_c$) or metal-to-insulator transition temperature ($T_{MIT}$) of manganite thin films with epitaxial strain produced by growing films of different thicknesses on different substrates.**

| Manganite thin films | Substrates | Strain (in-plane) Tensile (+ve) and Compressive (-ve) | $T_c$ or $T_{MIT}$ (K) | | References |
|---|---|---|---|---|---|
| | | | *No strain* | With strain | |
| $La_{0.67}Sr_{0.33}MnO_3$ | (001) LAO | -2.00% | 345 | 300 | [112] |
| $La_{0.67}Sr_{0.33}MnO_3$ | (001) STO | +0.80% | 345 | 320 | [112] |
| $La_{0.67}Sr_{0.33}MnO_3$ | (110) NGO | -0.30% | 345 | 340 | [112] |
| $La_{0.7}Sr_{0.3}MnO_3$ | (001) STO | +0.93% | 345 | 315 | [181] |
| $La_{0.67}Sr_{0.33}MnO_3$ | (001) LSAT | -0.40% | 345 | 345 | [338] |
| $La_{0.67}Sr_{0.33}MnO_3$ | (110) GSO | +2.30% | 345 | 335 | [338] |
| $La_{0.67}Sr_{0.33}MnO_3$ | SRO/(001) STO | +0.93% | 345 | 298 | [338] |
| $La_{0.67}Sr_{0.33}MnO_3$ | $LaSrFeO_3$ (LSFO)/(001)STO | +0.93% | 345 | 295 | [338] |
| $La_{0.67}Sr_{0.33}MnO_3$ | SRO/(110) LSAT | +0.40 | 345 | 275 | [338] |
| $La_{0.67}Sr_{0.33}MnO_3$ | LSFO/(110) GSO | +2. 3% | 345 | 250 | [338] |
| $La_{0.67}Sr_{0.33}MnO_3$ | (110) DSO | +2.0% | 345 | 200 | [338] |
| $La_{0.7}Sr_{0.3}MnO_3$ | (001) LAO | -2.30% | 345 | 290 | [332] |
| $La_{0.7}Sr_{0.3}MnO_3$ | (001) LSGO | -1.00% | 345 | 323 | [332] |
| $La_{0.7}Sr_{0.3}MnO_3$ | (110) NGO | -0.50% | 345 | 326 | [332] |
| $La_{0.7}Sr_{0.3}MnO_3$ | (100) LSAT | -0.40% | 345 | 330 | [332] |
| $La_{0.7}Sr_{0.3}MnO_3$ | (100) STO | +0.60% | 345 | 346 | [332] |
| $La_{0.7}Sr_{0.3}MnO_3$ | (110) GSO | +2.30% | 345 | 275 | [332] |
| $La_{0.7}Sr_{0.3}MnO_3$ | (110) SSO | +2.70% | 345 | 280 | [332] |
| $La_{0.7}Sr_{0.3}MnO_3$ | (110) NSO | +3.20% | 345 | 260 | [332] |
| $La_{0.7}Ca_{0.3}MnO_3$ | (001) STO | +1.20% | 250 | 80 | [181] |
| $La_{0.67}Ca_{0.33}MnO_3$ | (001) STO | +1.20% | 270 | 180 | [329] |
| $La_{0.67}Ca_{0.33}MnO_3$ | (100) STO | +0.09% | 270 | 248 | [348] |
| $La_{0.67}Ca_{0.33}MnO_3$ | (110) STO | +0.22% | 270 | 247 | [348] |
| $La_{0.67}Ca_{0.33}MnO_3$ | (111) STO | +0.32% | 270 | 245 | [348] |
| $La_{0.80}Ca_{0.20}MnO_3$ | (001) STO | +0.70% | 270 | 188 | [349] |
| $La_{0.80}Ca_{0.20}MnO_3$ | (001) STO | +0.10% | 270 | 200 | [349] |
| $La_{0.80}Ca_{0.20}MnO_3$ | (001) LAO | -1.00% | 270 | 170 | [349] |
| $La_{0.80}Ca_{0.20}MnO_3$ | (001) LAO | -0.40% | 270 | 225 | [349] |
| $La_{0.67}Ca_{0.33}MnO_3$ | (001) STO | +1.00% | 270 | 180 | [334] |
| $La_{0.67}Ca_{0.33}MnO_3$ | (110) STO | +0.80% | 270 | 255 | [334] |
| $La_{5/8-0.3}Pr_{0.3}Ca_{3/8}MnO_3$ | (100) SLGO | < 0.1% | 250 | 240 | [82] |
| $La_{5/8-0.3}Pr_{0.3}Ca_{3/8}MnO_3$ | (100) LAO | -1.40% | 250 | 230 | [82] |
| $La_{5/8-0.3}Pr_{0.3}Ca_{3/8}MnO_3$ | (100) STO | +1.60% | 250 | 220 | [82] |
| $Pr_{0.7}(Ca_{0.8}Sr_{0.2})_{0.3}MnO_3$ | (001) LAO | -1.55% | 95 ($T_{MIT}$) | 50 ($T_{MIT}$) | [351] |
| $Pr_{0.7}(Ca_{0.8}Sr_{0.2})_{0.3}MnO_3$ | (001) LAO | -1.10% | 95 ($T_{MIT}$) | 80 ($T_{MIT}$) | [351] |
| $(La_{1-y}Pr_y)_{0.67}Ca_{0.33}MnO_3$ with $y \leq 0.35$ | (110) NGO | +0.07% | 220 ($T_{MIT}$) | 165 ($T_{MIT}$) | [354] |
| $(La_{1-y}Pr_y)_{0.67}Ca_{0.33}MnO_3$ with $y \leq 0.35$ | (001) LSAT | -0.03%; +0.24% | 190 ($T_{MIT}$) | 155 ($T_{MIT}$) | [354] |
| $(La_{1-y}Pr_y)_{1-x}Ca_xMnO_3$ ($x$ = 0.33 & $y$ = 0.60) | (020) NGO | +0.25% | 185 ($T_{MIT}$) | 155 ($T_{MIT}$) | [279] |
| $(La_{1-y}Pr_y)_{1-x}Ca_xMnO_3$ ($x$ = 0.33 & $y$ = 0.60) | ($\bar{1}$01) NGO | +0.60% | 185 ($T_{MIT}$) | 130 ($T_{MIT}$) | [279] |

**Table 4: Comparison of magnetocaloric effects (change in entropy ΔS and RCP) with strain in selected manganite thin films.**

| Manganite thin films | Peak Temperature $T_o$ (K) | ΔS ($J K^{-1} kg^{-1} T^{-1}$) | RCP ($J kg^{-1}T^{-1}$) | Substrate | Strain (%) on film | References |
|---|---|---|---|---|---|---|
| $La_{0.67}Ca_{0.33}MnO_3$ | 265 | -1.2 | ~30 | (001) STO | 0.0 | [442] |
| $La_{0.7}Ca_{0.3}MnO_3$ | 225 | -0.7 | 13 | (001) BTO | +1.8 | [115] |
| $La_{0.7}Ca_{0.3}MnO_3$ | 190 | -9 | 14 | (001) BTO | -0.23 | [115] |
| $La_{0.67}Ca_{0.33}MnO_3$ | 250 | -0.5 | 40 | (100) LAO | 0.0 | [441] |
| $La_{0.67}Ba_{0.33}MnO_3$ | 295 | -0.26 | 20 | (100) LAO | 0.0 | [441] |
| $La_{0.67}Sr_{0.33}MnO_3$ | 345 | -0.32 | 28 | (100) LAO | 0.0 | [441] |
| $La_{0.67}Sr_{0.33}MnO_3$ | 321 | -0.95 | 22.83 | (001) LSAT | -0.05 | [444] |
| $La_{0.67}Sr_{0.33}MnO_3$ | 312 | -1.03 | 33.44 | (001) STO | +0.90 | [444] |
| $La_{0.67}Sr_{0.33}MnO_3$ | 283 | -1.95 | 14 | (001) BTO | +0.25 | [443] |
| $La_{0.67}Sr_{0.33}MnO_3$ | 210 | -1.0 | 10 | (001) BTO | -0.23 | [443] |
| $Sm_{0.55}Sr_{0.45}MnO_3$ | 120 | -1.67 | 113 | (001) STO | +1.7 | [116] |
| $Sm_{0.55}Sr_{0.45}MnO_3$ | 130 | -0.94 | 54 | (001) LSAT | +0.98 | [116] |
| $Sm_{0.55}Sr_{0.45}MnO_3$ | 165 | -0.35 | 41 | (001) LAO | -1.0 | [116] |
| $La_{0.67}Ca_{0.33}MnO_3$ | 210 | -1.0 | ~30 | (100) LAO | 0.0 | [445] |
| $La_{0.67}Ca_{0.33}Mn_{0.94}Cr_{0.06}O_3$ | 238 | -0.1 | ~15 | (100) LAO | 0.0 | [445] |
| $La_{0.7}Sr_{0.3}MnO_3/CoFe_2O_4$ | 80 | -0.63 | 73.6 | (100) MgO | +1.2 | [438] |
| $[La_{0.7}Sr_{0.3}MnO_3$ (20 u.c.) $/SrRuO_3$(1 u.c.)$]_{\times 15}$ | 325 | -1.175 | 62.5 | (001) STO | +0.9 | [446] |
| $[La_{0.7}Sr_{0.3}MnO_3$ (20 u.c.) $/SrRuO_3$(3 u.c.)$]_{\times 15}$ | 325 | -1.1 | 59.5 | (001) STO | +0.9 | [446] |
| $[La_{0.7}Sr_{0.3}MnO_3$ (20 u.c.) $/SrRuO_3$(6 u.c.)$]_{\times 15}$ | 325 | -0.76 | 50 | (001) STO | +0.9 | [446] |
| $[La_{0.7}Sr_{0.3}MnO_3$ (11 u.c.) $/SrRuO_3$(3 u.c.)$]_{\times 15}$ | 357 | -0.65 | 34 | (111) STO | -0.64 | [447] |
| [$SrRuO_3$(11 u.c.) / $La_{0.7}Sr_{0.3}MnO_3$ (3 u.c.) /$]_{\times 15}$ | 127 | -0.13 | ~6.5 | (111) STO | +0.64 | [447] |
| $(La_{1-y}Pr_y)_{1-x}Ca_xMnO_3$ ($x$ = 0.375, $y$ = 0.32) | 80 K | -0.25 | - | (100) STO | +1.0 | [448] |
| $(La_{1-y}Pr_y)_{1-x}Ca_xMnO_3$ ($x$ = 0.375, $y$ = 0.32) (500nm disk array) | 100 K | -1.0 | - | (100) STO | +1.0 | [448] |

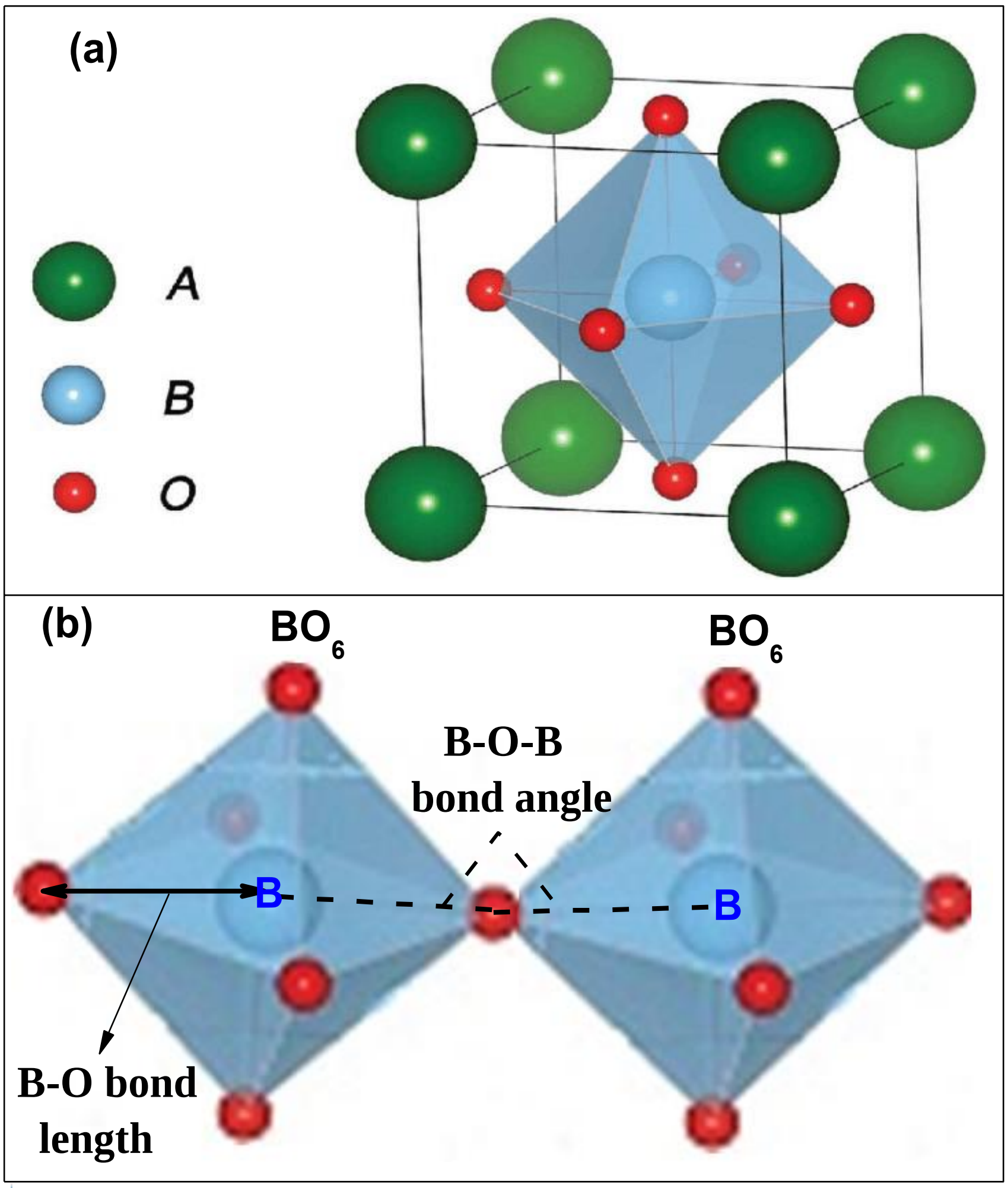

Fig.1: (a) Schematic of a unit cell of the perovskite ($ABO_3$) oxide structure with A sites atom (green spheres) and the oxygen (red spheres) octahedral containing the B site atom (blue sphere). (b) $BO_6$ octahedral showing B-O bond length and B-O-B bond angle in the perovskite structure, which contributes to two main lattice distortions of the octahedral.

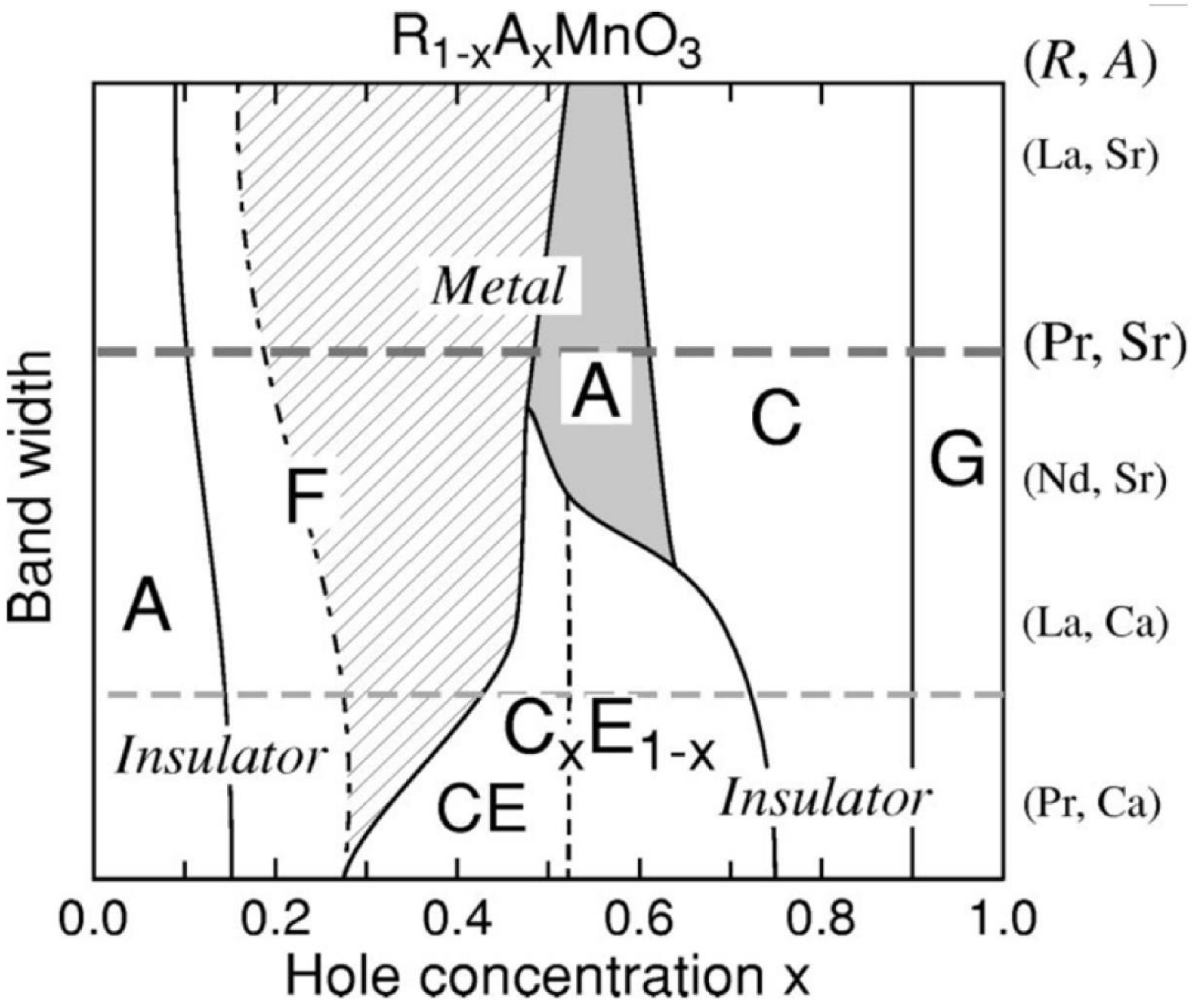

Fig. 2: Schematic phase diagram of $R_{1-x}A_xMnO_3$ (R: rare-earth cation, A: alkaline earth metal cation). F denotes FM state. A, CE, C, and G denote A-type, CE-type, C-type, and G-type AFM states, respectively. The $C_xE_{1-x}$ represents an incommensurate charge/orbital ordered state. From [155].

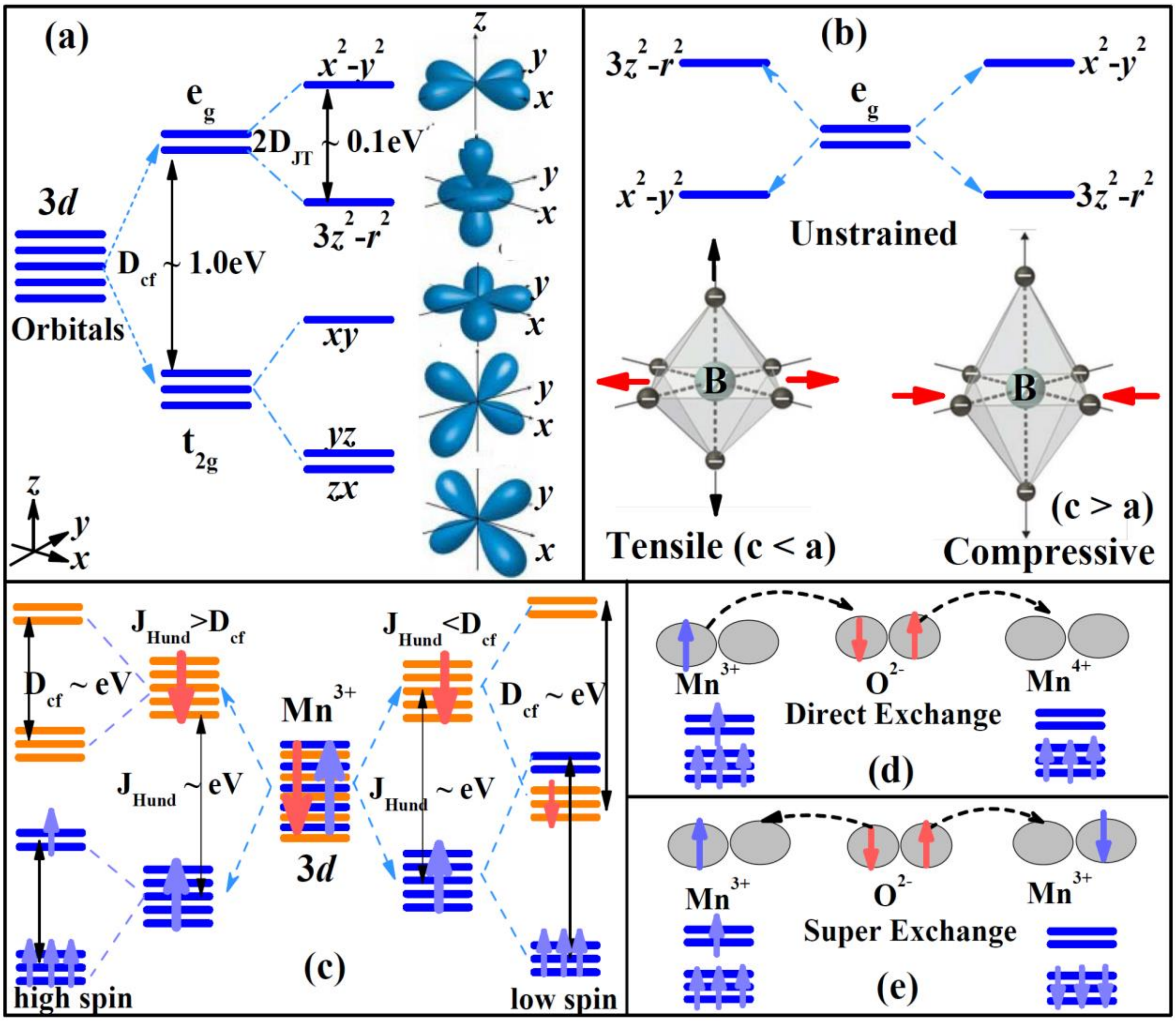

Fig. 3: (a) The schematic of energy splitting of 3*d* orbital in perovskites. Crystal field lifts the degeneracy of 3d orbital with $e_g$ (high energy) and $t_{2g}$ (lower energy) orbitals and the energy gap ($\Delta_{cf}$) is in the order of eV. The degeneracy can be further lifted by lattice distortion like the Jahn-teller effect with the energy gap ($\Delta_{JT}$) of order ~ 0.1 eV. The spatial distribution of the five *d* orbitals (*xy*, *yz*, *zx*, $x^2$-$y^{2,}$ and $3z^2$-$r^2$) is also shown. (b) The $e_g$ orbitals under unstrained and strained conditions. The lower panel of (b) represents the $MnO_6$ octahedral distortions under tensile and compressive strain. (c) spin configurations of a cation ($Mn^{3+}$) in perovskites. The spin-dependent degeneracy (degeneracy of spin-up and spin-down states) is lifted by Hund's energy ($J_{Hund}$). The relative magnitude of Hund's energy and crystal field determines the spin configuration with high spin ($\Delta_{cf} >> J_{Hund}$) and low spin ($\Delta_{cf} << J_{Hund}$) states. Schematic of double exchange (d) and superexchange (e) magnetic interaction, which favours the ferromagnetic and antiferromagnetic coupling, respectively.

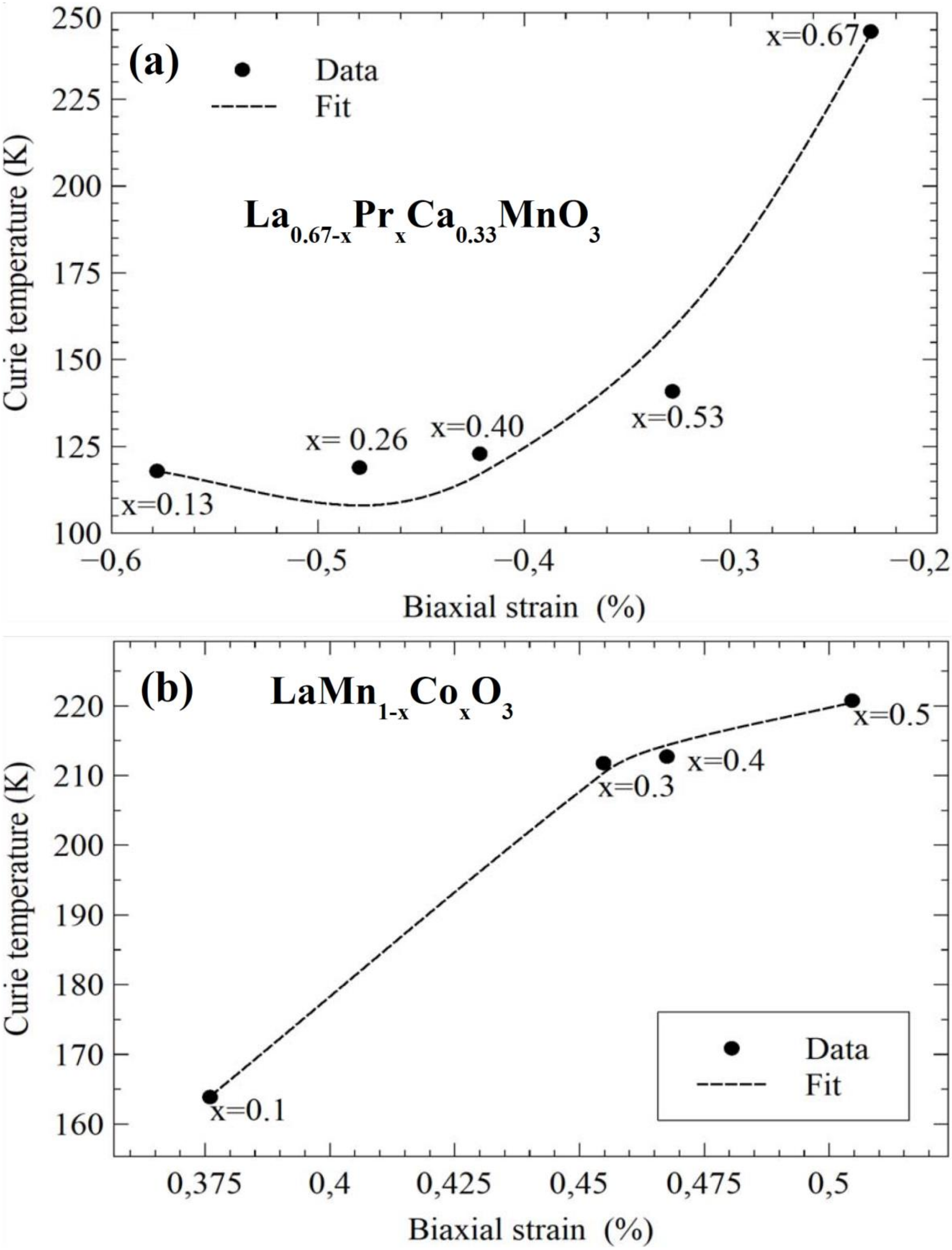

Fig. 4: Variation of biaxial strain with Curie temperature for different compositions of (a) $La_{0.67-x}Pr_xCa_{0.33}MnO_3$ manganites with $0.13 \leq x \leq 0.67$, (b) $LaMn_{1-x}Co_xO_3$ with $0.1 \leq x \leq 0.50$. The values of in-plane biaxial strains for different compositions of the manganites are estimated from XRD measurements. From [195].

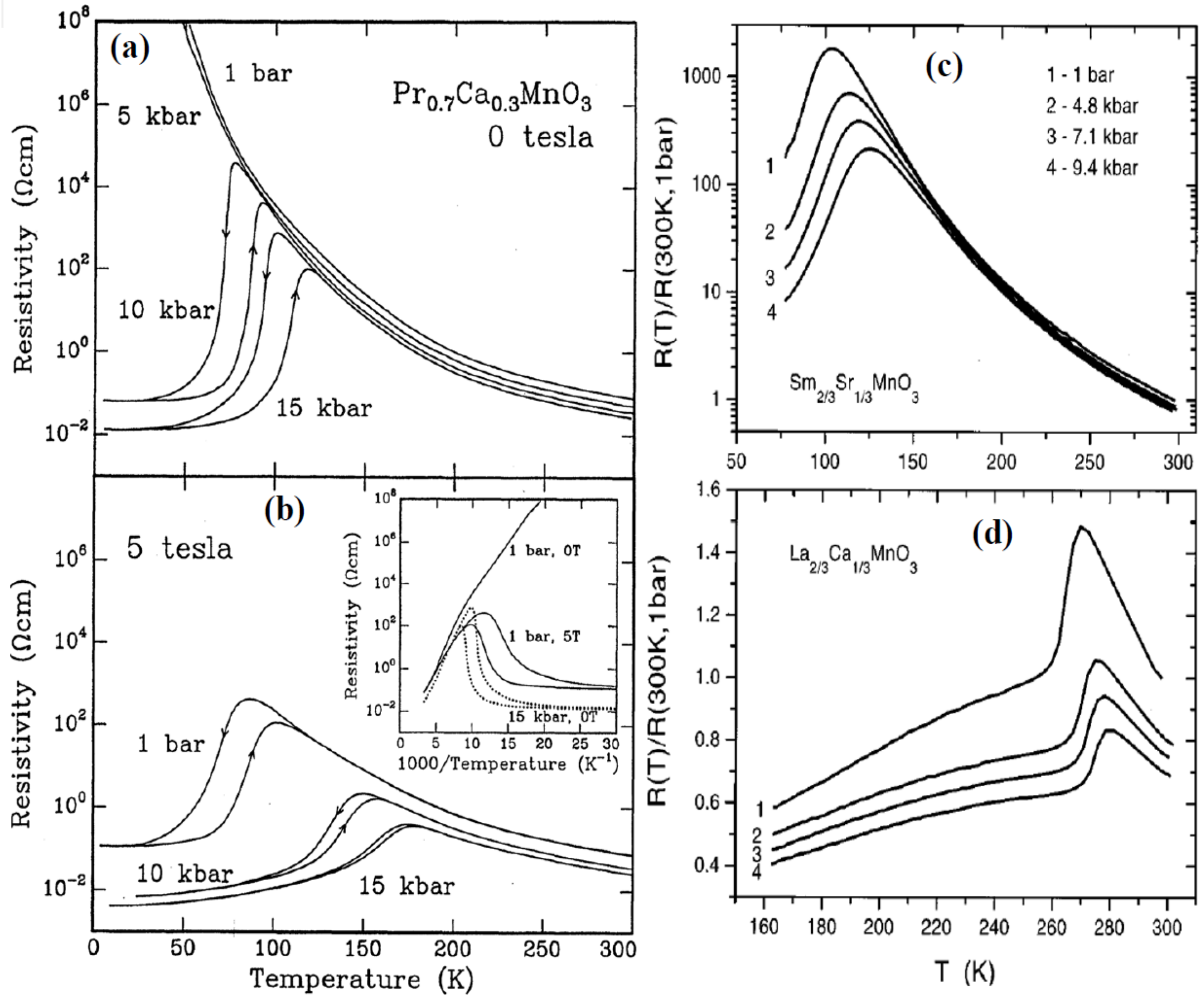

Fig. 5: Temperature-dependent resistivity at different hydrostatic pressures for PCMO ($x$ = 0.3) bulk sample at a magnetic field of (a) 0 and (b) 5 T. Inset of (b) shows a comparison of resistivity as a function of 1000/T for different pressure and magnetic field. Normalized resistance vs temperature at several pressures for (c) SSMO ($x$ = 0.33) and (d) LCMO ($x$ = 0.33) samples in the absence of a magnetic field. A shift in the MIT temperature to a higher temperature as well decrease in the peak resistivity at MIT was observed on the application of hydrostatic pressure. From [196].

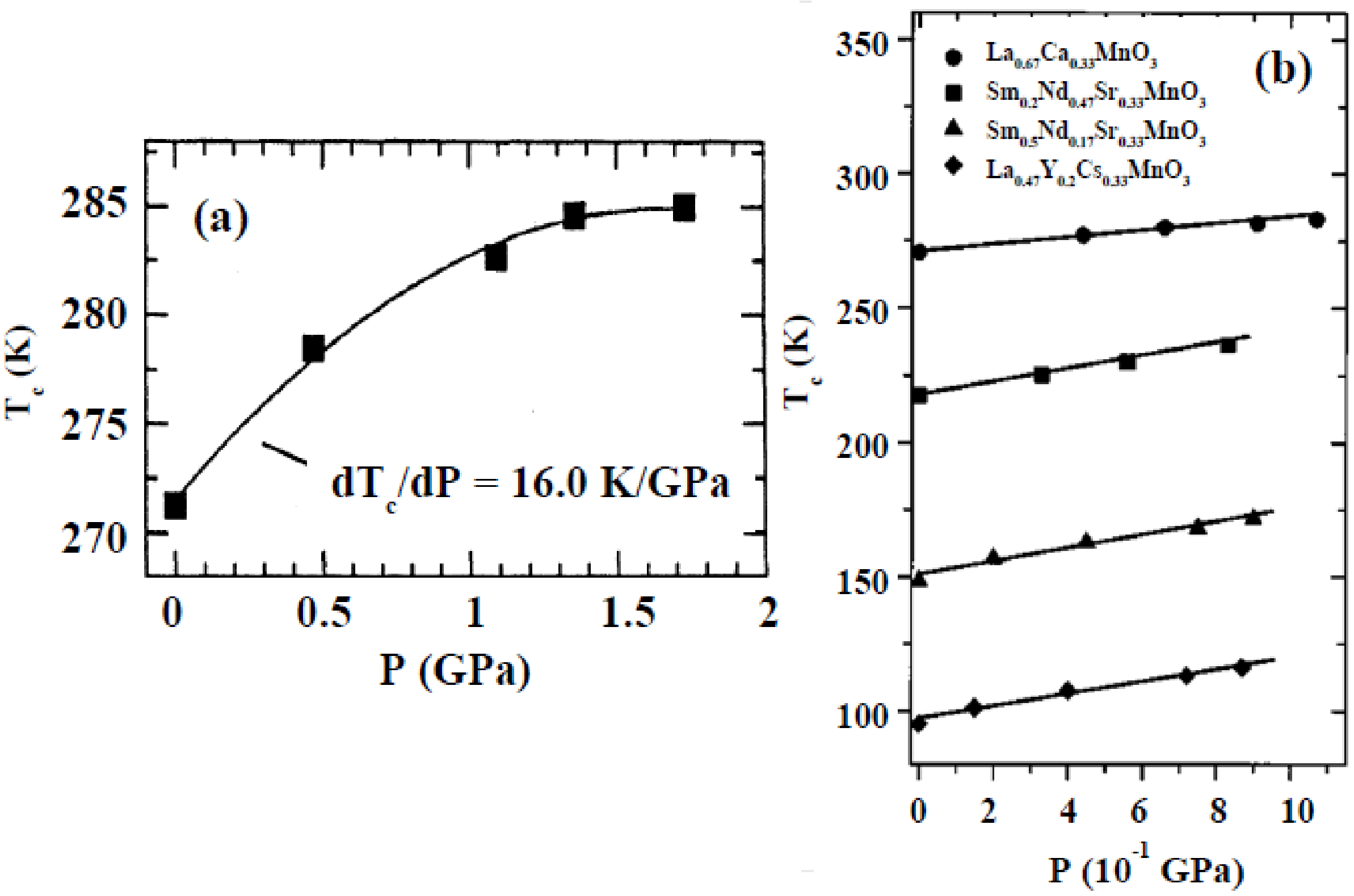

Fig. 6: (a) Variation of $T_c$ with applied hydrostatic pressure ($P$) for bulk LCMO ($x$ = 0.4). From [203]. (b) $T_c$ vs $P$ curves for some other manganites. From [197]. The positive value of d$T_c$ /d$P$ suggests the stabilization of the ferromagnetic/metallic state.

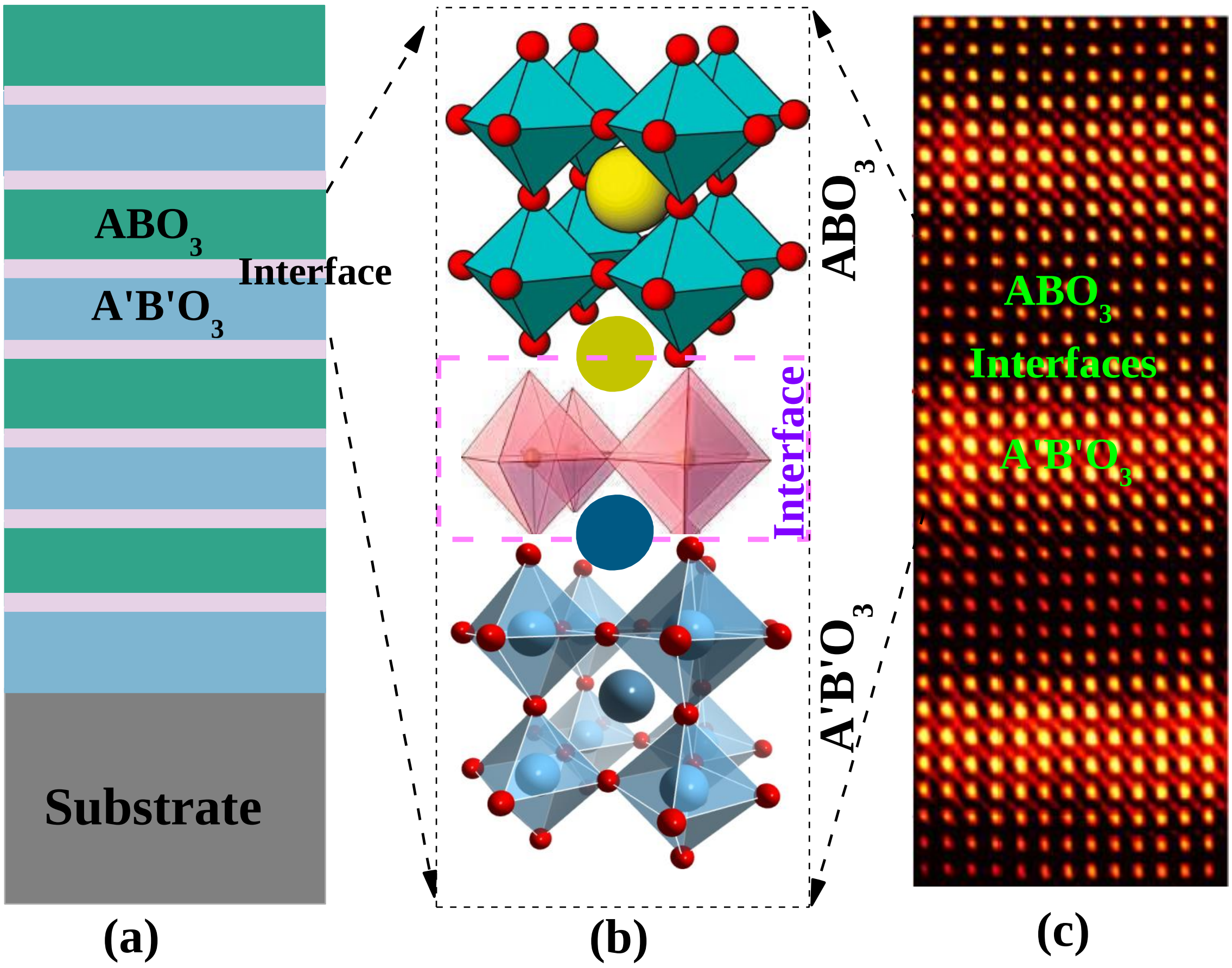

Fig. 7: (a) Schematic of oxide superlattice consisting of periodic alternating layers of different oxides forming different interfaces. (b) Schematic of a perovskite $ABO_3/A'B'O_3$ bilayer with the interface. (c) High-angle annular dark-field (HAADF) image showing the interfaces between two perovskite oxides $ABO_3/A'B'O_3$. Interfacial strain can modify the bond length and bond angle resulting in new phenomena at the interface of two complex perovskite structures.

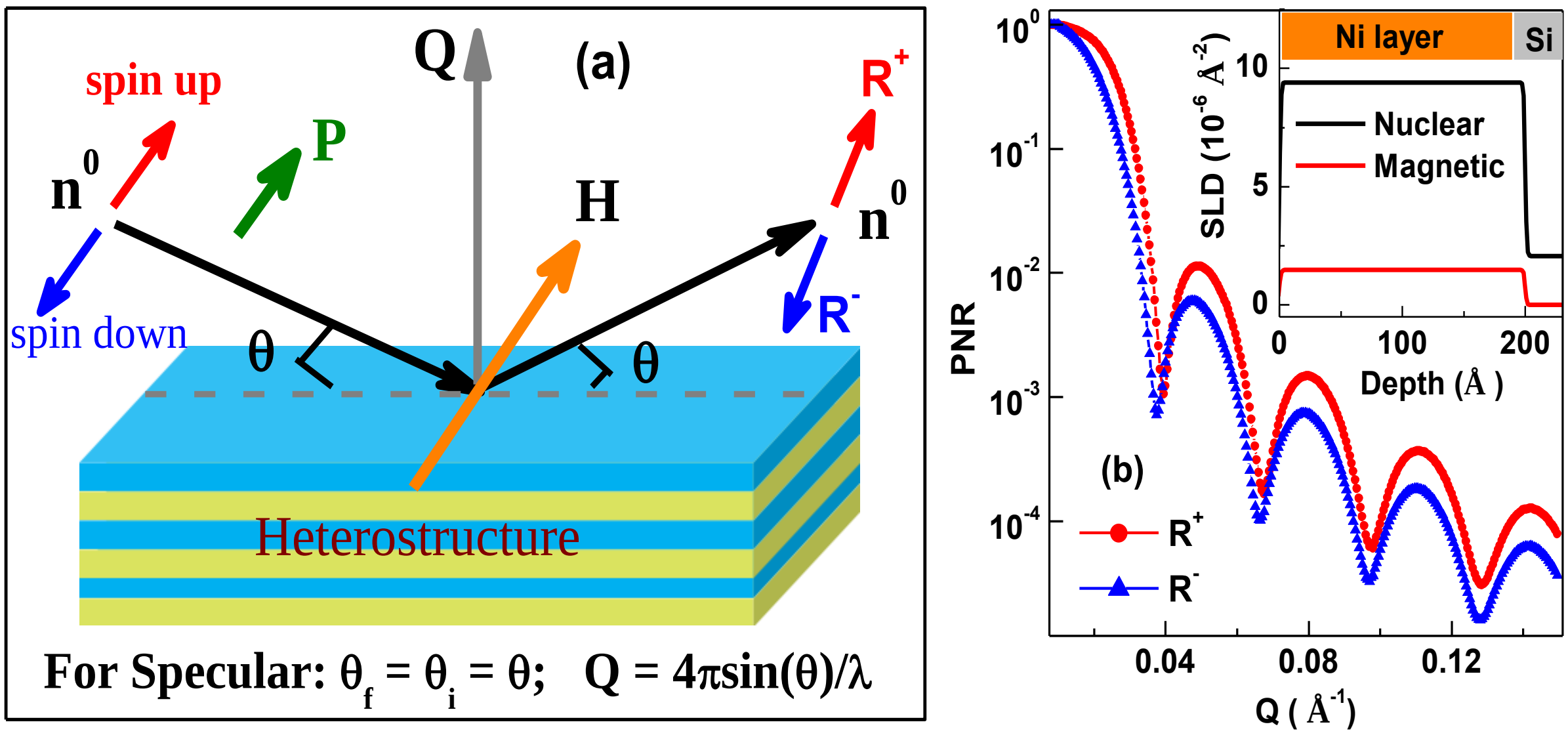

Fig. 8: (a) Schematic of specular PNR (angle of incidence = angle of reflection) experiments, where neutron ($n^0$) reflectivity is measured as a function of the wave-vector transfer $Q$. The polarization ($P$) direction of the neutron beam is either parallel or antiparallel to the applied field ($H$) direction and the corresponding reflectivity is termed as $R^+$ (spin-up) and $R^-$ (spin-down). (b) PNR profiles [$R^+$ and $R^-$] simulated for a Ni film (thickness = 200 Å) on Si substrate (see inset). The splitting of $R^+$ and $R^-$ is evident from (b), suggesting a magnetic contribution from Ni film. Inset (b) shows the nuclear and magnetic scattering length density (NSLD and MSLD) depth profile of the sample used to generate the PNR profile.

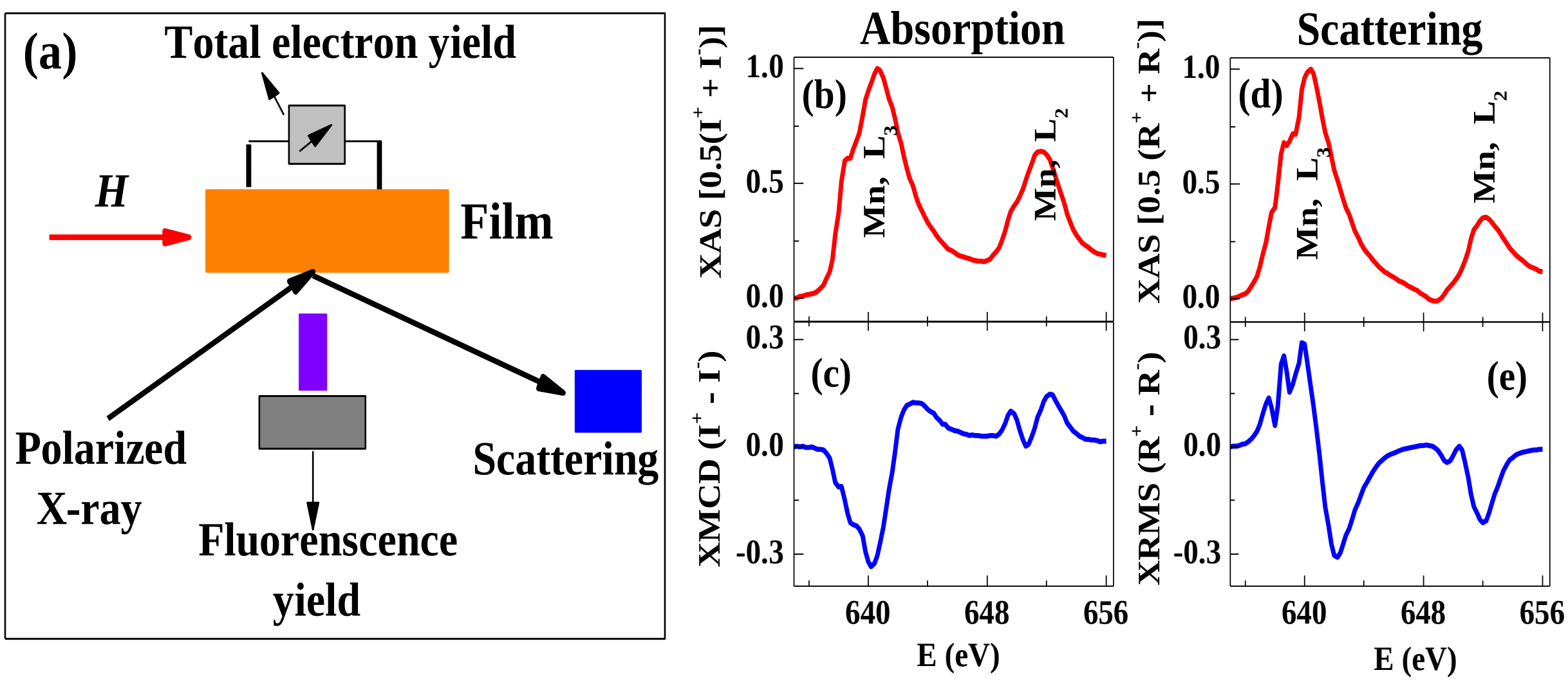

Fig. 9: (a) Schematic showing the experimental configuration for simultaneous measurement of x-ray absorption (XAS), near-surface (TEY) and bulk sensitive (FY) as well as resonant scattering while switching the polarization between left and right circular polarization (LCP and RCP). The energy spectrum of $(La_{1-y}Pr_y)_{1-x}Ca_xMnO_3$ (LPCMO) (x = 0.33 and y =0.60) film at the Mn edge was measured in different modes. (b) Near-surface XAS (sum of the intensity for LCP and RCP x-ray) and (c) XMCD (difference of the intensity for LCP and RCP x-ray) spectra for LPCMO film at 20 K. (d) XAS spectra in scattering mode and (e) X-ray resonant magnetic scattering (XRMS) spectra form LPCMO film at 20 K.

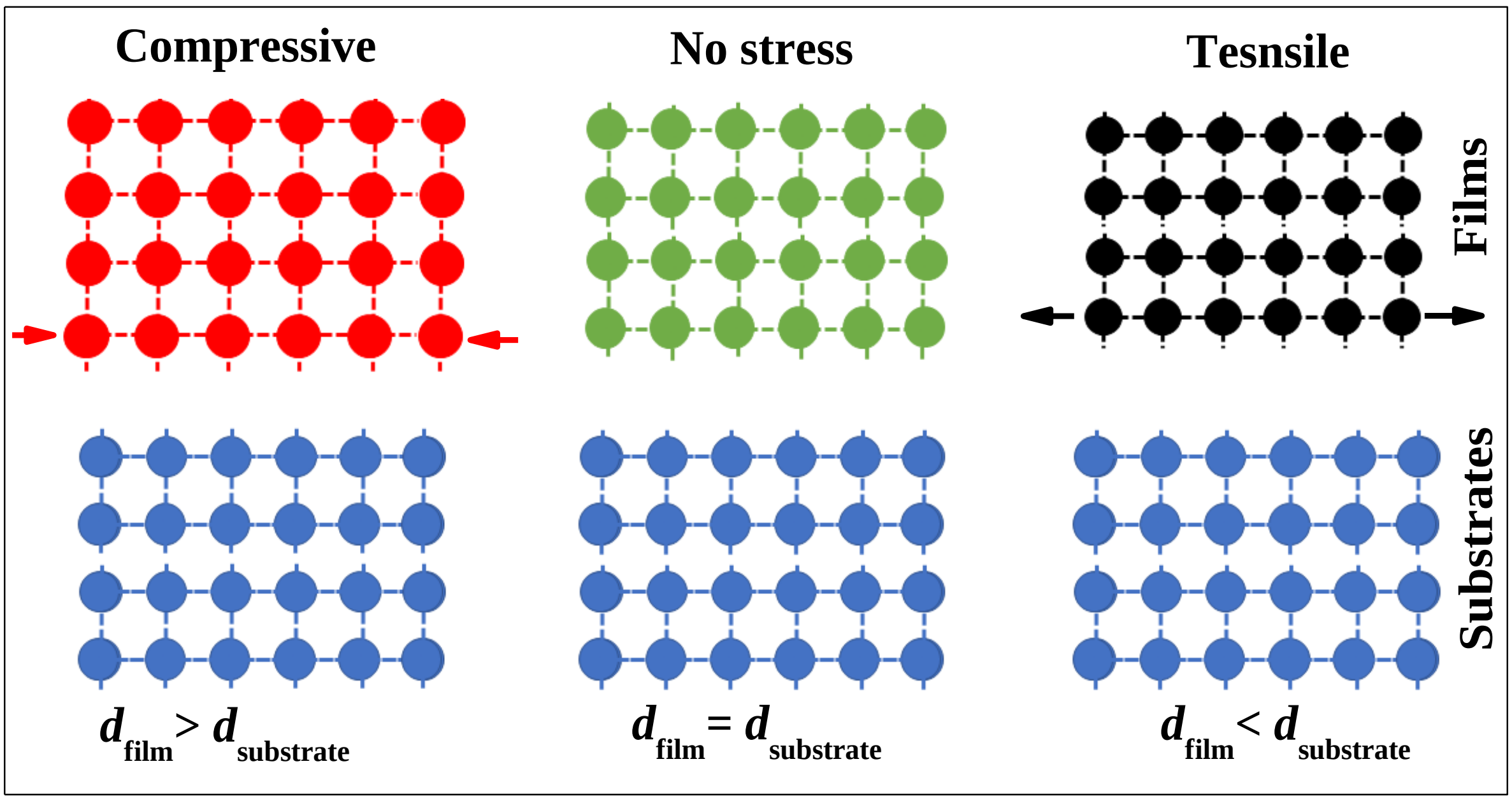

Fig. 10: Schematic of the epitaxial film without stress (centre panel) and with compressive (left panel) and tensile (right panel) stress induced on the film depending on the lattice misfit to the substrate.

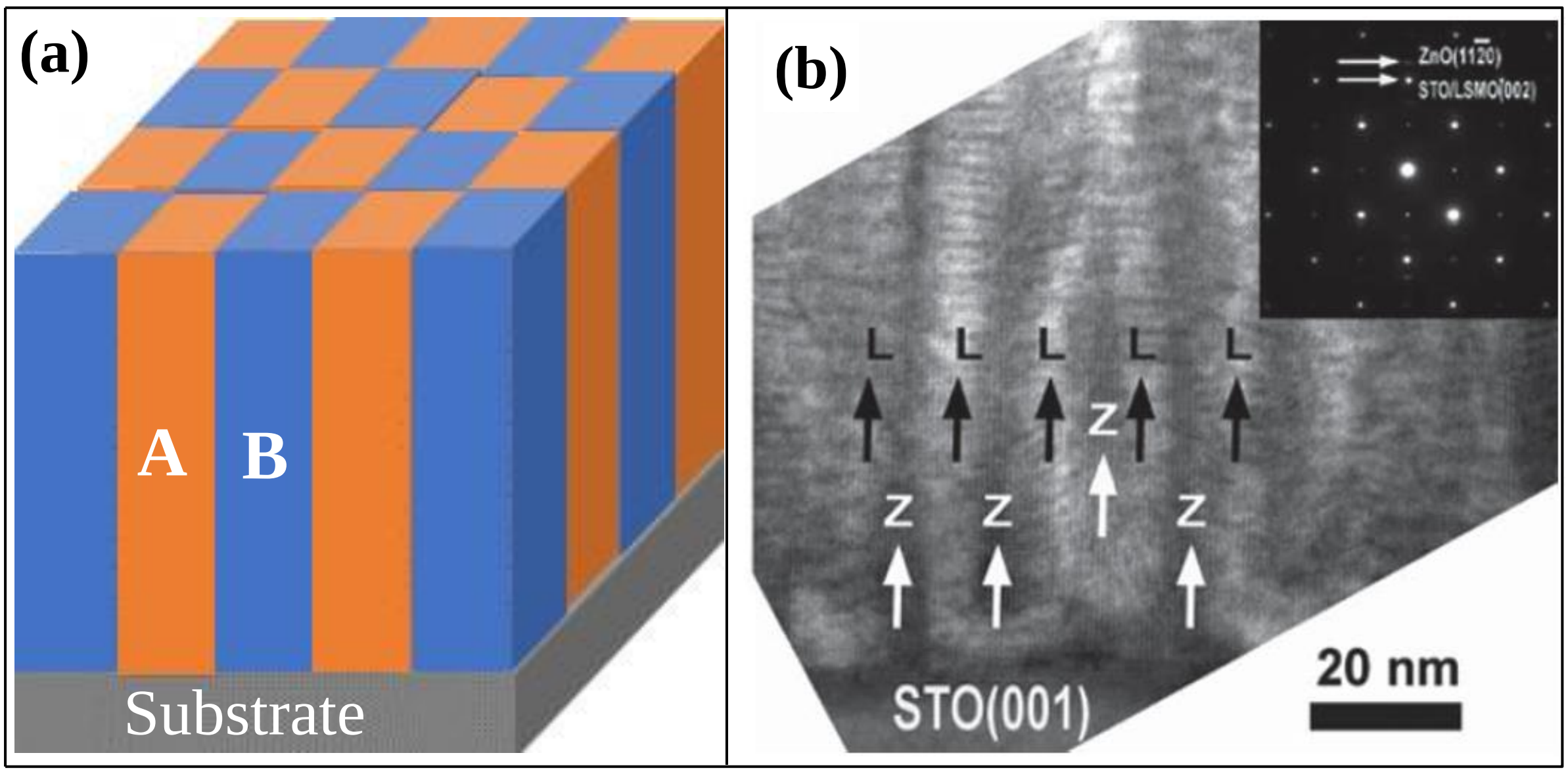

Fig. 11: (a) Schematic of a self-assembled vertically aligned nanocomposite (VAN) film on a substrate with complex oxide A and B nanocolumns. (b) A cross-sectional TEM image of an LSMO: ZnO VAN on an STO substrate with LSMO (white and denoted by L) and ZnO (dark and denoted by Z) phases. Inset shows the corresponding selected-area electron diffraction pattern from the film [283]. Copyright 2011, Wiley.

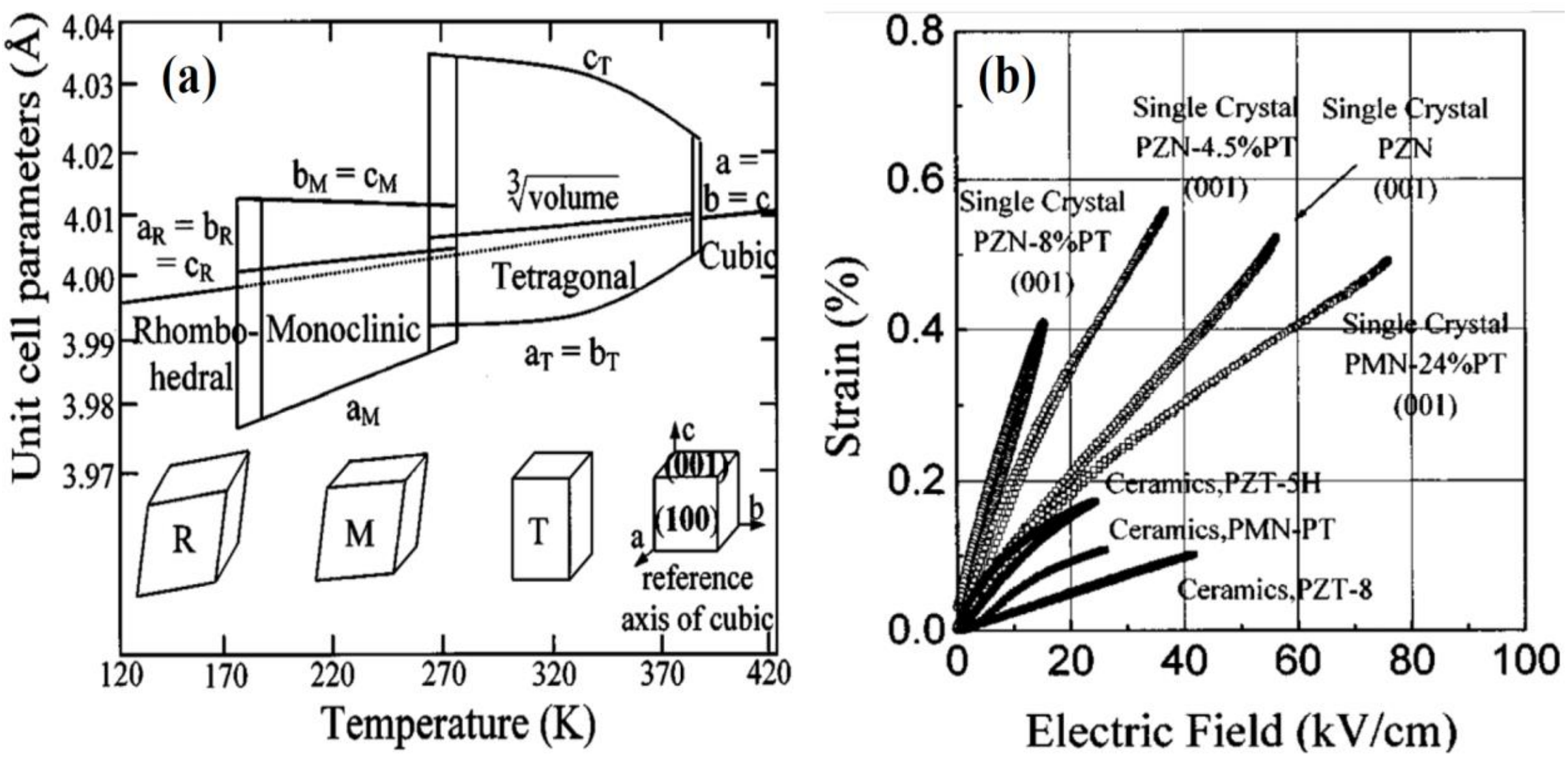

Fig. 12: (a) Variation of lattice parameters as a function of the temperature of $BaTiO_3$ (BTO) crystal. The R, M, and T in the figure, represent tetragonal, monoclinic, and rhombohedral phases, respectively. From [83]. Change in lattice parameter on phase transition in BTO is used to apply strain on the oxide films grown on this substrate. (b) Strain vs *E*-field behaviour for (001) oriented rhombohedral crystals of PZN-PT and PMN-PT and for various electromechanical ceramics. From [288].

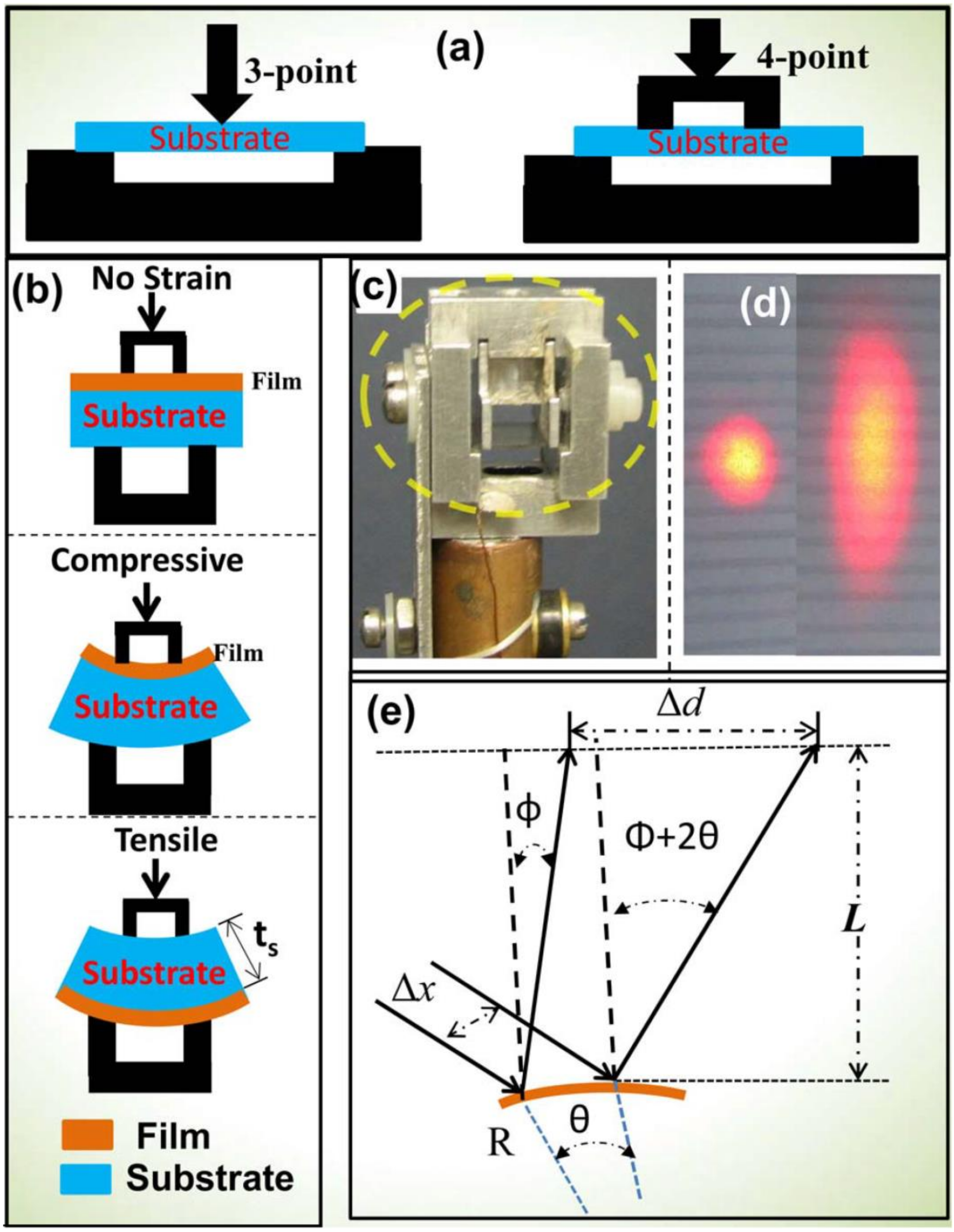

Fig. 13**:** (a) Schematic for three-point beam (left) and four-point beam (right) bending technique for applying compressive and tensile strain to a thin film. (b) side-view schematic representation of applied bending stress (four-point beam) for a thin film under different conditions of strains i.e., no strain, compressive and tensile. (c) End-view image of the four-point (mechanical) jig (circled) mounted on a cryostat. (d) The reflected image of the laser beams from unstrained and strained (using a four-point jig) thin film. (e) The ray diagram of the reflected laser beams from the curved surface for measuring the radius of curvature ($R$) of the curved surface.

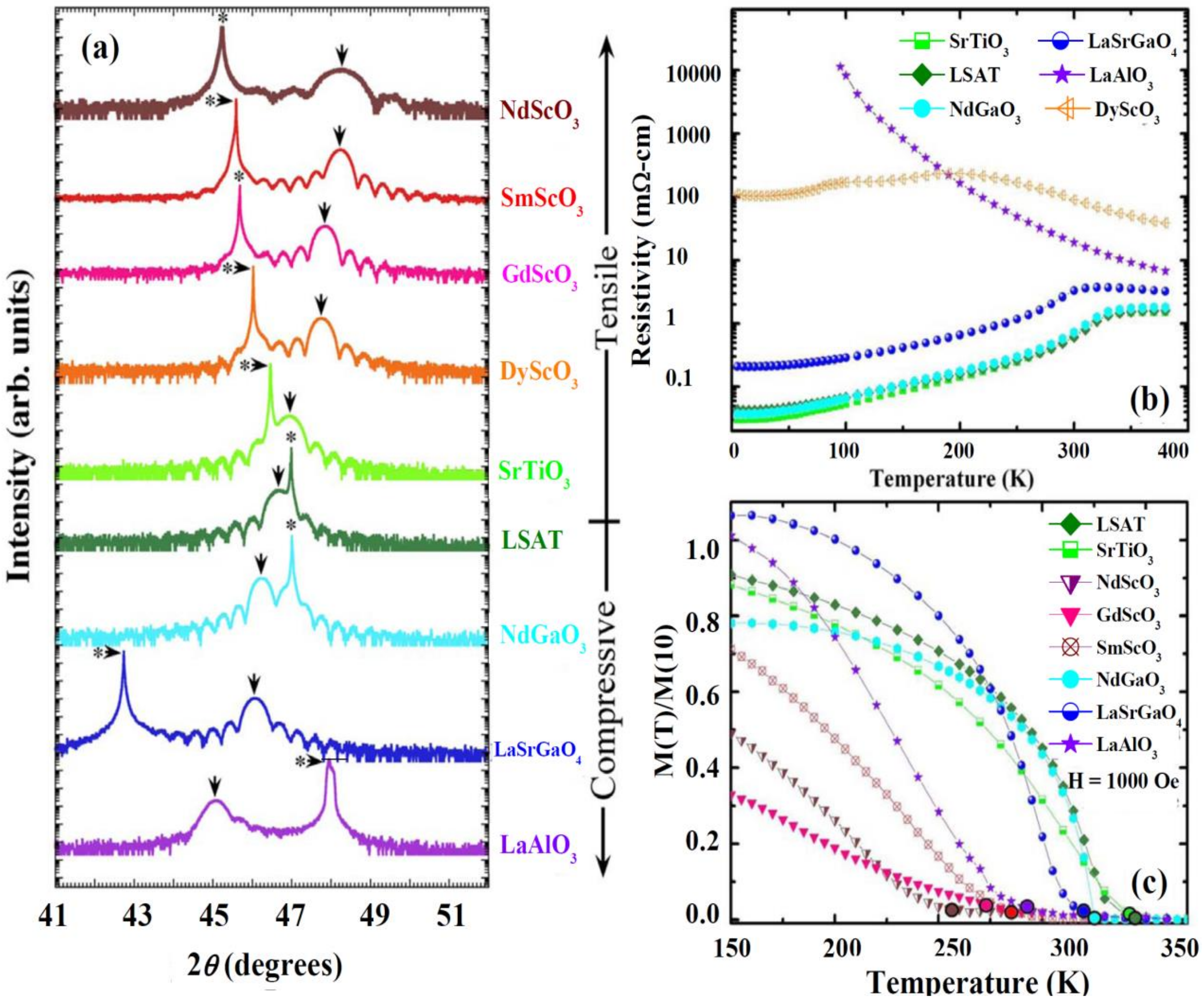

Fig. 14: (a) The θ-2θ XRD scans of LSMO ($x$ = 0.3) films grown on different substrates. The arrows (↓) indicate the (002) peak of LSMO film. (b) Resistivity vs temperature plot, at zero applied magnetic fields, of LSMO films on different substrates. (c) The temperature dependence of the magnetization normalized at 10 K of LSMO film on different substrates cooled in a 1000 Oe field. Closed circles indicate the Curie temperature, $T_C$. From [332].

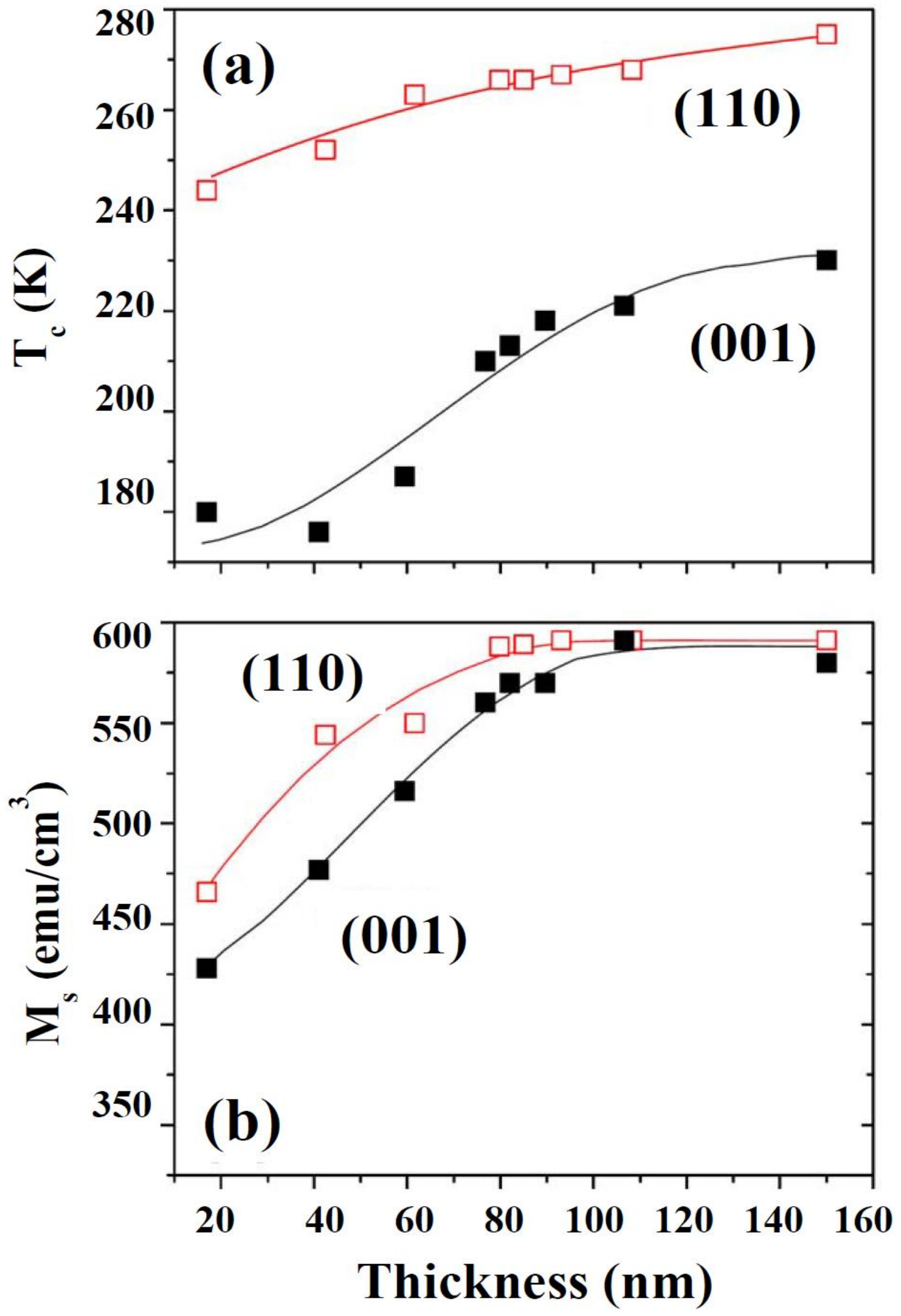

Fig. 15: (a) Curie temperature, $T_c$ and (b) saturation magnetization, $M_s$ of (001) (closed symbols) and (110) (open symbols) LCMO films deposited on (001) and (110) STO substrates as function of film thickness. These figures represent that the (110) LCMO films have higher $T_c$ and $M_s$ than (001) LCMO films of similar thickness. From [334].

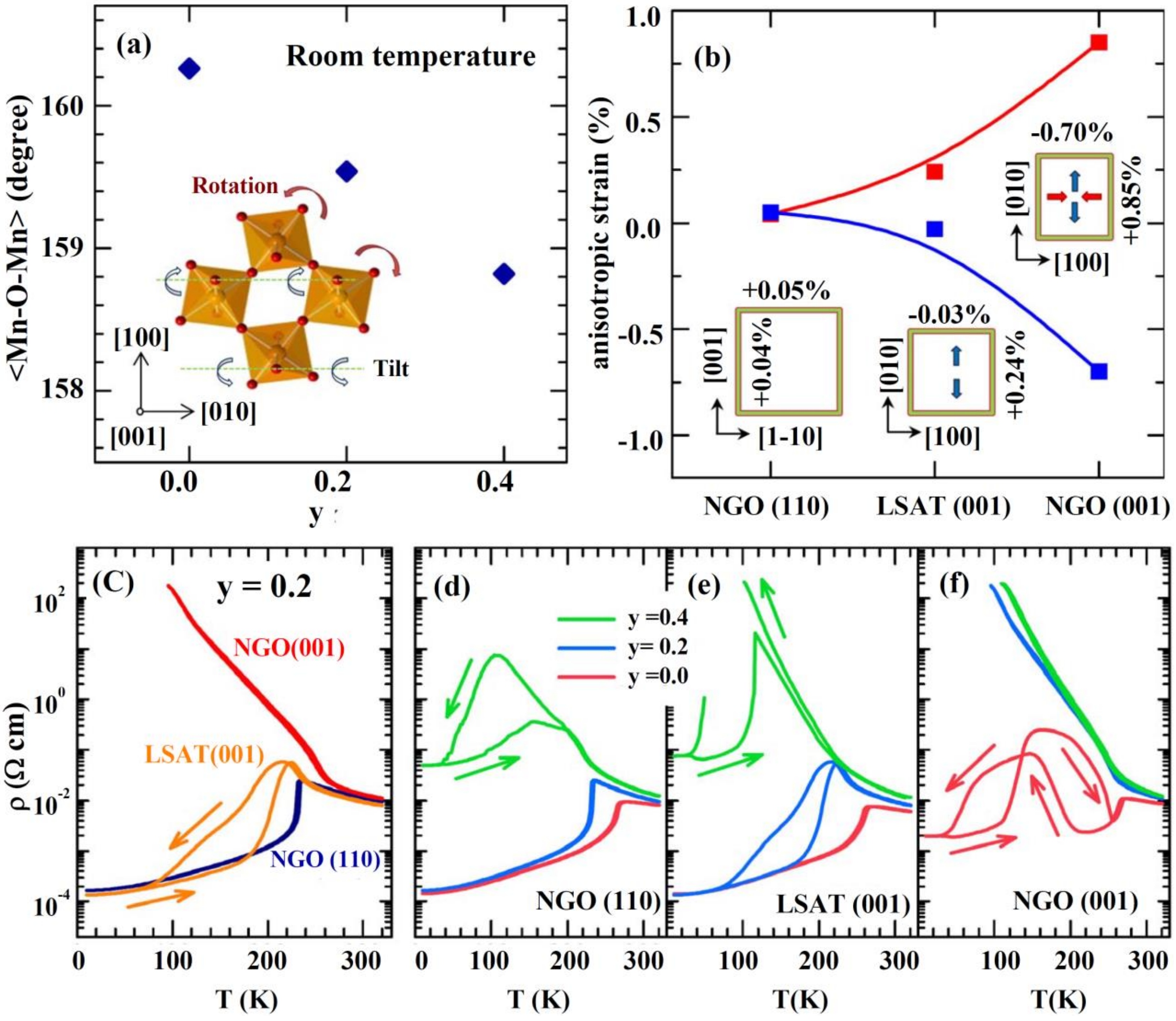

Fig. 16: (a) The average <Mn–O–Mn> bond angle at various Pr doping of bulk LPCMO ($x$ = 0.33 and $0.0 \leq y \leq 0.4$). Inset shows the octahedral framework of orthorhombic perovskite oxides with the *Pbnm* space group). (b) LCMO films on various substrates with progressively increased anisotropic epitaxial strain (AES). The inset shows schematic strain states with in-plane views of the commensurate LCMO films. It is noted that LPCMO films will also show similar AES as their lattice constants are similar. (c) resistivity ($\rho$) vs temperature ($\rho$(T)) curves measured from LPCMO ($x$ = 0.33 and $y$ = 0.20) films on (110) NGO, (001) LSAT, and (001) NGO substrates with progressively increased AES states. $\rho$(T) curves were from LPCMO ($x$ =0.33 and $y$ = 0.0, 0.2, and 0.4) thin films on (d) (110) NGO, (e) (001) LSAT, and (d) (001) NGO substrates, respectively. The transport measurements suggested that the AES and chemical doping can cooperatively facilitate the COI phase in LPCMO films. From [356].

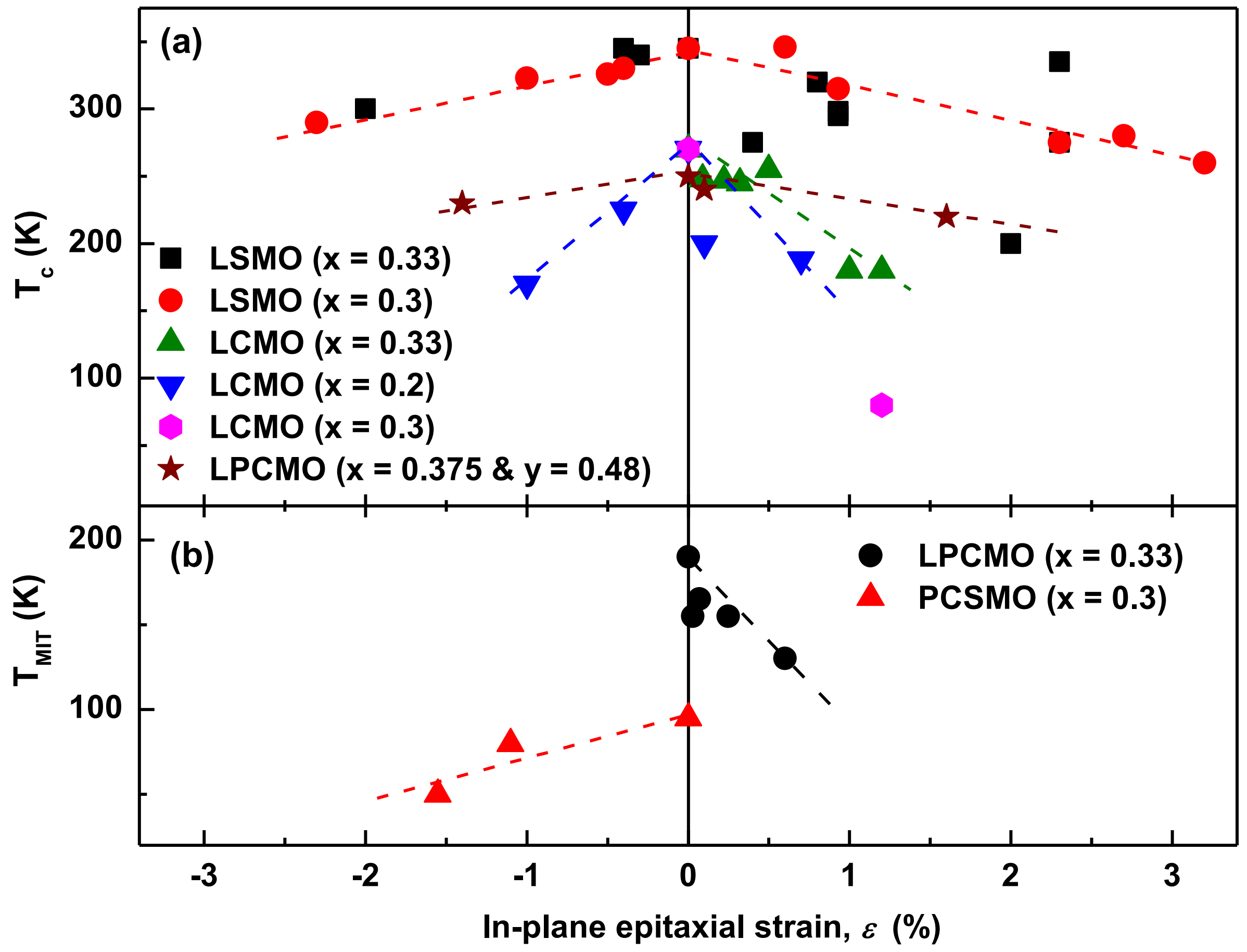

Fig. 17: Effect of in-plane epitaxial strain on (a) magnetic transition temperature ($T_c$) and (b) metal-to-insulator transition temperature ($T_{MIT}$) of different manganite thin films grown on different substrates. The results suggested a mixed report on strain-dependent magnetic properties. Data were taken from [82, 112, 181, 279, 329, 332, 334, 338, 348, 349, 351, 354].

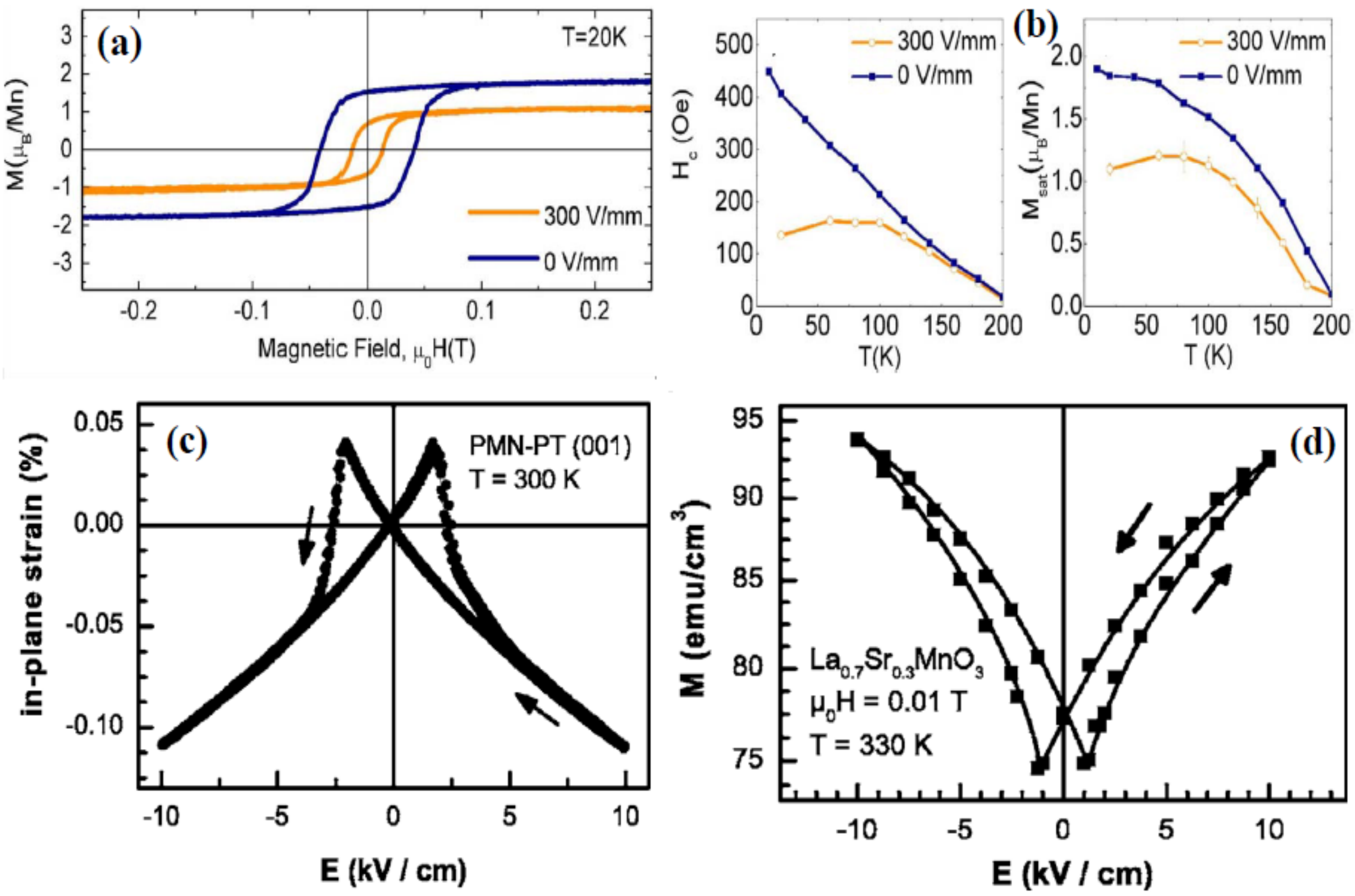

Fig. 18: (a) Magnetic hysteresis loops at 20 K from 100 Å thick LCMO film grown on BTO substrate with no electric field and with a field of 3kV/cm applied perpendicular to the thin film. (b) Variation of coercive fields (left) and saturation moments (right) with temperature, extracted from magnetic hysteresis loops from the film with and without applied electric field. From [363]. (c) In-plane piezoelectric strain vs. the applied electric field, E, || (001) recorded along a (100) edge of a PMN-PT(001) substrate. (d) Magnetization, M || (100) vs electric field E || (001) applied to the substrate for a LSMO ($x$ = 0.3)/PMN-PT(001) heterostructure. From [294].

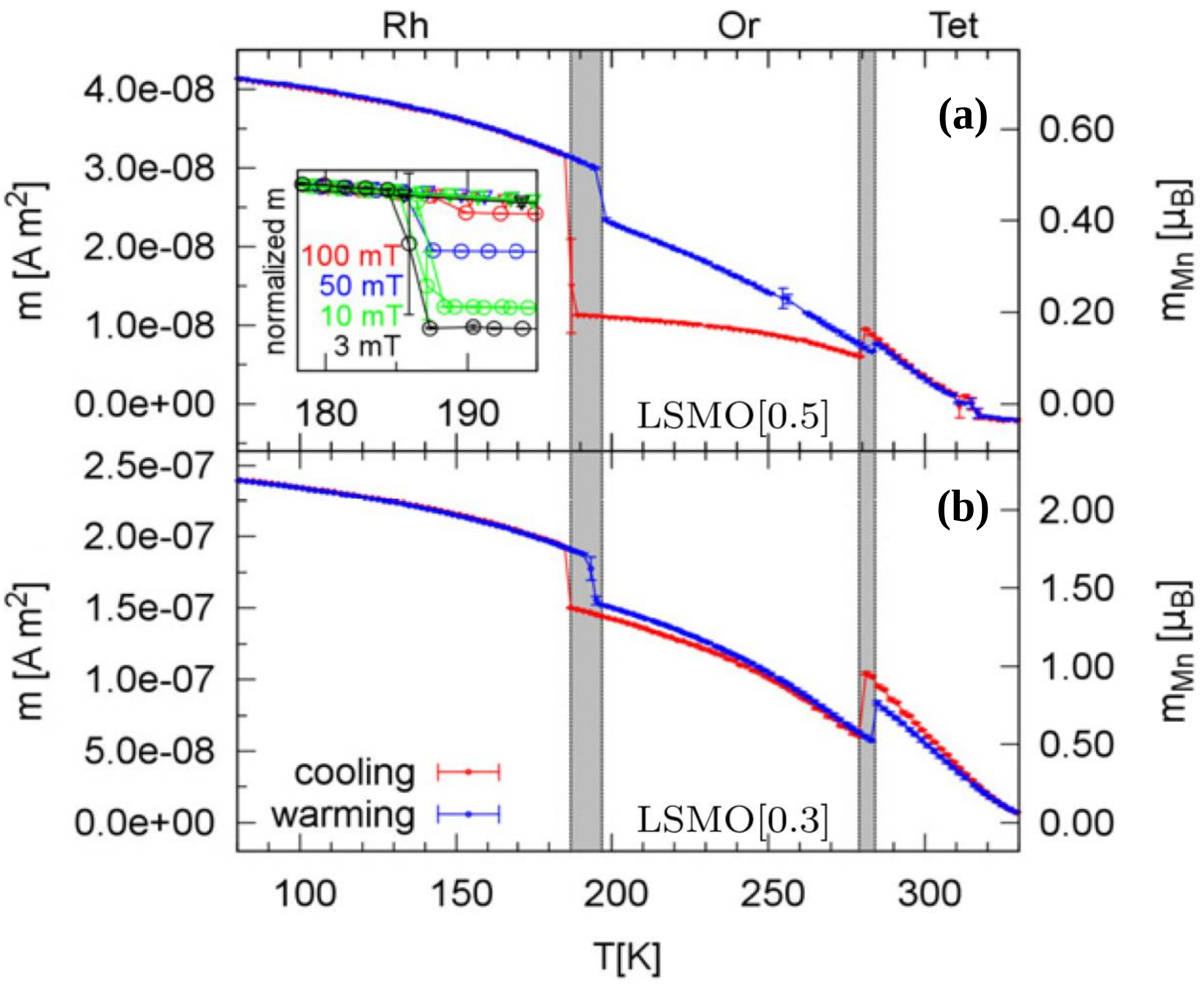

Fig. 19: Variation of in-plane magnetization/magnetic moment of LSMO films, with (a) $x$ = 0.5 and (b) $x$ = 0.3 grown on BTO substrate, with temperature across different structural phase transition of BTO substrates during field cooled cooling (FC) and field cooled warming (FW) in a magnetic field of $\mu_0 H$ = 10 mT. From [285]. The inset (a) shows the evolution of magnetization steps at the Orthorhombic (Or) → rhombohedral (Rh) transition with respect to the external magnetic field.

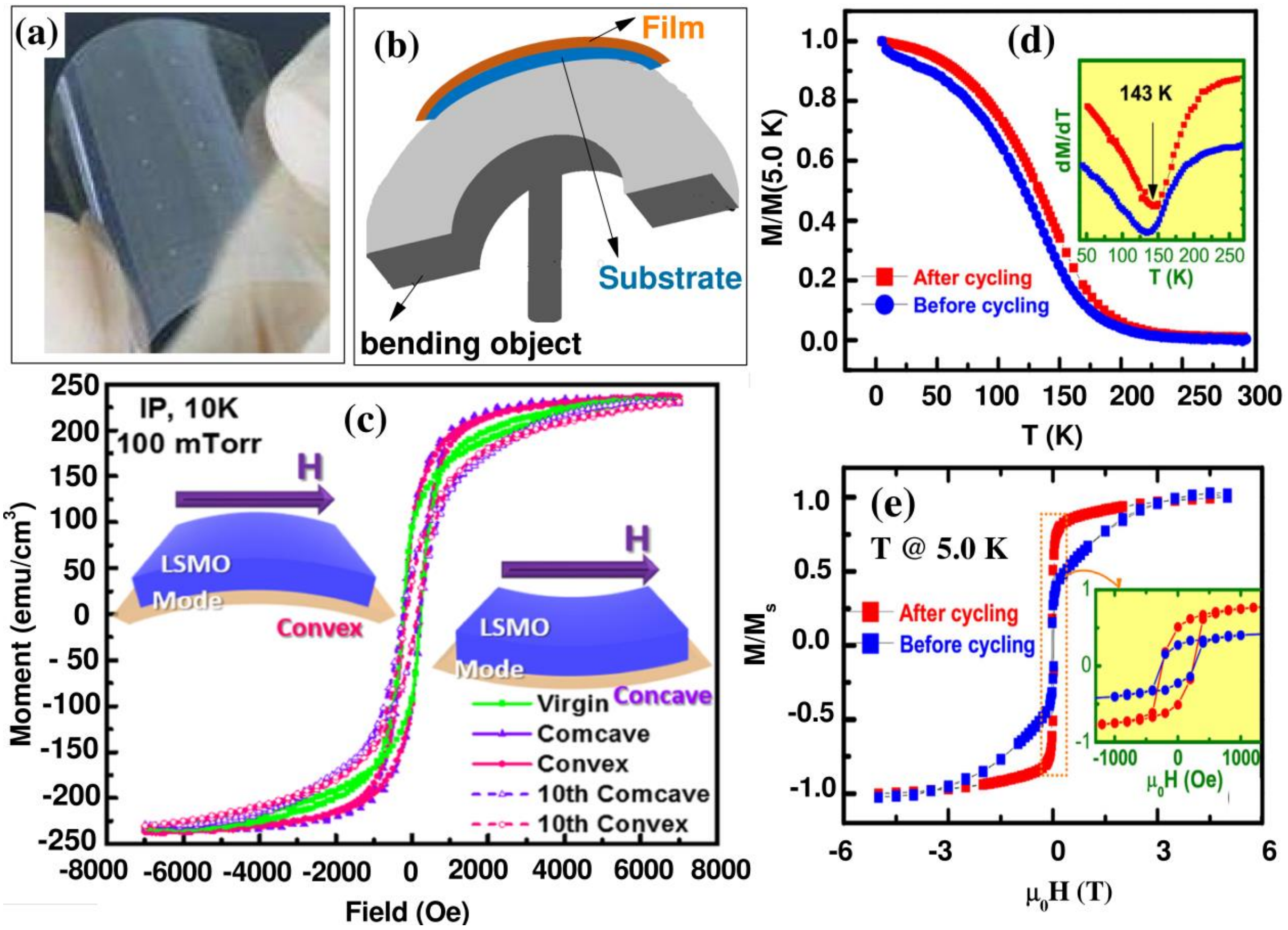

Fig. 20: (a) Image of a flexible plastic substrate. (b) Schematic of a bending object on which the flexible magnetic film is fixed for bending experiment. (c) comparison of M-H hysteresis loops at 10 K with applied field along in-plane (IP) direction of the pristine LSMO ($x$ = 0.33) film on flexible (mica) substrate and with bending [tensile (convex) and compressive (concave)] once and 10 times. Adapted from [391]. Comparison of (d) $M$(T) and (e) $M$(H) curves for BTO/LSMO ($x$ =0.2) heterostructure on mica flexible substrate before and after 10 k times cycling fatigue test. Adapted from [392]. The inset of (d) and (e) shows the plots of d$M$/d$T$ versus $T$ and the magnified magnetic hysteresis loop in the low field region, respectively.

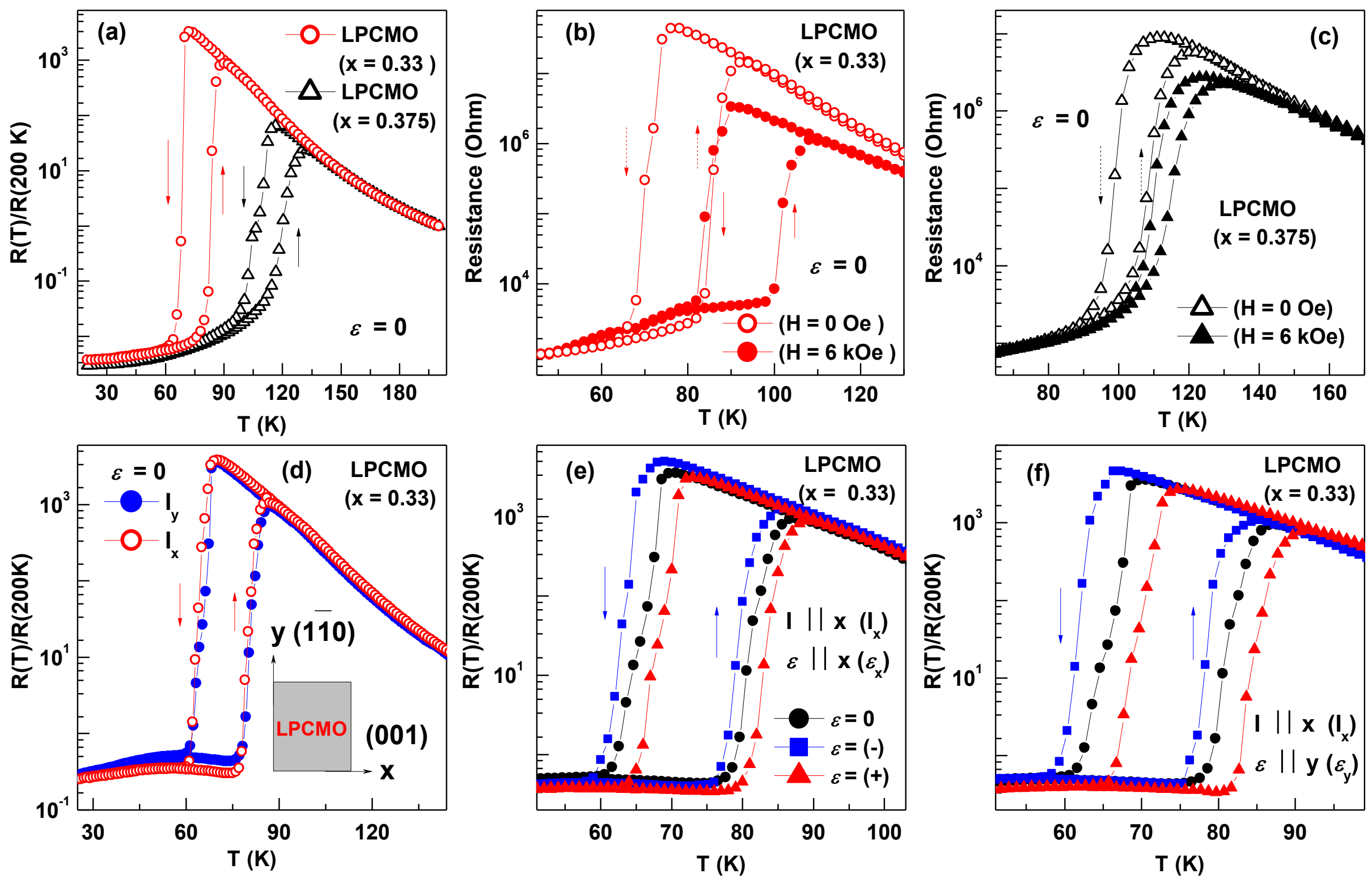

Fig. 21: (a) Normalized resistance [$R(T)/R(200\ \mathrm{K})$] data for LPCMO ($y$ = 0.60 and $x$ = 0.33 and 0.375) films on (110) NGO single crystal substrate. Resistance as a function of temperature at the different applied fields for LPCMO films with $x$ = 0.33 (b) and 0.375 (c). (d) Normalized resistance as a function of temperature for LPCMO films with $x$ = 0.33 measured along two in-plane perpendicular directions (shown in the inset). Normalized resistance as a function of temperature for LPCMO films with $x$ = 0.33 at different applied strains (ε = 0.0%, compressive ε = -0.006% and tensile ε = +0.006%) when both transport and applied strain were along (001) NGO direction (e) and when transport was measured along (001) NGO and the strain was applied along $(1\bar{1}0)$ NGO direction (f). Adapted from [87].

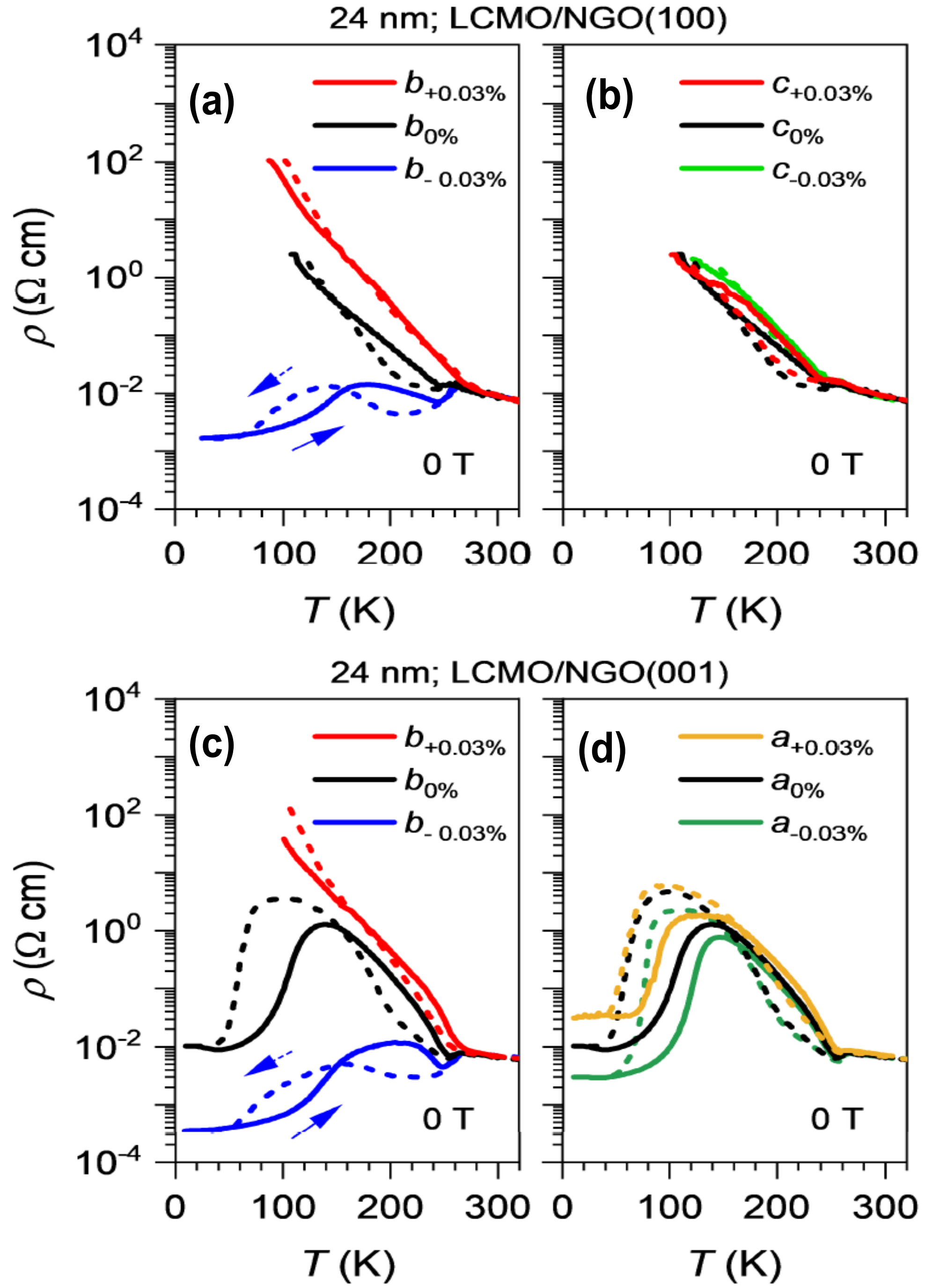

Fig. 22: Temperature-dependent resistivities of the LCMO ($x$ = 0.34) film and thickness 24 nm grown on (100) NGO (a and b) and (001) NGO (c and d) substrate measured under various bending strain states. From [397].

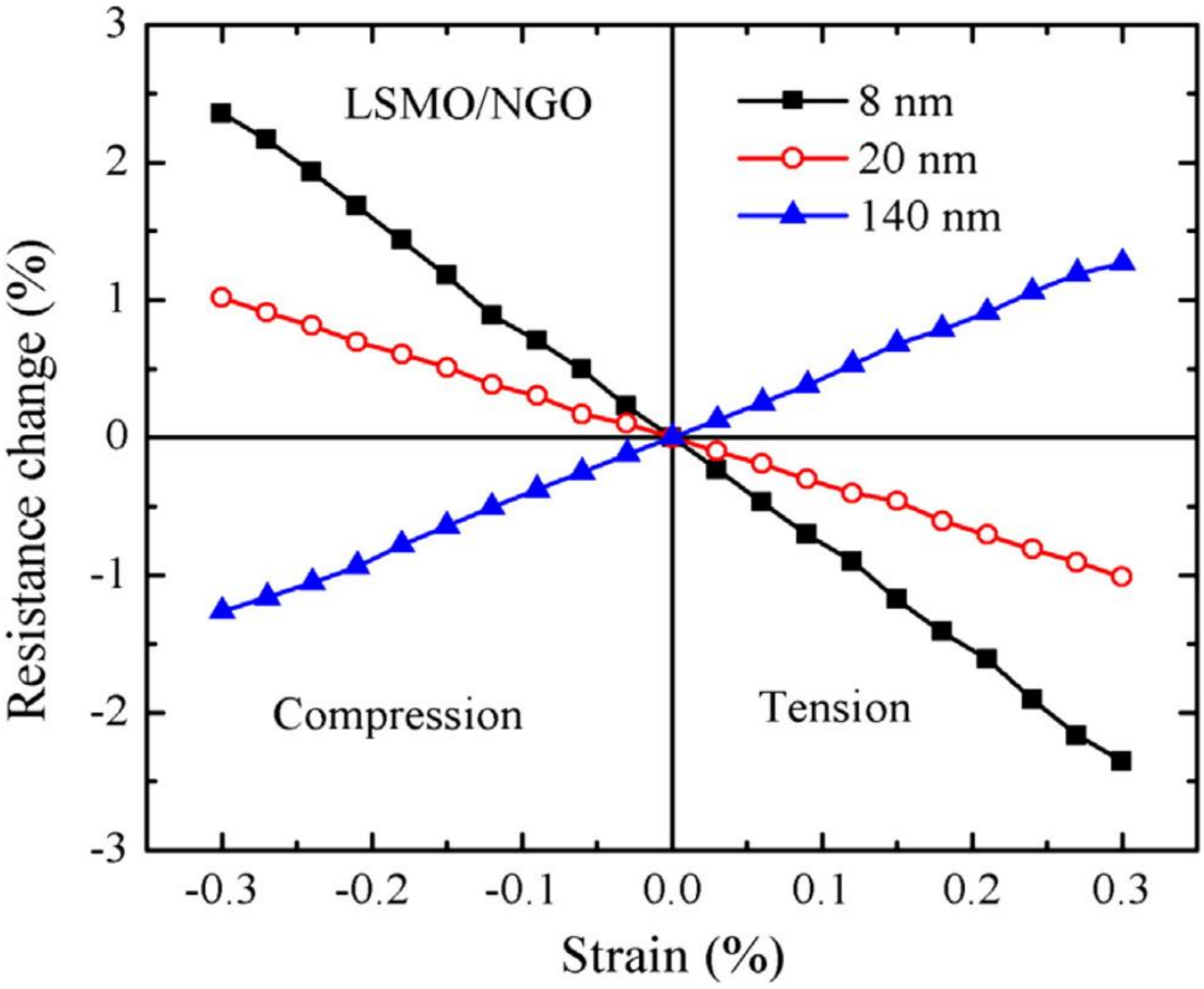

Fig. 23: The resistance change $(R_{str} - R)/R$ of LSMO films of different thicknesses grown on NGO substrates when subjected to external uniaxial compression and tension along the [100] direction and measured at 290 K. Here $R_{str}$ is the resistance of the film under stress; R is the resistance when it was not subjected to external stress. The resistance change exhibited linear dependence on the strain in all the investigated directions for both the compressive and tensile strains. From [413].

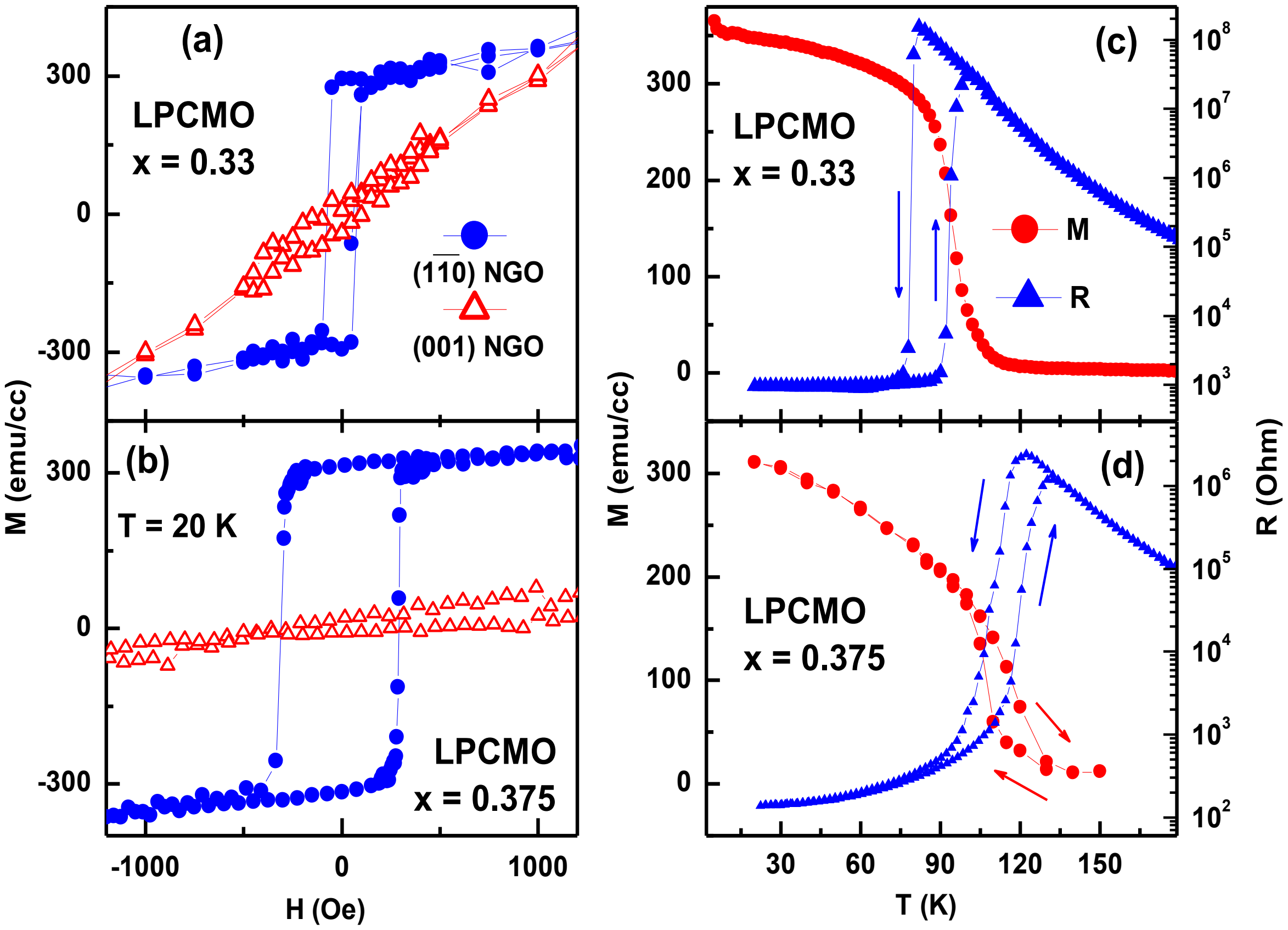

Fig. 24: $M$ ($H$) curves along two in-plane perpendiculars (001) and (110) directions of NGO substrate for LCMO films with $y = 0.6$ and (a) $x = 0.33$ and (b) $x = 0.375$ on (110) NGO single crystal. Adapted from [39]. (c) and (d) represent the $M(T)$ and $R(T)$ curves as a function of temperature for these LPCMO films. $R(T)$ curves show a prominent hysteresis in temperature whereas $M$ ($T$) shows small hysteresis only for LPCMO film with x = 0.375.

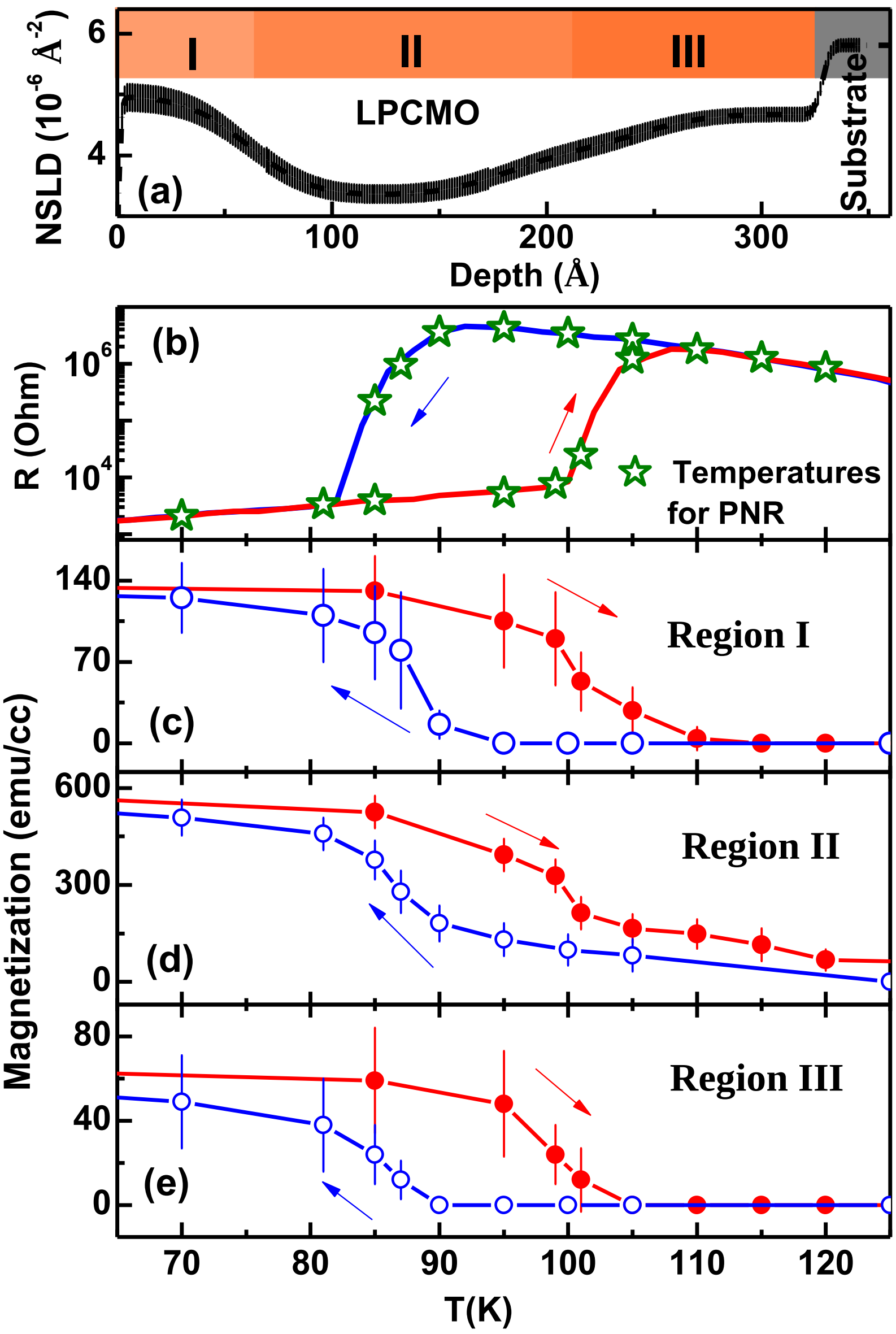

Fig. 25: (a) Nuclear scattering length density (NSLD) depth profile obtained from PNR data for the LPCMO ($x$ = 0.33 and $y$ = 0.60) film (thickness ~ 325 Å) grown on (110) NGO single crystal substrate. PNR data suggested a variation in NSLD (change in chemical composition) at the interfacial region and a three layers structure was obtained as shown in the inset of (a) as regions I (air-film interface), II (film bulk) and III (substrate film interface). (b) $R(T)$ data from the film while cooling (blue line) and warming (red line). The open stars (green) represent the value of temperatures at which PNR measurements were carried out for determining the temperature-dependent magnetization depth profile of the film. Magnetization at different temperatures on cooling (blue open circles) and warming (red closed circles) cycles for three different regions along the thickness of the film, e.g., regions I (c), II (d), and III (e). Adapted from [39]

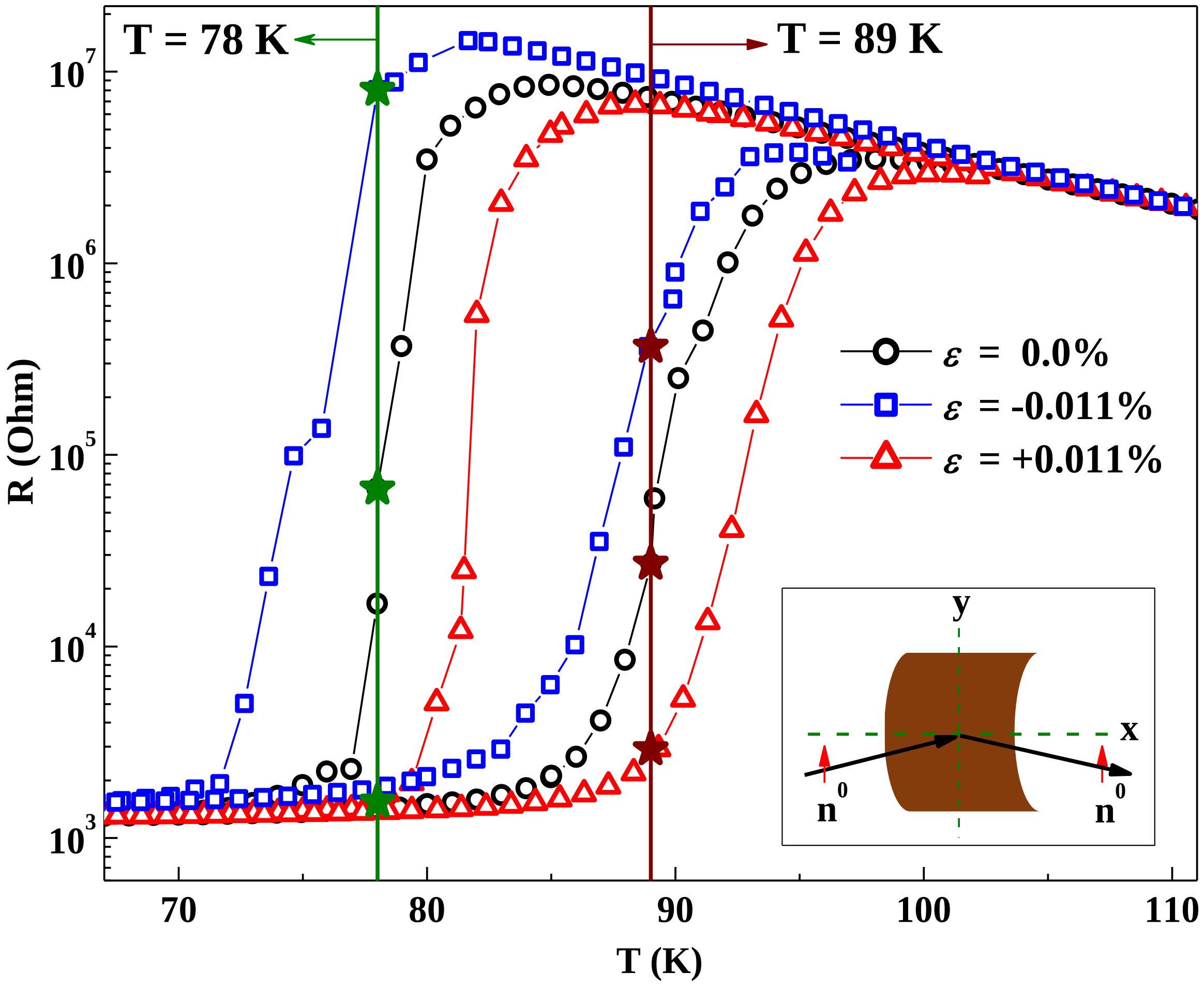

Fig. 26: (a) Transport data for LPCMO (*x* = 0.33 and *y* = 0.60) grown on (110) NGO substrate on the application of applied bending strain ($\varepsilon$) of 0% (without strain), +0.011% (tensile strain) and -0.011% (compressive strain). Adapted from [85]. Transport data were measured simultaneously with PNR measurements. PNR data were measured near MIT (T = 78 K while cooling and T = 89 K while warming and shown as a vertical line in (a)) for different cases of applied bending strain. The closed star symbols at 78 and 89 K for three cases of strain where PNR measurements were carried out. Inset show the schematic of the axis of bending of the LPCMO film and neutron reflectivity measurement. Transport measurements were performed along the y-axis which is the easy axis for LPCMO film.

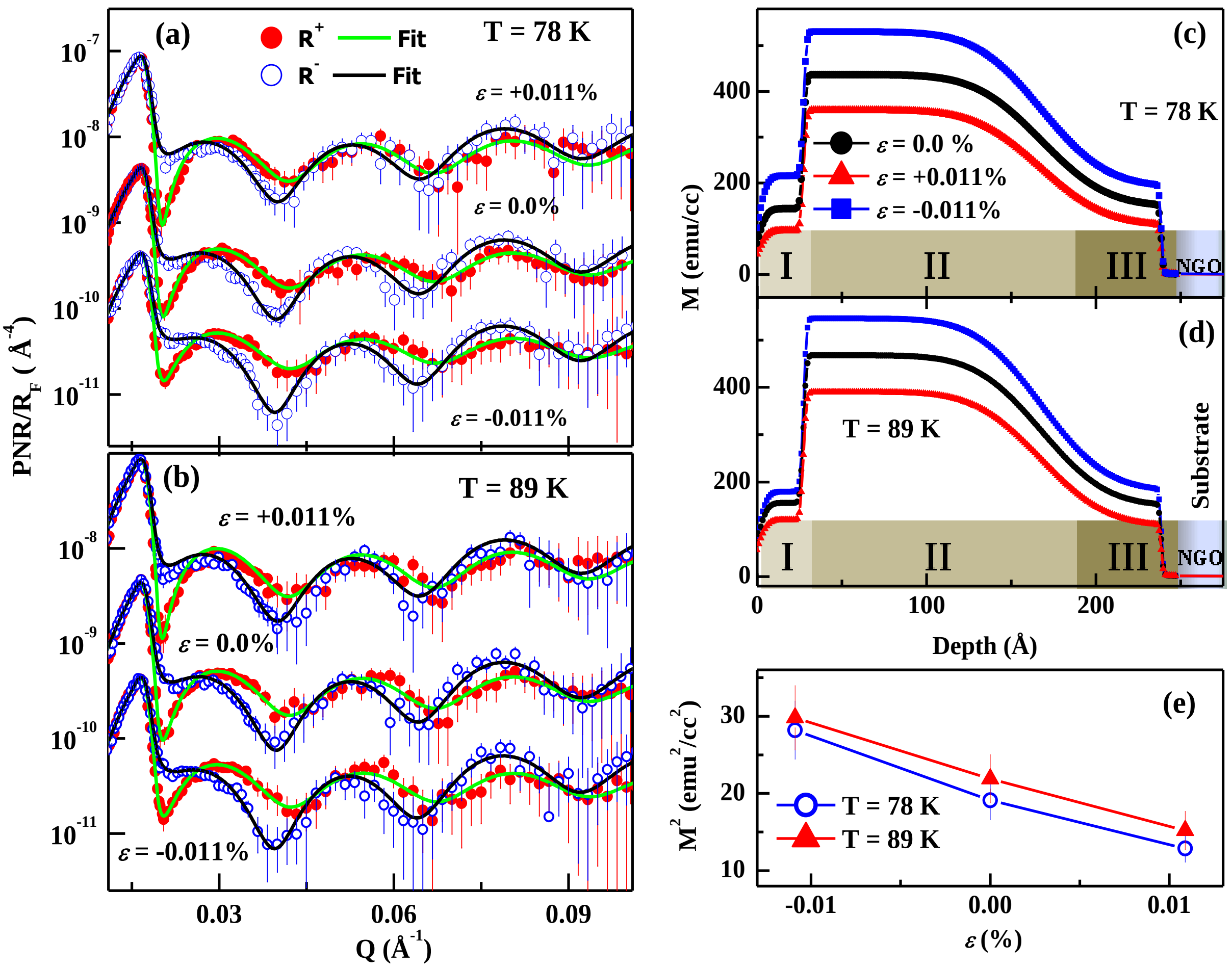

Fig. 27: PNR data for the LPCMO (x = 0.33 and y = 0.60) film on the application of applied bending strain ($\varepsilon$) of 0% (without strain), +0.011% (tensile strain) and -0.011% (compressive strain) at 78 K (a) and 89 K (b), during cooling and warming cycle, respectively (see Fig. 25 (a)). Magnetization depth profiles for the LPCMO film at T = 78 K (c) and 89 K (d) for different cases of strain as obtained from PNR data. $M^2$ vs $\varepsilon$ curve for LPCMO film (region II), which was used to estimate the coupling coefficient. Adapted from [85].

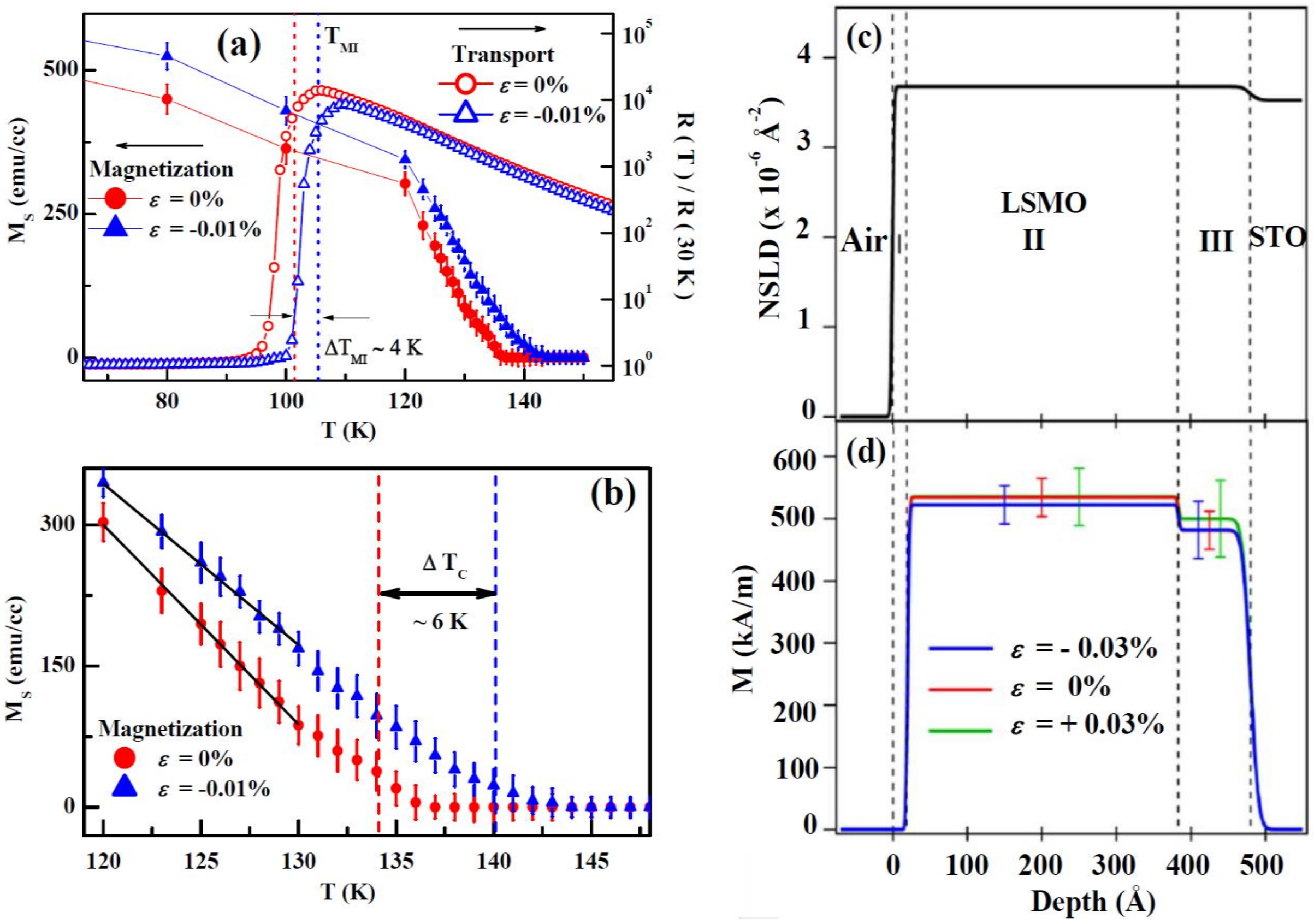

Fig. 28: $M_s$ ($T$) curves obtained from the polarized neutron reflectivity for LPCMO ($x$ = 0.33 and $y$ = 0.6) film under an applied bending strain ($\varepsilon$) of -0.01% (compressive strain) and 0% (without strain). Superimposed is the resistance of the film measured during the neutron experiment without (red circles) and with (blue triangles) applied bending stress. The dotted lines correspond to the metal-insulator transitions during warming for the two states of stress. (b) Shows the $M_s$ ($T$) close to the ordering temperature. Extrapolation of linear fits of $M_s$ ($T$) to $M_s$ = 0 yields estimates for $T_c$. Adapted from [86]. (c) The nuclear scattering length density (NSLD) depth profile of LSMO ($x$ = 0.3) film (thickness ~ 500 Å) grown on the STO substrate. (d) the magnetization depth profile of LSMO film under the influence of applied bending stress. From [88].

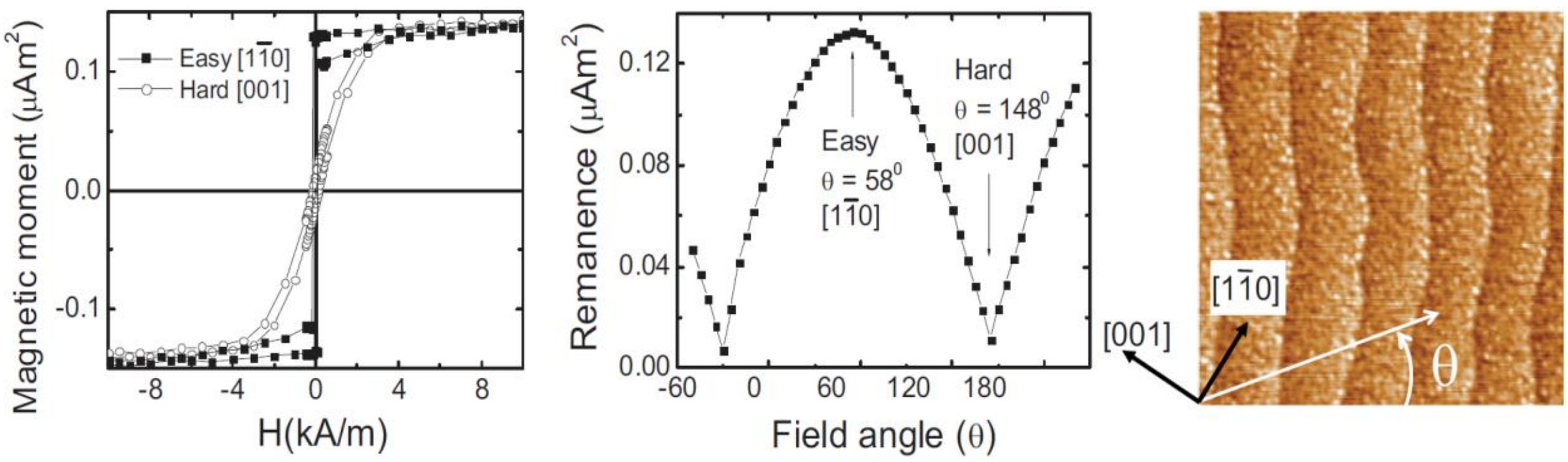

Fig. 29: (a) Magnetic moment as a function of applied magnetic field *H* at room temperature for an LSMO ($x$ = 0.3) film (thickness = 180 Å) grown on (110) NGO, measured along the two in-plane crystal directions [(001) and (1$\bar{1}$0)] as indicated. (b) Remanence magnetic moment as a function of in-plane field angle for the same film. (c) AFM image (1 μm ×1 μm) of the same film, with the two in-plane crystal directions indicated by black arrows, and the angle $\theta$ at which the magnetic field is applied with respect to the edge of the substrate, as indicated in white. From [420].

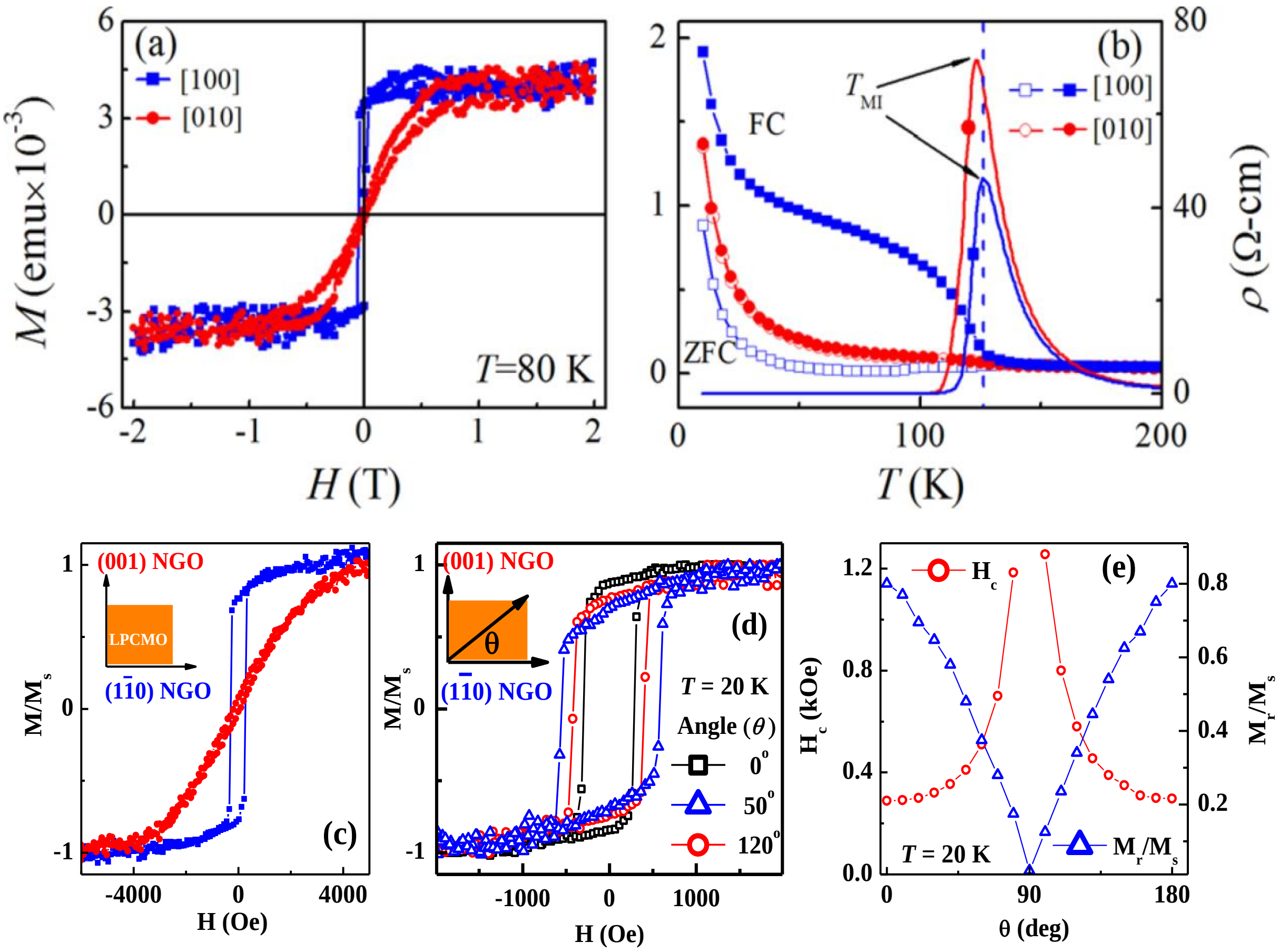

Fig. 30: (a) Hysteresis loops [M(H) curves] of the PCSMO ($x$ = 0.35 and $y$ = 0.3) film, grown on (110) NGO substrate, measured along the two in-plane directions at T = 80 K. (b) Temperature dependence of magnetization (left axis, for both the field-cooled and the zero-field cooled conditions) and resistivity (right axis) measured along the two in-plane directions of the PCSMO film. From [422]. (c) Reduced M(H) curves for the LPCMO film ($x$ = 0.375 and $y$ = 0.5) at 20 K measured by applying an in-plane magnetic field (H) along the (001) and (1$\bar{1}$0) NGO directions, suggesting (1$\bar{1}$0) NGO is an easy axis. (d) Reduced M(H) curves at 20 K along different angles from the in-plane easy axis. The inset shows the direction of the angle at which the field was applied. (e) Variation of $H_c$ and $M_r/M_s$, at 20 K with the in-plane angle of applied filed from the easy axis. Adapted from [35].

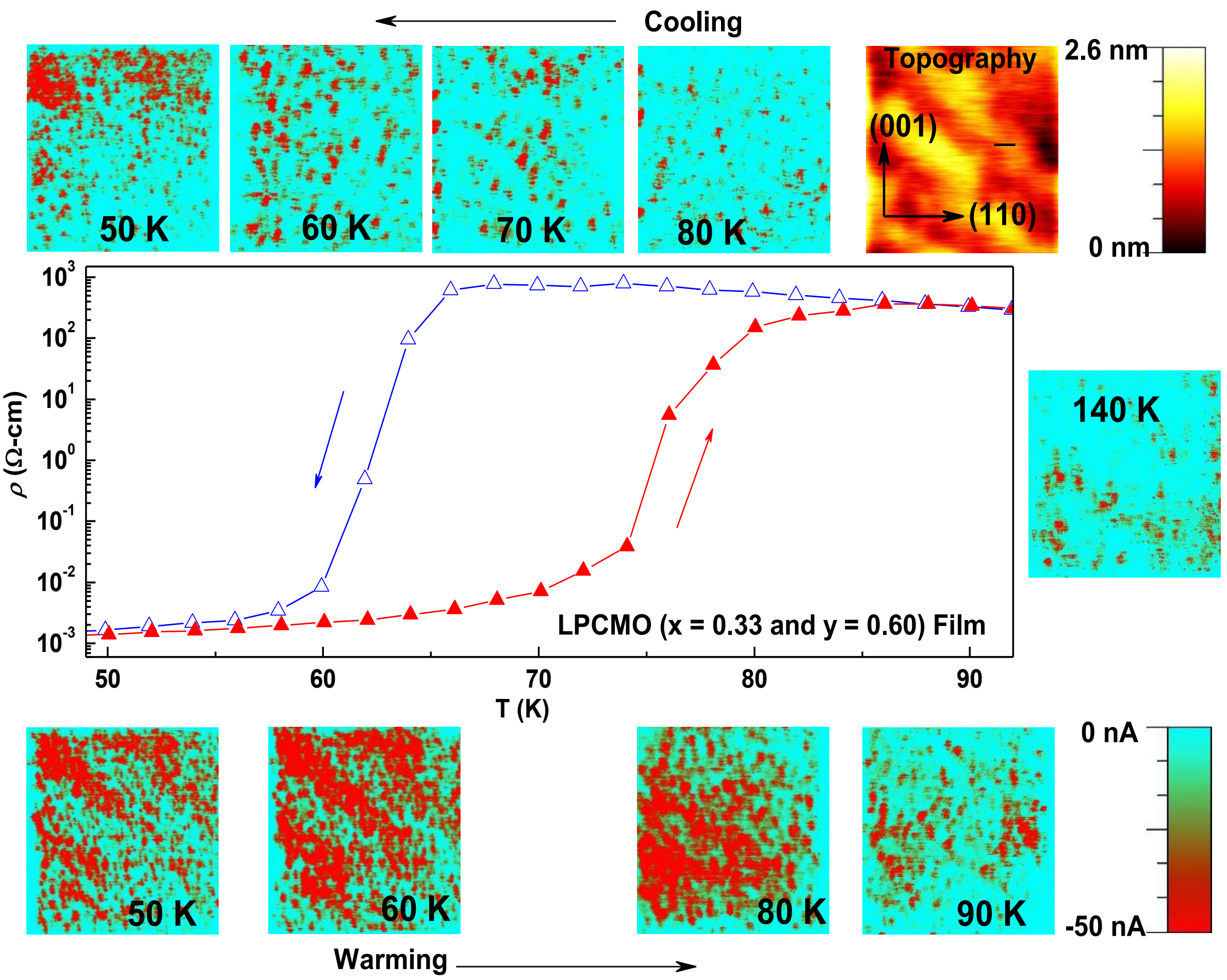

Fig. 31: The conducting atomic force microscope (cAFM) images of conductivity at different temperatures while cooling (upper panel) and warming (lower panel) the LPCMO ($x$ = 0.33, $y$ = 0.60) film grown on (110) NGO substrate. The upper panel also shows a topographic image of the film at 140 K along with the two in-plane perpendicular directions ($1\bar{1}0$) and (001) of NGO substrate. The topography of the film remains similar across the temperature scan. The graph in the middle panel shows the transport measurement from the film. The coexistence of metallic (conducting regions in red) and non-metallic (insulating region in cyan) phase across the MIT is evident. Adapted from [26].

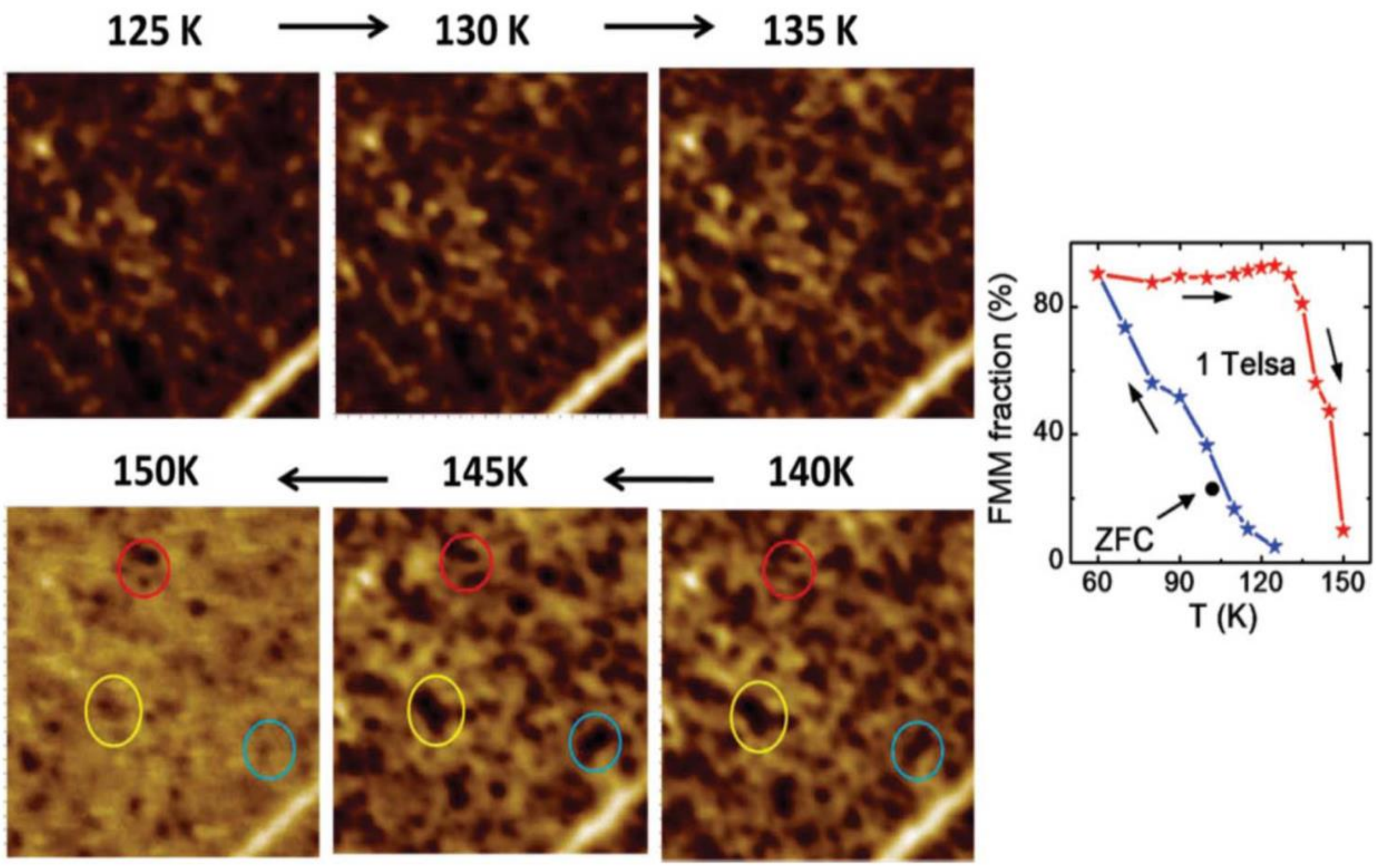

Fig. 32: MFM images (5 μm × 5 μm) of LPCMO ($x$ = 0.375 and $y$ = 0.72) film on NGO substrate at different temperatures in a magnetic field of 1 T taken during the warming of the film after field cooling to 60 K. Phase coexistence of FMM (dark) and AFI (light) phases at different temperatures shows an increase in AFI phase at a higher temperature. The scale for images from 125 K to 150 K is 40, 39, 30, 21, 9, and 4 degrees, respectively. The FMM phase fraction was estimated from MFM images collected during 1 T field cooling (blue star) and subsequent warming (red star), and at 102 K/1 T after zero-field cooling (black circle). From [429].

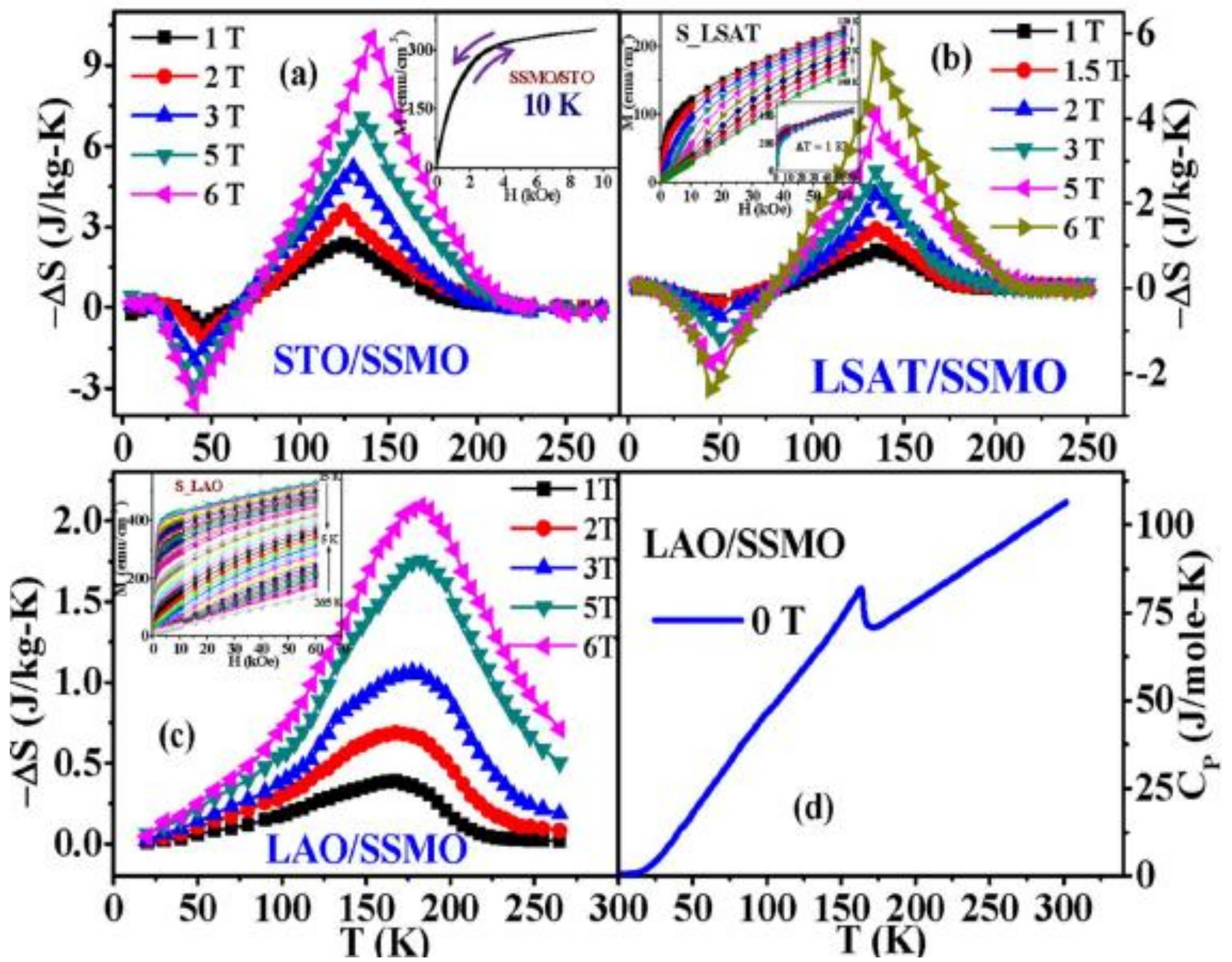

Fig. 33: Temperature-dependent change in magnetic entropy -|Δ$S$| at different magnetic fields calculated from M-H data for (a) STO/SSMO, (b) LSAT/SSMO, and (c) LAO/SSMO heterostructures. Insets of (a), (b), and (c) show the corresponding $M$ ($H$) plots. (d) $C_P(T)$ curve at H = 0 T for LAO/SSMO heterostructure. From [116].

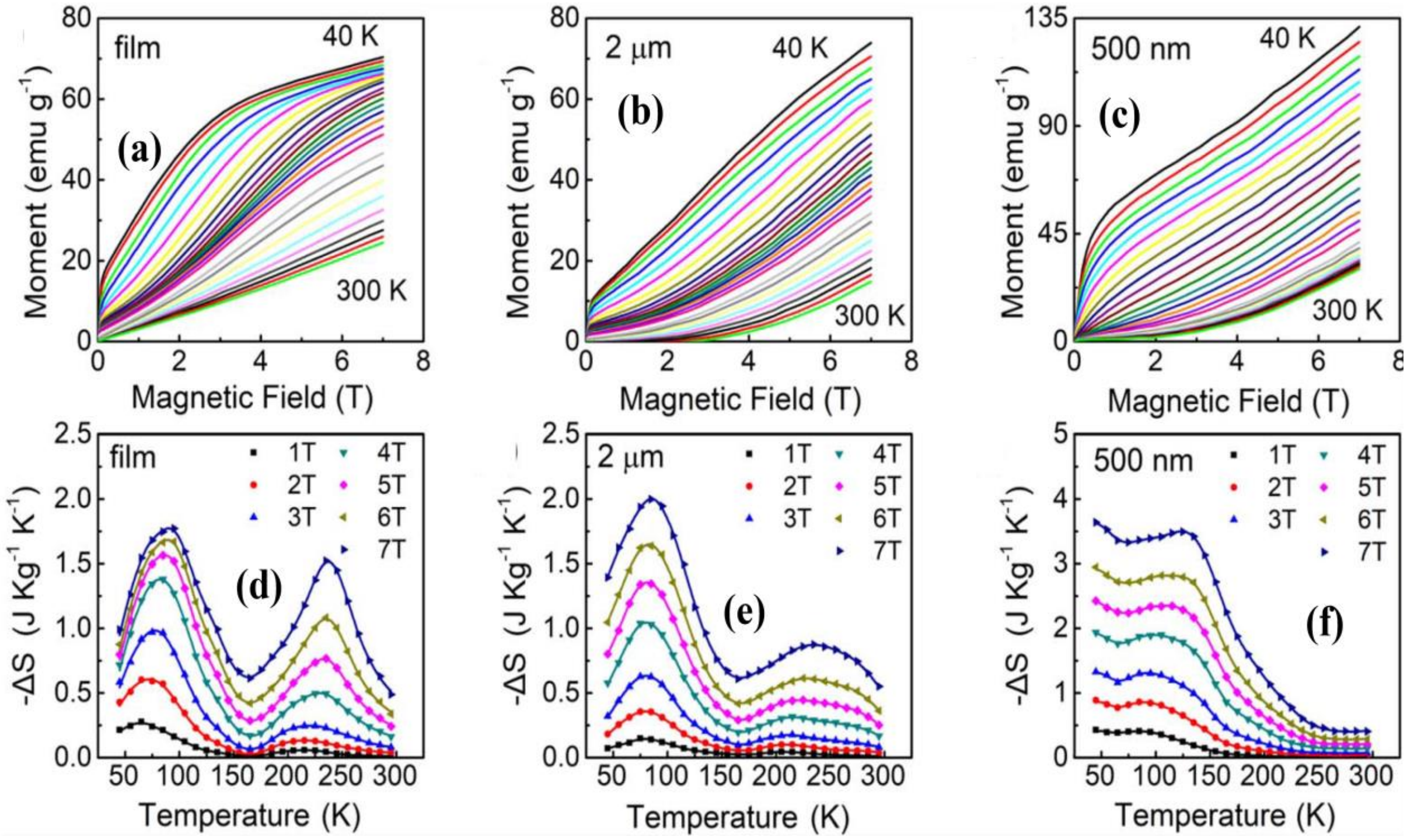

Fig. 34: (a) - (c) M (H) curves measured in the temperature range of 40–300 K with a 10 K interval for the LPCMO ($x$ = 0.375 and $y$ = 0.32) film, 2-μm disk array, and the 500-nm disk array of LPCMO, respectively. (d)–(f) depicts temperature-dependent magnetic entropy changes (−ΔS) at different magnetic fields for the LPCMO film, 2-μm disk array, and 500-nm disk array of LPCMO, respectively. The results suggest that the MCE in LPCMO can be tuned by changing the size of nanodisk arrays through standard nanofabrication. From [448].

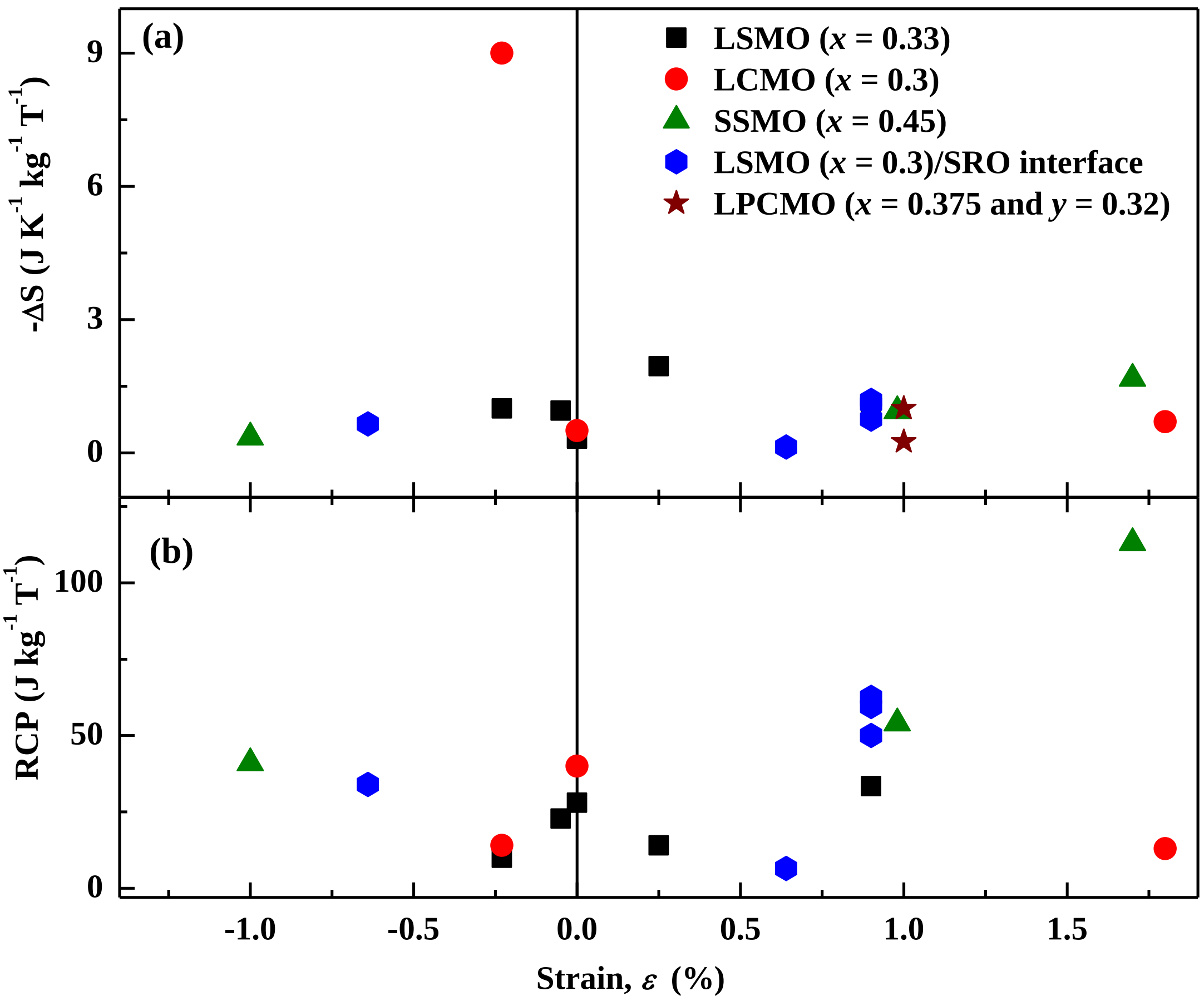

Fig. 35: Effect of strain on the change of magnetic entropy (ΔS) (a) and relative cooling power (RCP) of the manganite thin films, grown on different substrates. Data from [115, 116, 438, 441-448].